# A Prediction Divergence Criterion for Model Selection

Stéphane Guerrier[¶] & Maria-Pia Victoria-Feser[†]

[¶]Department of Statistics
University of Illinois at Urbana-Champaign, USA
[†]Research Center for Statistics
Geneva School of Economics and Management
University of Geneva, Switzerland

**Abstract**

The problem of model selection is inevitable in an increasingly large number of applications involving partial theoretical knowledge and vast amounts of information, like in medicine, biology or economics. The associated techniques are intended to determine which variables are "important" to "explain" a phenomenon under investigation. The terms "important" and "explain" can have very different meanings according to the context and, in fact, model selection can be applied to any situation where one tries to balance variability with complexity. In this paper, we introduce a new class of error measures and of model selection criteria, to which many well know selection criteria belong. Moreover, this class enables us to derive a novel criterion, based on a divergence measure between the predictions produced by two nested models, called the Prediction Divergence Criterion (PDC). Our selection procedure is developed for linear regression models, but has the potential to be extended to other models. We demonstrate that, under some regularity conditions, it is asymptotically loss efficient and can also be consistent. In the linear case, the PDC is a counterpart to Mallow's $C_p$ but with a lower asymptotic probability of overfitting. In a case study and by means of simulations, the PDC is shown to be particularly well suited in "sparse" settings with correlated covariates which we believe to be common in real applications.

## 1 Introduction

Model selection is an important and challenging problem in statistics. Indeed, it becomes unavoidable in more and more applications involving incomplete theoretical knowledge about the phenomenon under investigation and important amounts of available information, like in medicine, biology, economics, etc. Very often model selection is about choosing among a set of predictors, the subset that best predicts or explains a response variable.

In general, in regression problems, the task of model selection is performed using either stepwise approaches or through penalized estimation based methods. The former optimizes a criterion such as for example the AIC on a given sequence of models. The latter produces simultaneously a sequence on which a criterion is employed to select a model, the lasso being a popular example. More details on model selection methods are provided a literature review given in Appendix A.

One of the main challenges in model selection is the development of simple and rapid techniques that can deal with large data sets especially in sparse settings, i.e. when only a small minority of the variables at hand are significant predictors. This is for example the case in areas such as genetics or economics and the available information can now easily contain millions of observations and hundreds of thousands of potential predictors. Even moderately large dataset can lead to very large ones when considering for instance all potential interactions. In this context, stepwise techniques can often provide improvements in term of computational speed and/or prediction accuracy (see e.g. Foster and Stine, 2004; Lin et al., 2011). Moreover, a stepwise approach is also often considered as more suitable when there is a natural sequence of potential predictors. Such a sequence can arise from an priori theoretical insight and is the one of interest for the scientist, like in experimental settings where main effects are more important than interactions (for mixed models, see e.g. Müller et al., 2013). Another example comes from natural sciences were a sequence of models corresponds to a sequence of null hypotheses of interest that need to be tested and upon which a model selection approach is used (see e.g. Johnson and Omland, 2004).

In this paper, we propose a new class of model selection criteria that are suited for stepwise approaches or can be used as selection criteria in penalized estimation based methods. This new class, called the $d$-class of error measure, generalizes Efron's $q$-class (see Efron, 1986). This class not only contains classical criteria such as Mallow's $C_p$ or the AIC, but also enables one to define new criteria that are more general. Within this new class, we propose a model selection criterion based on a prediction divergence between two nested models' predictions that we call the Prediction Divergence Criterion (PDC). The PDC provides a different measure of prediction error than a criterion, say $C$, associated to each potential model within a sequence and for which the selection decision is based on the sign of $\Delta C$. The PDC directly measures the prediction error divergence between two nested models and provides different criteria than $\Delta C$. As an example, we consider linear regression models and propose a PDC criterion that is the direct counterpart of Mallow's $C_p$. We show that a selection procedure based on the PDC, compared to the $C_p$, has a smaller probability of overfitting and a negligible asymptotic probability of selecting a larger model for models with more than one additional non significant covariate (which is not the case for selection procedures based on the $C_p$).

In practice, the PCD can be applied to a known or estimated sequence of models to select candidate models. Compared to other criteria used in stepwise approaches the PDC appears to often perform better in finite samples, especially in sparse settings. Both in simulations and a real example we find that the PDC tends to select far smaller models with generally better or comparable out-of-sample predictive performances. Compared to penalized estimation based methods, the PDC, even based on a rather simple ordering rule, appears to selects at least as many of the significant variables but with far fewer of the non-significant ones hence providing smaller models. In terms of prediction accuracy the PDC seems to achieve comparable results to adaptive penalized estimation based methods in sparse settings.

This point is illustrated through the analysis of a dataset on childhood malnutrition in Zambia in Section 2. We compare the performance of several model selection procedures and conclude that different procedures can lead to very different selected models, which in practice rises the problem of choosing which model selection procedure to use. The simulation study in Section 5, which compares the finite sample performance of the PDC's estimator with other model selection techniques (including adaptive ones), reveals that in sparse and correlated covariates settings, the PDC tends to select as many as the significant variables than the other best performing methods, but far less of the noisy ones, with comparable out-of-sample prediction errors. This leads then to conclude, for the analyzed dataset, that the PDC is the most reliable selection criterion that provides the smallest model including most of the explanatory variables.

The reminder of the paper is organized as follows. In Section 3 we introduce the $d$-class of error measures which generalizes Efron's $q$-class and we derive the associated optimism theorem. We also present a new class of model selection criteria, which is based on the $d$-class of error measures (i.e. the PDC), and derive a corresponding unbiased estimator. In Appendix C, for the linear model and $L_2$ divergence, we compare the PDC to Mallows' $C_p$ and also show that the latter is a particular case of the the PDC class. In Section 4, we consider the PDC for variable selection in linear regression models and derive the asymptotic properties, namely the probabilities of overfitting, consistency, asymptotic loss efficiency and the Signal to Noise Ration (SNR) (see e.g. McQuarrie and Tsai, 1998). We also propose, for these models, a suitable sequence of nested models that we use in our simulation exercise. The use of the PDC to other models such as smoothing splines or for the order of autoregressive models can be found in Guerrier (2013).

# 2 An Illustrative Example: Childhood Malnutrition in Zambia

Childhood malnutrition is considered to be one of the worst health problems in developing countries (e.g. United Nations Children's Fund, 1998). Both a manifestation and a cause of poverty, malnutrition is thought to contribute to over a third of death in children under five years old globally (United Nations Children's Fund, 2012). Moreover, it is well established in the medical literature that maternal and child undernutrition have considerable consequences for adult health and human capital (see e.g. Victora et al., 2008 and the references therein). Such conditions are, for example, associated with less schooling, reduced economic productivity, and for women lower offspring birthweight. It has also been reported that lower birthweight and undernutrition in childhood have an influence on cancer occurrence and are risk factors for high glucose concentrations, blood pressure, and harmful lipid profiles.

The evolution of (childhood) malnutrition is monitored through regular demographic and health surveys conducted by Macro International in cooperation with the World Health Organization. These data have been made publicly available and can be obtained from `www.measuredhs.com`. We will consider here a dataset obtained from a demographic and health survey conducted in Zambia in 2007 (see Zambia DHS, 2007 for details).

Undernutrition is generally assessed by comparing anthropometric indicators such as height or weight at a certain age to a reference population. A well established measurement for the study of acute malnutrition is given by (see World Health Organisation, 1995 for details):

$$Y_i = \frac{H_{i,j} - \mu_j}{\sigma_j} \tag{1}$$

TABLE 1: *Childhood Malnutrition in Zambia:* MSPE *and* MAPE *obtained by* 10*-fold cross-validation as defined in (2) for the models selected by the lasso, the adaptive lasso (a-lasso), MCP, SCAD, the forward stepwise AIC, BIC and PDC. The numbers in parentheses are standard error estimates obtained by bootstrap with* $B = 500$ *resampling. NbReg denotes the number of parameters in each model.* $n = 1927$

| | Without interactions ($p = 31$) | | | With interactions ($p = 496$) | | |
|---|---|---|---|---|---|---|
| | MAPE | MSPE | NbReg | MAPE | MSPE | NbReg |
| AIC | 1.19 (0.02) | 2.52 (0.12) | 12 | 1.21 (0.03) | 2.58 (0.16) | 32 |
| BIC | 1.21 (0.04) | 2.61 (0.19) | 8 | 1.19 (0.02) | 2.44 (0.13) | 9 |
| lasso | 1.19 (0.03) | 2.50 (0.14) | 17 | 1.30 (0.04) | 2.87 (0.25) | 61 |
| a-lasso | 1.18 (0.03) | 2.48 (0.16) | 13 | 1.18 (0.02) | 2.42 (0.06) | 14 |
| MCP | 1.20 (0.03) | 2.55 (0.08) | 18 | - | - | - |
| SCAD | 1.19 (0.03) | 2.55 (0.09) | 19 | - | - | - |
| PDC | 1.20 (0.02) | 2.54 (0.06) | 3 | 1.17 (0.03) | 2.32 (0.08) | 5 |

where $H_{i,j}$, $\mu_j$ and $\sigma_j$ denote, respectively, the height of the $i^{\text{th}}$ child at age $j$, the median height of a child of the same age in the reference population and the associated standard deviation. Several variables are assumed to have a determinant influence on undernutrition. These are generally linked to education, income, and nutritional situation of the parents, access to clean water and sanitation, and primary health care, and immunization facilities (see e.g. Madise et al., 1999). Therefore, taking into account the available information in the considered dataset, we selected the covariates given in Appendix B to explain $Y_i$.

After removing missing data, the dataset contains $n = 1927$ observations and $p = 31$ covariates. In a first step, we compare the performance of the lasso, the adaptive lasso, MCP, SCAD, and the forward stepwise AIC, BIC (see Appendix A for more detailed explanations) and PDC (based on the $L_2$ divergence, with $\lambda_n = 2$ and using the ordering rule given in (25), see Sections 3 and 4). The selection criteria as well as the specifications for the lasso and adaptive methods are presented in Appendix F. The selected models are compared using the mean squared and median absolute error of prediction (MSPE and MAPE), as measured by 10-fold cross-validation. That is, we split the data into ten roughly equal-sized parts and for each part, we carry out model selection using the other nine parts of the data and calculate the MSPE and MAPE of the chosen model when predicting the $k^{\text{th}}$ part of the data, i.e.

$$\begin{aligned} \text{MSPE} &= \frac{1}{10}\sum_{k=1}^{10} ||\boldsymbol{y}^{(k)} - \hat{\boldsymbol{y}}^{(k)}||_2^2 \\ \text{MAPE} &= \frac{1}{10}\sum_{k=1}^{10} ||\boldsymbol{y}^{(k)} - \hat{\boldsymbol{y}}^{(k)}||_1 \end{aligned} \tag{2}$$

where $|| \cdot ||_p$ denotes the $L_p$ norm, $\boldsymbol{y}^{(k)}$ the $k^{\text{th}}$ split of the vector of observation $\boldsymbol{y}$ and $\hat{\boldsymbol{y}}^{(k)}$ the prediction of $\boldsymbol{y}^{(k)}$ made without $\boldsymbol{y}^{(k)}$. For all methods, the data were split in the same way. In a second analysis, we include in the model all possible first order interactions between the regressors, hence leading to $p = 496$ potential regressors.

In Table 1 the MSPE and MAPE for the different selection methods are presented. One first notices that the performances, as measured by the MSPE or by the MAPE, are the same across the different selection procedures, for both performance measures. We also see that the PDC is the selection procedure producing overall and by far the smallest model. Regularization methods need to included many more variables in the model to achieve the same prediction accuracy as the PDC and other stepwise procedures considered here. With the addition of interactions as potential explanatory variables, two adaptive methods fail in providing a selected model (MCP and SCAD).

However, the analysis of a real example does not necessarily provide the whole story. In particular, it is important for the model selection procedures not to miss important information (i.e. significant variables). In the simulation study provided in Section 5, we find that with correlated covariates and in sparse settings, the PDC (with a simple ordering rule), compared to other methods, not only chooses at least comparable proportions of the significant variables, but also selects less of the noisy ones.

# 3 The $d$-Class Error Measures and an Associated Prediction Divergence Criterion

In this section, we develop a general framework that provides model selection criterion estimators in a rather straightforward manner. This framework generalizes Efron's $q$-class of error measures (see Efron, 1986).

Consider a random variable $Y$ distributed according to model $F_{\boldsymbol{\theta}}$, possibly conditionally on a set of fixed covariates $\mathbf{x} = [x_1 \dots x_p]$. We observe a random sample $\boldsymbol{Y} = (Y_i)_{i=1,\dots,n}$ supposedly generated from $F_{\boldsymbol{\theta}}$, possibly together with a non-random $n \times K, K \geq p$ matrix of inputs $\mathbf{X}$. Given a prediction function $\hat{\boldsymbol{Y}}$ that depends on the chosen model, Efron (1986) uses a function $Q(\cdot,\cdot)$ based on the $q$-class error measure to build an out-of-sample "performance" measure of a given prediction rule. The $q$-class of error measures is given by

$$Q(u, v) = q(v) + \dot{q}(v)(u - v) - q(u)$$

where $\dot{q}(v)$ is the derivative of $q(\cdot)$ evaluated at $v$. The particular choice of $q(u) = u(1 - u)$ gives the squared loss function $Q(u, v) = (u - v)^2$. The prediction error measure is quantified by the (out-of-sample) expected prediction error

$$\text{EPErr} = \frac{1}{n}\sum_{i=1}^{n} \text{EPErr}_i \quad \text{where} \quad \text{EPErr}_i = \mathbb{E}\left[\mathbb{E}_0\left[Q(Y_i^0, \hat{Y}_i)|\boldsymbol{Y}\right]\right] \tag{3}$$

with $\boldsymbol{Y}^0 = (Y_i^0)_{i=1,\dots,n}$ a random variable distributed as $\boldsymbol{Y}$, which can be interpreted as an out-of-sample version of $\boldsymbol{Y}$. As throughout this article, $\mathbb{E}[\cdot]$ and $\mathbb{E}_0[\cdot]$, denote expectations under the distribution of $Y_i|\boldsymbol{x}_i$, respectively $Y_i^0|\boldsymbol{x}_i$, the correct model. The expectations are, depending on the context, simple or multiple.

Efron's optimism theorem (see Efron, 2004) demonstrates that

$$\text{EPErr}_i = \mathbb{E}\left[\mathbb{E}_0\left[Q(Y_i^0, \hat{Y}_i)|\boldsymbol{y}\right]\right] = \mathbb{E}\left[Q(Y_i, \hat{Y}_i) + \Omega_i\right]$$

with $\Omega_i = \text{cov}\left(\dot{q}(\hat{Y}_i), Y_i\right)$. Hence, an estimator of EPErr is obtained as

$$\widehat{\text{EPErr}} = \frac{1}{n}\sum_{i=1}^{n}\left(Q(y_i, \hat{y}_i) + \widehat{\text{cov}}\left(\dot{q}(\hat{Y}_i), Y_i\right)\right) \tag{4}$$

where, depending on the distribution of $Y_i|\mathbf{x}_i$, $\widehat{\text{cov}}(\cdot,\cdot)$ is obtained analytically up to a value of $\boldsymbol{\theta}$, the model's parameters, which is then replaced by $\hat{\boldsymbol{\theta}}$, or by resampling methods (see e.g. Efron, 2004).

We may extend this methodology and the $q$-class of error measures by changing $Q(Y_i^0, \hat{Y}_i)$ in (3) to a more general class of error measures $D(\cdot,\cdot)$ between two equidimentional vector valued functions $\boldsymbol{g}_1\left(\boldsymbol{Y}^0, \hat{\boldsymbol{\theta}}_1\right)$ and $\boldsymbol{g}_2\left(\boldsymbol{Y}, \hat{\boldsymbol{\theta}}_2\right)$ where $\hat{\boldsymbol{\theta}}_1$ and $\hat{\boldsymbol{\theta}}_2$ denote the estimated parameter vectors associated, respectively, to the models $F_{\boldsymbol{\theta}_1}$ and $F_{\boldsymbol{\theta}_2}$. Such a criterion can be defined without loss of generality as

$$\text{C} = \mathbb{E}\left[\mathbb{E}_0\left[D\left(\boldsymbol{g}_1\left(\boldsymbol{Y}^0, \hat{\boldsymbol{\theta}}_1\right), \boldsymbol{g}_2\left(\boldsymbol{Y}, \hat{\boldsymbol{\theta}}_2\right)\right)\right]\right] \tag{5}$$

where the expectation is multidimensional. The divergence $D(\boldsymbol{g}_1(\boldsymbol{Y}^0, \hat{\boldsymbol{\theta}}_1), \boldsymbol{g}_2(\boldsymbol{Y}, \hat{\boldsymbol{\theta}}_2))$ is said to belong to the *d-class of error measures* if the function $D(\cdot,\cdot)$ is a valid Bregman divergence and if the functions $\boldsymbol{g}_j(\hat{\boldsymbol{\theta}}_j, \boldsymbol{Y})$ are equidimensional and associated, respectively, to the models $F_{\boldsymbol{\theta}_j}$, $j = 1, 2$. In some sense, Bregman divergences are the multivariate equivalent of Efron's $q$-class (see Bregman, 1967 for more details). This divergence encompasses squared error, relative entropy, logistic loss, Mahalanobis distance and other error measures. It has often been used as loss function in the context of model selection (see e.g. Zhang, 2008; Zhang et al., 2009, 2010). The Bregman divergence between two equidimensional vectors $\mathbf{u}$ and $\mathbf{v}$ (of the same dimension) is defined as

$$D(\mathbf{u}, \mathbf{v}) = \psi(\mathbf{u}) - \psi(\mathbf{v}) - (\mathbf{u} - \mathbf{v})^T \nabla\psi(\mathbf{v}) \tag{6}$$

where $\psi(\cdot)$ is a scalar and $\nabla\psi(\mathbf{v})$ represents the gradient vector of $\psi(\cdot)$ evaluated at $\mathbf{v}$. The function $\psi(\cdot)$ is strictly convex and differentiable. For example, a squared loss function $D(\mathbf{u}, \mathbf{v}) = ||\mathbf{u} - \mathbf{v}||_2^2$ is obtained when $\psi(\mathbf{v}) = \mathbf{v}^T\mathbf{v}$.

Based on the $d$-class of error measure and for the criterion defined in (5) we can derive the following "optimism" theorem whose proof is presented in Appendix D.1.

Theorem 1: *Let the divergence $D(\cdot,\cdot)$ be a valid d-class error measure based on $\psi(\cdot)$ and assume that*

$$\mathbb{E}\left[\left(\boldsymbol{g}_1\left(\boldsymbol{Y}, \hat{\boldsymbol{\theta}}_1\right) - \mathbb{E}_0\left[\boldsymbol{g}_1\left(\boldsymbol{Y}^0, \hat{\boldsymbol{\theta}}_1\right)\right]\right)^T \left(\boldsymbol{g}_1\left(\boldsymbol{Y}, \hat{\boldsymbol{\theta}}_1\right) - \mathbb{E}_0\left[\boldsymbol{g}_1\left(\boldsymbol{Y}^0, \hat{\boldsymbol{\theta}}_1\right)\right]\right)\right] < \infty$$

$$\mathbb{E}\left[\left(\nabla\psi\left(\boldsymbol{g}_2\left(\boldsymbol{Y}, \hat{\boldsymbol{\theta}}_2\right)\right)\right)^T \left(\nabla\psi\left(\boldsymbol{g}_2\left(\boldsymbol{Y}, \hat{\boldsymbol{\theta}}_2\right)\right)\right)\right] < \infty.$$

*Then*

$$\begin{aligned}\mathbb{E}\left[\mathbb{E}_0\left[D\left(\boldsymbol{g}_1\left(\boldsymbol{Y}^0, \hat{\boldsymbol{\theta}}_1\right), \boldsymbol{g}_2\left(\boldsymbol{Y}, \hat{\boldsymbol{\theta}}_2\right)\right)\right]\right] &= \mathbb{E}\left[D\left(\boldsymbol{g}_1\left(\boldsymbol{Y}, \hat{\boldsymbol{\theta}}_1\right), \boldsymbol{g}_2\left(\boldsymbol{Y}, \hat{\boldsymbol{\theta}}_2\right)\right)\right] \\ &\quad + \text{tr}\left\{\text{cov}\left[\boldsymbol{g}_1\left(\boldsymbol{Y}, \hat{\boldsymbol{\theta}}_1\right), \nabla\psi\left(\boldsymbol{g}_2\left(\boldsymbol{Y}, \hat{\boldsymbol{\theta}}_2\right)\right)\right]\right\}.\end{aligned}$$

The direct consequence of Theorem 1 is that for any criterion C as defined in (5) one can construct an estimator, say $\widehat{\mathrm{C}}$ similarly to (4) for the $q$-class of error measures, using (3). Indeed a "natural" (and unbiased) estimator of C is

$$\widehat{\mathrm{C}} = D\left(\boldsymbol{g}_1\left(\boldsymbol{y}, \hat{\boldsymbol{\theta}}_1\right), \boldsymbol{g}_2\left(\boldsymbol{y}, \hat{\boldsymbol{\theta}}_2\right)\right) + \mathrm{tr}\left\{\widehat{\mathrm{cov}}\left[\boldsymbol{g}_1\left(\boldsymbol{y}, \hat{\boldsymbol{\theta}}_1\right), \nabla\psi\left(\boldsymbol{g}_2\left(\boldsymbol{y}, \hat{\boldsymbol{\theta}}_2\right)\right)\right]\right\} \tag{7}$$

where as in (4), depending on the distribution of $\boldsymbol{Y}|\mathbf{X}$, $\widehat{\mathrm{cov}}(\cdot,\cdot)$ is obtained analytically up to a value of $\boldsymbol{\theta}_1$ and $\boldsymbol{\theta}_2$, the models' parameters, which are then replaced by $\hat{\boldsymbol{\theta}}_1$ and $\hat{\boldsymbol{\theta}}_2$, or by resampling methods (see e.g. Efron, 2004). We shall refer to the first and second terms of (7) as the *apparent divergence* and the *divergence optimism*, respectively.

The class (7) encompasses many selection criteria such as Mallows' $C_p$ which is presented as an example in Appendix C together with meaningful properties a criterion should satisfy for the task of model selection.

As a model selection criterion to choose between two models, say $\mathcal{M}_j$ nested in $\mathcal{M}_k$, we propose to consider the class C in (5) with $\boldsymbol{g}_1\left(\boldsymbol{Y}^0, \hat{\boldsymbol{\theta}}_j\right) = \hat{\boldsymbol{Y}}_j^0$ and $\boldsymbol{g}_1\left(\boldsymbol{Y}, \hat{\boldsymbol{\theta}}_k\right) = \hat{\boldsymbol{Y}}_k$. This class *directly* compares the out-of-sample prediction computed in the smaller model $\hat{\boldsymbol{Y}}_j^0$ with the in-sample prediction in the larger model $\hat{\boldsymbol{Y}}_k$, quantified by the Bregman divergence $D(\cdot,\cdot)$ (based on $\psi(\cdot)$).

More formally, a criterion that compares the prediction divergence between two models belongs to the following class:

$$\mathrm{PDC}_{j,k} = \mathbb{E}\left[\mathbb{E}_0\left[D\left(\hat{\boldsymbol{Y}}_j^0, \hat{\boldsymbol{Y}}_k\right)|\boldsymbol{Y}\right]\right] \tag{8}$$

where the expectation is multidimensional. By using Theorem 1 as in (7) we obtain $D(\hat{\boldsymbol{y}}_j, \hat{\boldsymbol{y}}_k) + \mathrm{tr}\left\{\widehat{\mathrm{cov}}\left[\hat{\boldsymbol{y}}_j, \nabla\psi(\hat{\boldsymbol{y}}_k)\right]\right\}$ as an unbiased estimator of the $\mathrm{PDC}_{j,k}$. Although this estimator can be computed analytically or using resampling methods for any valid Bregman divergence, we shall only consider here the squared loss function (the same as the one for the $C_p$) to get

$$\widehat{\mathrm{PDC}}_{j,k} = ||\hat{\boldsymbol{y}}_j - \hat{\boldsymbol{y}}_k||_2^2 + 2\,\mathrm{tr}\left[\widehat{\mathrm{cov}}(\hat{\boldsymbol{y}}_j, \hat{\boldsymbol{y}}_k)\right]. \tag{9}$$

In some sense, $\widehat{\mathrm{PDC}}_{j,k}$ is comparable (but not equal) to Mallow's $C_p$ in the PDC class since both criteria are based on the same loss function. It will thus be of particular interest to compare (9) with the $C_p$ to understand the differences between the PDC approach and the classical model selection approach. This comparison is presented in Section 4 in the context of the linear regression model.

The PDC has an intuitive interpretation. Indeed, if the smaller model is not correct, the additional elements in the larger model create differences in the predictions and therefore should be accounted for. For a "suitable" sequence of nested models with increasing complexity and such that model $\mathcal{M}_j$ is nested in model $\mathcal{M}_{j+1}$, $\widehat{\mathrm{PDC}}_{j,j+1}$ is expected to be minimal when $j = \jmath_0 < K$, $K$ being the number of potential sequentially nested models, and $\mathcal{M}_{\jmath_0}$ denotes the correct (or closest to the correct) model. Indeed, while $j < \jmath_0$, we expect $\widehat{\mathrm{PDC}}_{j,j+1}$ to be relatively large since model $\mathcal{M}_j$ is missing some elements of the correct model $\mathcal{M}_{\jmath_0}$ which are included in model $\mathcal{M}_{j+1}$. This is also true with $\widehat{\mathrm{PDC}}_{j,j+m}$, $j < j+m \leq K$ or $\widehat{\mathrm{PDC}}_{j-m,j}$, $1 < j-m < j$. On the other hand, if $j \geq \jmath_0$, $\widehat{\mathrm{PDC}}_{j,j+1}$ (or indeed $\widehat{\mathrm{PDC}}_{j,j+m}, m > 0$) is relatively small compared to when $j < \jmath_0$ because both models include the correct one. Among all models $j \geq \jmath_0$, $\widehat{\text{PDC}}_{j,j+m}$ should be minimised at $j = \jmath_0$ and $m = 1$ since $\widehat{\text{PDC}}_{\jmath_0,\jmath_0+1}$ compares the prediction of the correct model with the least overfitted one.

In the case of the linear regression model, we derive in Section 4 the (asymptotic) properties of the $\widehat{\text{PDC}}_{j,j+1}$. In particular, we show in Theorem 2 that for sufficiently large sample size $n$ we have that $\mathbb{E}\left[\widehat{\text{PDC}}_{\jmath_0,\jmath_0+1}\right] \leq \mathbb{E}\left[\widehat{\text{PDC}}_{j,j+m}\right]$ for $j$ and $m$ such that $0 < j < K$, $m > 0$ and $j + m \leq K + 1$. We also have that $\mathbb{E}\left[\widehat{\text{PDC}}_{\jmath_0,\jmath_0+1}\right] = \mathbb{E}\left[\widehat{\text{PDC}}_{j,j+m}\right]$ if and only if $j = \jmath_0$ and $m = 1$. This confirms the intuitive explanation given above.

Many authors (see e.g. Bhansali and Downham, 1977) have examined the penalty function of the AIC (and of other criteria) and defined, for example, the $\text{AIC}_\alpha$ in which the term 2 (see Table A-5 in Appendix F) of the conventional AIC is replaced by $\alpha$. We follow this strategy and define

$$\widehat{\text{PDC}}^{\lambda_n}_{j,k} = ||\hat{\boldsymbol{y}}_j - \hat{\boldsymbol{y}}_k||_2^2 + \lambda_n \operatorname{tr}\left[\widehat{\text{cov}}\left(\hat{\boldsymbol{y}}_j, \hat{\boldsymbol{y}}_k\right)\right] \tag{10}$$

where $\lambda_n$ is a constant depending possibly on the sample size $n$.

Hence, assuming that there exist $K$ competing nested models (and that the largest model is not the correct one) for describing the behavior of $\boldsymbol{Y}$, we propose to choose the model $\mathcal{M}_{\hat{\jmath}_{\lambda_n}}$ satisfying

$$\hat{\jmath}_{\lambda_n} = \underset{j=1,\dots,K-1}{\operatorname{argmin}} \ \widehat{\text{PDC}}^{\lambda_n}_{j,j+1} \,. \tag{11}$$

If a clear sequence of competing nested models does not exist, one can build one prior to applying the selection rule (11). This will be explained when treating the linear regression model in Section 4.4 (in particular see iterative rule (25)).

# 4 Stepwise Forward Selection for the Linear Regression Model

We consider in this section the model given in the following setting.

Setting A: *Consider a random variable $Y$ distributed according to a model that generates random samples $\boldsymbol{y} = (y_i)_{i=1,\dots,n}$ through*

$$\boldsymbol{y} = \boldsymbol{\mu} + \boldsymbol{\varepsilon},\ \boldsymbol{\varepsilon} \sim \mathcal{N}(\mathbf{0}, \sigma_\varepsilon^2 \mathbf{I}),\ 0 < \sigma_\varepsilon^2 < \infty \tag{12}$$

*In the linear regression setting, we also observe a non-random $n \times K$ full rank matrix of inputs $\mathbf{X}$, such that $\lim_{n\to\infty} 1/n\mathbf{X}^T\mathbf{X} = \mathbf{M}$ with $\mathbf{M}$ a positive definite matrix. We try to approximate (12) by means of the linear model*

$$\boldsymbol{y} = \mathbf{X}\boldsymbol{\beta} + \boldsymbol{\varepsilon},\ \boldsymbol{\varepsilon} \sim \mathcal{N}(\mathbf{0}, \sigma_\varepsilon^2 \mathbf{I}),\ 0 < \sigma_\varepsilon^2 < \infty \tag{13}$$

*with $\boldsymbol{\theta} = \boldsymbol{\beta} \in \mathcal{B} \subseteq \mathbb{R}^K$. We actually consider in this setting the case where the covariates are "ordered". We hence rewrite $\boldsymbol{\beta} := \boldsymbol{\beta}_j = [\beta_1, ..., \beta_j, 0, ..., 0]$ and define $\boldsymbol{\beta}_0 = [0, ..., 0]$. The columns of $\mathbf{X}$ are ordered accordingly.*

*The Least Squares Estimator (LSE) of $\boldsymbol{\beta}$ is given by*

$$\hat{\boldsymbol{\beta}} = \underset{\boldsymbol{\beta}\in\mathcal{B}}{\operatorname{argmin}} \; ||\boldsymbol{y} - \mathbf{X}\boldsymbol{\beta}||_2^2 \,. \tag{14}$$

*For each possible model, say $\mathcal{M}^o$ constructed from (13), we associate (sample) predictions $\hat{\boldsymbol{y}}^o = \mathbf{S}^o\boldsymbol{y}$ where $\mathbf{S}^o = \mathbf{X}^o((\mathbf{X}^o)^T\mathbf{X}^o)^{-1}(\mathbf{X}^o)^T$ denotes the "hat" matrix of model $\mathcal{M}^o$ in which $\mathbf{X}^o$ is a column's subset of $\mathbf{X}$.*

The $\widehat{\text{PDC}}_{j,k}^{\lambda_n}$ defined in (10) can be simplified for two linear nested candidate models. Indeed, let model $\mathcal{M}_j$ be nested within model $\mathcal{M}_k$ and so that $\dim(\boldsymbol{\beta}_j) = j < \dim(\boldsymbol{\beta}_k) = k$. Then, the $\widehat{\text{PDC}}_{j,k}^{\lambda_n}$ based on the squared loss function defined by $\psi(\mathbf{z}) = \mathbf{z}^T\mathbf{z}$, for $\lambda_n = 2$ (then $\widehat{\text{PDC}}_{j,k}^{2} \equiv \widehat{\text{PDC}}_{j,k}$), is equal to

$$\widehat{\text{PDC}}_{j,k} = ||\hat{\boldsymbol{Y}}_j - \hat{\boldsymbol{Y}}_k||_2^2 + 2\sigma_\varepsilon^2 \operatorname{tr}(\mathbf{S}_j\mathbf{S}_k) = ||\hat{\boldsymbol{Y}}_j - \hat{\boldsymbol{Y}}_k||_2^2 + 2\sigma_\varepsilon^2 j \,\cdot \tag{15}$$

Note that (15) is not equal (or proportional) to the difference in Mallows' $C_p$ computed at respectively models $\mathcal{M}_j$ and $\mathcal{M}_k$.

We obtain the modified version $\widehat{\text{PDC}}_{j,k}^{\lambda_n}$ as:

$$\widehat{\text{PDC}}_{j,k}^{\lambda_n} = ||\hat{\boldsymbol{y}}_j - \hat{\boldsymbol{y}}_k||_2^2 + \lambda_n\sigma_\varepsilon^2 j. \tag{16}$$

When the value of $\sigma_\varepsilon^2$ is unknown, one may replace $\sigma_\varepsilon^2$ by a consistent estimator, say $\hat{\sigma}_\varepsilon^2$, like the LSE at the full (possible) model. In Theorem 2 below, we show that $\widehat{\text{PDC}}_{j,j+m}^{\lambda_n}$ is expected, for sufficiently large sample size, to reach its smallest value for $j = \jmath_0$ and $m = 1$, where $\jmath_0$ is the number of covariates in the correct underlying model $\mathcal{M}_{\jmath_0}$. This motivates the selection rule defined in (11). The proof of Theorem 2 is presented in Appendix D.2.

Theorem 2: *We assume Setting A and that $\sigma_\varepsilon^2$ is either known or estimated by a consistent estimator, say $\hat{\sigma}_\varepsilon^2$. Then, for $j$ and $m$ such that $0 < j < K$, $m > 0$ and $j + m \leq K + 1$ we have that for sufficiently large $n$*

$$\mathbb{E}\left[\widehat{\text{PDC}}_{\jmath_0,\jmath_0+1}\right] \leq \mathbb{E}\left[\widehat{\text{PDC}}_{j,j+m}\right] \tag{17}$$

*We also have that $\mathbb{E}\left[\widehat{\text{PDC}}_{\jmath_0,\jmath_0+1}\right] = \mathbb{E}\left[\widehat{\text{PDC}}_{j,j+m}\right]$ if and only if $j = \jmath_0$ and $m = 1$.*

Given Setting A, the selection problem hence amounts at selecting the number $j$. Let the class of models $\mathscr{J}$ correspond to all possible models that can be constructed. Then we define the "best" model $\jmath_0 \in \mathscr{J}$ as

$$\jmath_0 = \max\left\{ j \in \{1, ..., K\} \;:\; \underset{n\to\infty}{\operatorname{plim}} \left|\hat{\beta}_j^\star\right| > 0 \right\} \tag{18}$$

where $\hat{\beta}_j^\star$ denotes the $j^{\text{th}}$ element of $\hat{\boldsymbol{\beta}}_K$. We also define the class of models $\mathscr{J}_0$ which "includes" $\jmath_0$ and we say that model $j \in \mathscr{J}_0$ if $j \geq \jmath_0$.

Remark A: *The definition of $\jmath_0$ in (18) may not always be a suitable definition of the model we should select. Indeed, in some situations it could be possible that some elements of the vector $\boldsymbol{\beta}$* *are decreasing functions of $n$ which vanish as $n \to \infty$. Therefore, in such a setting, (18) should not be employed and a more complex definition of $\jmath_0$ should be used. However, in this article we will assume that none of the elements of $\boldsymbol{\beta}$ depend on the sample size $n$.*

Suppose that $\jmath_0$ is estimated using (16) together with the selection rule given in (11), we obtain

$$\hat{\jmath}_0 = \underset{0 \le j \le K-1}{\text{argmin}} \; \widehat{\text{PDC}}^{\lambda_n}_{j,j+1} \; \text{ where } \; \widehat{\text{PDC}}^{\lambda_n}_{j,j+1} = \|\hat{\boldsymbol{y}}_j - \hat{\boldsymbol{y}}_{j+1}\|_2^2 + \lambda_n j \sigma_\varepsilon^2. \tag{19}$$

To derive the asymptotic property of the selection procedure given in (19), we will consider the following assumptions:

(A.1) When $\sigma_\varepsilon^2$ is unknown it can be replaced by a consistent estimator $\hat{\sigma}_\varepsilon^2$.

(A.2) The scalars $\lambda_n$ and $\jmath_0$ are such that $\lambda_n \, \jmath_0 = \mathcal{O}\left(\sqrt{n}\right)$ and $\lambda_n > 0$.

Remark B: *Assumption (A.1) will be employed in nearly all the results considered in this article. In practice, a popular choice of $\hat{\sigma}_\varepsilon^2$ is given by $\tilde{\sigma}_\star^2$ (i.e. the LSE at the full model) which is consistent under model (12) with $\boldsymbol{\mu} = \mathbf{X}\boldsymbol{\beta}_{\jmath_0}$. However, this estimator is not consistent if $\boldsymbol{\mu} \neq \mathbf{X}\boldsymbol{\beta}_{\jmath_0}$. An alternative could for instance be to use of the nearest neighbor method (see e.g. Stone, 1977). A more detailed discussion on this matter can be found in Section 4 of Shao (1997).*

Remark C: *Assumption (A.2) imposes some conditions on $\lambda_n$ (as presented in (19)) and on $\jmath_0$. In most situations, it is quite reasonable to assume that $\jmath_0$ does not depend on the sample size $n$. However, this is not always the case (see e.g. the "one-mean versus k-mean" example of Shao (1997)). The penalty term $\lambda_n$ can also be an increasing function of $n$ and we will see in Theorem 5 that this is a required condition for the selection procedure (19) to be consistent. However the product of $\lambda_n$ and $\jmath_0$ can never increase at a faster rate than $\sqrt{n}$.*

In the following subsections we study in turn the probabilities of underfitting and overfitting as well as the conditions for consistency and asymptotic loss efficiency for $\widehat{\text{PDC}}^{\lambda_n}_{j,k}$. We also study the SNR of $\widehat{\text{PDC}}^{\lambda_n}_{j,k}$ and compare it to other model section criteria in Theorem A-8 given in Appendix E.

## 4.1 Underfitting

The least we can expect from any reliable model selection procedure is that for sufficiently large $n$ the probability of selecting a model $\hat{\jmath}$ that does not belong to $\mathscr{J}_0$ (i.e. the probability of underfitting) tends to zero. Under reasonable conditions (see e.g. Remark A) almost all model selection criteria such as the AIC, Mallow's $C_p$ or the BIC have a nil asymptotic probability of underfitting. Theorem 3 examines the asymptotic probability of selecting a model that does not belong to $\mathscr{J}_0$ using $\hat{\jmath}_{\lambda_n}$ as defined in (19). The proof of Theorem 3 is presented in Appendix D.3.

Theorem 3: *Let $j \notin \mathscr{J}_0$, then under Setting A and Assumptions: (A.1) and (A.2) we have that*

$$\lim_{n \to \infty} \Pr\left(\hat{\jmath}_{\lambda_n} \notin \mathscr{J}_0\right) = 0.$$

## 4.2 Probability of Overfitting

The next theorem (whose proof is presented in Appendix D.4) determines the asymptotic probability of overfitting in the case where we only consider choosing between models $\jmath_0$ and $j \in \mathscr{J}_0 \setminus \{\jmath_0\}$. It is based on the variance-gamma distribution (see e.g. Kotz et al., 2001) which is described in Appendix G. In this situation, model $\jmath_0$ will be selected if the following inequality is true:

$$||\hat{\boldsymbol{y}}_{\jmath_0} - \hat{\boldsymbol{y}}_{\jmath_0+m}||_2^2 + \lambda_n \jmath_0 \sigma^2 \leq ||\hat{\boldsymbol{y}}_j - \hat{\boldsymbol{y}}_{j+m}||_2^2 + \lambda_n j \sigma^2 \tag{20}$$

where $m = j - \jmath_0$ and $j \in \mathscr{J}_0 \setminus \{\jmath_0\}$. In general, we only consider the case where $m = 1$ as in the selection rule given in (19), but we provide here the results for any $m \in \mathbb{N}^+$ such that $j + m \leq K + 1$.

Theorem 4: *Let $j \in \mathscr{J}_0 \setminus \{\jmath_0\}$ and let $\mathcal{G}_m$ be a random variable following a modified variance-gamma distribution with parameter $m = j - \jmath_0$. We assume that $j + m \leq K + 1$ and under Setting A and Assumption (A.1) we have that*

$$\lim_{n\to\infty} \Pr\left(\widehat{\text{PDC}}_{j,j+m}^{\lambda_n} \leq \widehat{\text{PDC}}_{\jmath_0,\jmath_0+m}^{\lambda_n}\right) = \lim_{n\to\infty} \Pr\left(\mathcal{G}_m \leq \lambda_n\, m\right).$$

Remark D: *In the case where $\sigma_\varepsilon^2$ is known the result of Theorem 4 is no longer asymptotic and the limit can therefore be removed. This directly follows from (A-25) in Appendix D.4.*

Theorem 4 enables to compare the asymptotic probabilities of overfitting by $m$ variables of $\widehat{\text{PDC}}_{j,j+1}^{\lambda_n}$ with other approaches such as Mallow's $C_p$. The results are presented in Table 2 and are based on the derivation presented in McQuarrie and Tsai (1998, Section 2.5). It can be observed that $\widehat{\text{PDC}}_{j,j+1}$ (i.e. $\lambda_n = 2$ resulting from Theorem 1) will overfit less than the $C_p$ (and other asymptotically equivalent methods such as the AIC or the FPE) but more than the AICu. Clearly, $\widehat{\text{PDC}}_{j,j+1}^{\lambda_n}$ based on a $\lambda_n$ that tends to infinity with $n$ has, similarly to the BIC, HQ and HQc, nil asymptotic probabilities of overfitting.

In Appendix D.5 we derive the asymptotic probability of overfitting for $\widehat{\text{PDC}}_{j,j+1}^{\lambda_n}$ (see Theorem A-7). Woodroofe (1982) and Zhang (1992) showed that the $C_p$ (and asymptotically equivalent methods) are such that

$$\lim_{n\to\infty} \Pr\left(\hat{\jmath}_{C_p} = \jmath_0\right) \geq \lim_{K-\jmath_0\to\infty}\ \lim_{n\to\infty} \Pr\left(\hat{\jmath}_{C_p} = \jmath_0\right) \approx 0.712. \tag{21}$$

For $\widehat{\text{PDC}}_{j,j+1}$, the counterpart of Mallow's $C_p$ in the PDC class , we have that

$$\lim_{n\to\infty} \Pr\left(\hat{\jmath}_2 = \jmath_0\right) \geq \lim_{K-\jmath_0\to\infty}\ \lim_{n\to\infty} \Pr\left(\hat{\jmath}_2 = \jmath_0\right) \approx 0.894.$$

The $\widehat{\text{PDC}}_{j,j+1}$ has therefore a larger probability of selecting the "best" model compared to the $C_p$ or the AIC. Figure 1 presents the limiting overfitting probabilities for Mallow's $C_p$ and $\widehat{\text{PDC}}_{j,j+1}$. It can be observed that unlike the $C_p$, $\widehat{\text{PDC}}_{j,j+1}$ is not much affected by the number of models larger than $\jmath_0$ (i.e. $K - \jmath_0$) and tends to select less overfitted models. Indeed, when there is only one model larger than $\jmath_0$, $\widehat{\text{PDC}}_{j,j+1}$ adds on average about 0.10 additional variables in the model, and when the number of models tends to infinity, the number of additional variables only increases to about 0.11. Moreover, regardless of the number of candidate models, the asymptotic probability for $\widehat{\mathrm{PDC}}_{j,j+1}$ to select a model with three or more additional variables is virtually nil. From Corollary A-3 in Appendix D.6 (see Remark G), we show that $\widehat{\mathrm{PDC}}_{j,j+1}$ selects on average 0.11 more variables, while for the $C_p$ (and the AIC) this number is 0.946 (see also Woodroofe, 1982 and Zhang, 1992 for details).

TABLE 2: *Asymptotic probabilities of overfitting by m variables. Probabilities refer to selecting one particular overfitting model over the true model. Values for the $C_p$ and the* AICu *are taken from McQuarrie and Tsai (1998, Table 2.3). Note that the $C_p$ is asymptotically equivalent to the* AIC, *the* AICc *and the* FPE. *The* AICu *is asymptotically equivalent to the* FPEu. *The values for $\widehat{\mathrm{PDC}}_{j,j+1}$ are computed using Theorem 4 and the integration of (A-39) using (A-40).*

| $m$ | $C_p$ | AICu | $\widehat{\mathrm{PDC}}_{j,j+1}$ |
|---|---|---|---|
| 1 | 15.73 % | 8.33 % | 10.45 % |
| 2 | 13.53 % | 4.98 % | 6.77 % |
| 3 | 11.16 % | 2.93 % | 4.30 % |
| 4 | 9.16 % | 1.74 % | 2.75 % |
| 5 | 7.52 % | 1.04 % | 1.77 % |
| 6 | 6.20 % | 0.62 % | 1.15 % |
| 7 | 5.12 % | 0.38 % | 0.75 % |
| 8 | 4.24 % | 0.23 % | 0.49 % |
| 9 | 3.52 % | 0.14 % | 0.32 % |
| 10 | 2.93 % | 0.09 % | 0.21 % |

## 4.3 Consistency and Asymptotic Loss Efficiency

A selection procedure which uses $\hat{\jmath}$ to estimate $\jmath_0$ is said to be *consistent* if

$$\Pr\left(\hat{\jmath} = \jmath_0\right) \xrightarrow{\mathcal{P}} 1. \tag{22}$$

Note that (22) implies $\Pr\left(L_n\left(\hat{\jmath}\right) = L_n\left(\jmath_0\right)\right) \xrightarrow{\mathcal{P}} 1$ where

$$L_n\left(j\right) = \frac{||\boldsymbol{\mu} - \hat{\boldsymbol{y}}_j||_2^2}{n}$$

the squared error loss for a given $\mathcal{M}_j$ producing a response prediction $\hat{\boldsymbol{y}}_j$ and $\boldsymbol{\mu}$ is the vector of true expected response $\mathbb{E}[\boldsymbol{Y}]$ (see Setting A).

When the true model is finite, the BIC is consistent (see e.g. Haughton, 1988), while the AIC (and the $C_p$ which are asymptotically equivalent as shown e.g. in Nishii, 1984) have a non-nil (asymptotic) probability of overfitting. In finite samples, consistency is not necessarily a good property since consistent selection criteria may tend to underfit, so that the chosen models could have larger prediction errors.

In some cases, a selection procedure does not fulfil (22) but $\hat{\jmath}$ is still "close" to $\jmath_0$ in the following sense that is weaker than (22):

$$\frac{L_n\left(\hat{\jmath}\right)}{L_n\left(\jmath_0\right)} \xrightarrow{\mathcal{P}} 1. \tag{23}$$

A selection procedure satisfying (23) is said to be *asymptotically loss efficient* (see Shao, 1997).

The next theorem enables to relate $\widehat{\text{PDC}}_{j,j+1}^{\lambda_n}$ to properties (22) and (23) which are without surprise linked to the choice of $\lambda_n$. The proof of this theorem is presented in Appendix D.6.

THEOREM 5: *Under Setting A and Assumptions (A.1) and (A.2), we have that*

$$\frac{L_n\left(\hat{\jmath}_{\lambda_n}\right)}{L_n\left(\jmath_0\right)} \xrightarrow{\mathcal{P}} 1.$$

*If in addition* $\lambda_n$ *is such that* $\lim_{n\to\infty} \lambda_n = \infty$ *then we also have that*

$$\Pr\left(\hat{\jmath}_{\lambda_n} = \jmath_0\right) \xrightarrow{\mathcal{P}} 1.$$

## 4.4 Ordering rule

In order to apply rule (19), one needs a sequence of competing nested models that includes model $\jmath_0$. In practice, this sequence might not always exist naturally, and we have instead a set of $K$ predictors which can generate a very large number of nested sequences. We propose here an algorithm for finding such a sequence, and will show in Theorem 6 (below), that this algorithm orders the variable in a suitable sequence for sufficiently large $n$.

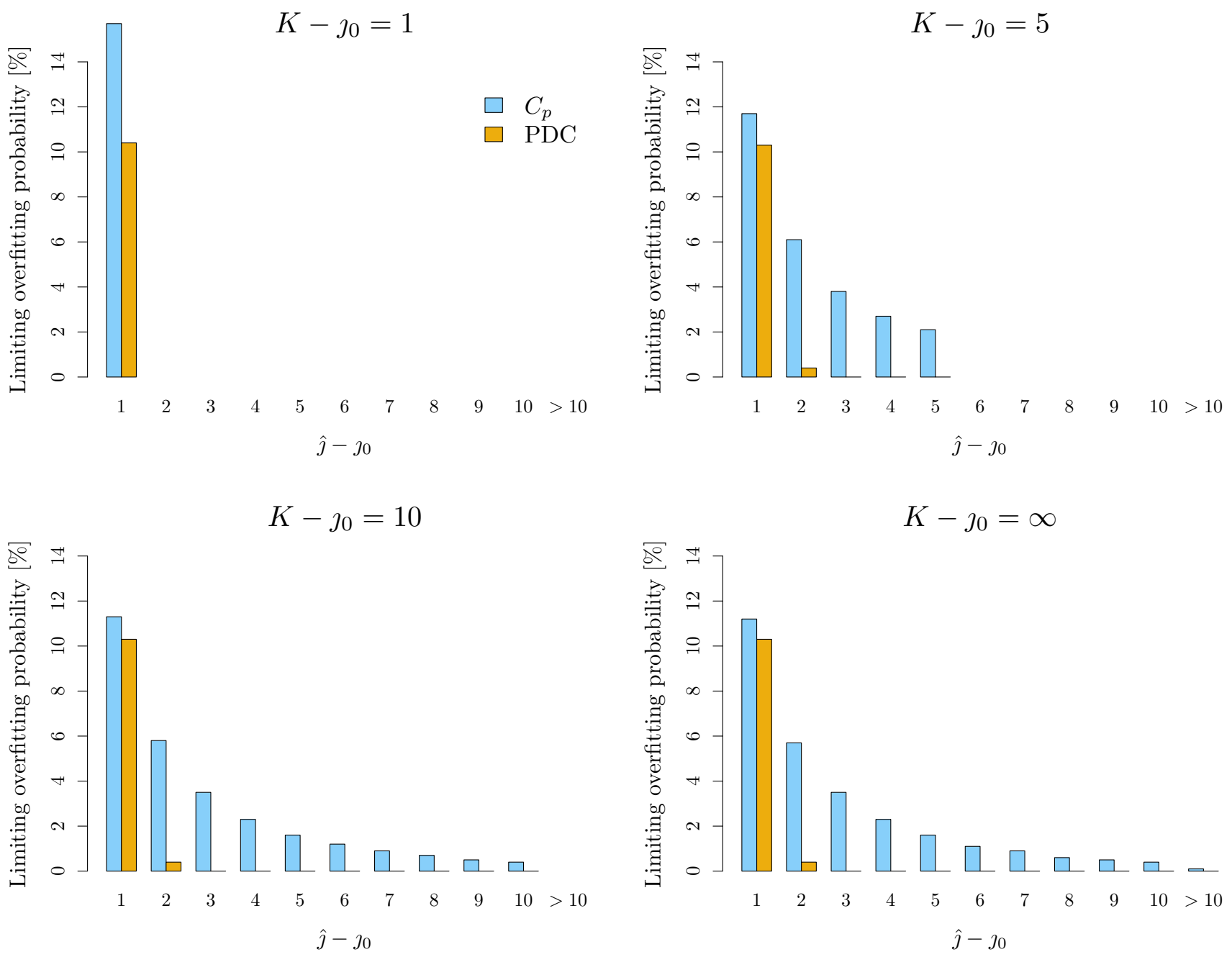

FIGURE 1: *Limiting overfitting probability. The values of Mallow's* $C_p$ *(and asymptotically equivalent methods) are taken from (Woodroofe, 1982, Table 1) while the values for* $\widehat{\text{PDC}}_{j,j+1}$ *were obtained numerically based on* $10^6$ *Monte Carlo simulations to evaluate (A-26).*

We suppose that a "null" model $\mathcal{M}_0$ of say size $0 \leq r < K$ which represents the smallest possible model nested in the correct one is available. Such a model is typically the model $\boldsymbol{Y} = \boldsymbol{\varepsilon}$ (in that case, $r = 0$). From $\mathcal{M}_0$, we choose model $\mathcal{M}_1$ that includes a single additional predictor chosen among the $K - r$ available ones, such that the $\widehat{\text{PDC}}^{\lambda_n}$ between the null model and the model with the chosen predictor is maximal. Then, starting from $\mathcal{M}_1$, one applies again the same rule until a sequence of $K - r$ nested models is obtained. This allows one to have a convex optimization function for the $\widehat{\text{PDC}}^{\lambda_n}$.

Let $\widehat{\text{PDC}}_{j,j+1}^{\lambda_n\,(k)}$ denote the PDC estimator between $\mathcal{M}_j$ and $\mathcal{M}_{j+1}$ with the latter containing the additional predictor $k$ among the $K - j$ available ones, then at model $\mathcal{M}_j$, the rule is

$$\underset{k=1,\ldots,K-j}{\text{argmax}} \; \widehat{\text{PDC}}_{j,j+1}^{\lambda_n\,(k)}. \tag{24}$$

It can be noted that for the squared loss function, maximizing $\widehat{\text{PDC}}_{j,j+1}^{\lambda_n\,(k)}$ in (19) is equivalent to simply maximizing the apparent divergence since all models have the same divergence optimism (i.e. $\lambda_n \sigma_\varepsilon^2 j$). Therefore, the rule given in (24) is equivalent to

$$\underset{k=1,\,\ldots,\,K-j}{\text{argmax}} \; ||\hat{\boldsymbol{Y}}_j - \hat{\boldsymbol{Y}}_{j+1}^{(k)}||_2^2. \tag{25}$$

Using (25) instead of (24) has the advantage of allowing to "order" the covariates without an estimate of $\sigma_\varepsilon^2$.

Moreover, rule (25) actually provides the same order as the one based on the $C_p$, i.e the residual sum of squares. Indeed, we have that

$$||\hat{\boldsymbol{Y}}_j - \hat{\boldsymbol{Y}}_{j+1}||_2^2 \;=\; \boldsymbol{Y}^T \left(\mathbf{S}_j - \mathbf{S}_{j+1}\right) \boldsymbol{Y} = \boldsymbol{Y}^T \left(\mathbf{I} - \mathbf{S}_{j+1}\right) \boldsymbol{Y} - \boldsymbol{Y}^T \left(\mathbf{I} - \mathbf{S}_j\right) \boldsymbol{Y}$$

which leads to the same optimum in $\hat{\boldsymbol{Y}}_{j+1}^{(k)}$ using (25) as optimising $\boldsymbol{Y}^T \left(I - \mathbf{S}_{j+1}\right) \boldsymbol{Y} = ||\boldsymbol{Y} - \hat{\boldsymbol{Y}}_{j+1}||_2^2$ or equivalently $||\boldsymbol{Y} - \hat{\boldsymbol{Y}}_{j+1}||_2^2 + 2\sigma^2(j+1)$.

In Theorem 6 (below) we verify that (25) enables to correctly order the covariates. The proof of this result is given in Appendix D.7.

Theorem 6: *In Setting A, let $\mathbf{X}^\star$ denote the matrix $\mathbf{X}$ whose columns are reorganised according to the iterative rule (25). Then the first $\jmath_0$ columns of $\mathbf{X}^\star$ contain all significant elements of $\boldsymbol{\beta}$ in the sense that (see (18))*

$$\jmath_0 = \max\left\{ j \in \{1, ..., K\} \;:\; \underset{n\to\infty}{\text{plim}} \left|\hat{\beta}_j^\star\right| > 0 \right\}.$$

Remark E: *In finite samples, the iterative rule (25) together with the $\widehat{\text{PDC}}_{j,j+1}^{\lambda_n}$ selection criterion defined in (19) is likely to produce "sparse" models. Indeed, the rule (25) maximizes at each step the apparent divergence while (19) adds at each step a larger penalty (since the number of parameters increases). Thus, the apparent divergence is expected to decrease at each iteration while the optimism penalty increases. In finite samples, there certainly exists, in some situations, another permutation of the columns of the design matrix $\mathbf{X}$, say $\mathbf{X}^\bullet$ (which is different from $\mathbf{X}^\star$) that leads to a model with more covariates than the model that would be obtained by applying (25) and then (19).*

Remark F: *Other ordering rules can in principle be applied to obtain an ordered design matrix* $\mathbf{X}^\star$ *whose first* $j_0$ *columns contain (asymptotically) all significant elements of* $\boldsymbol{\beta}$*. The lasso sequence is such a possibility (see Donoho and Elad, 2002; Donoho and Huo, 2002; Donoho, 2006 and Zou, 2006). The comparison with our procedure is however left for future research.*

# 5 Simulation Study

In this section we consider three different simulation settings that corresponds to three different possible situations encountered in practice. The aim is to compare the behavior of different selection rules to, on the one hand, study how the asymptotic properties apply in finite samples, and, on the other hand, to study the effect of different settings on the performance of the different selection rules.

We consider the $\widehat{\text{PDC}}_{j,j+1}^{\lambda_n}$ (defined in (19)) using three different $\lambda_n$, namely $\lambda_n = 2$ ($\widehat{\text{PDC}}_{j,j+1}$), $\lambda_n = \log(n)$ ($\widehat{\text{PDC}}_{j,j+1}^{\bullet}$) and $\lambda_n = 2\log(\log(n))$ ($\widehat{\text{PDC}}_{j,j+1}^{\star}$). We compare $\widehat{\text{PDC}}_{j,j+1}^{\lambda_n}$ with other selection methods in the linear regression framework. The latter include the stepwise forward approach using explicit model selection criteria such as the AIC, AICc, AICu, FPE, FPEu, BIC, HQ, HQc (see Table A-5 in Appendix F), which are applied to a prior ordering of the variables. This ordering processes use rule (24) in which $\widehat{\text{PDC}}_{j,j+1}^{\lambda_n}$ is replaced by the corresponding criterion (i.e. AIC, AICc, etc.). We also include the lasso which provides an ordered list of variables and the model is chosen by means of the $C_p$ statistic, as well more sophisticated penalized estimation methods such as the elastic net (enet), the adaptive lasso (a-lasso), the MCP and the SCAD. We also compute the LSE on the complete model as a benchmark. The packages used and the chosen settings for these methods, as well as for the lasso, are provided in Appendix F. Adaptive methods are expected to perform better because they make extensive use of the model assumptions. It should be noted that the PDC procedure could also be extended to an adaptive one, but this is left for further research.

The performance is measured by means of the indicators provided in Table 3 as well as distributions (boxplot) of the mean squared error of prediction (estimated on a test set), i.e

$$\text{PE}_{\boldsymbol{y}} = \frac{1}{n^\star}||\boldsymbol{y}_{test} - \mathbf{X}_{test}\hat{\boldsymbol{\beta}}||_2^2 \tag{26}$$

and the estimation precision assessed using the average mean squared error of estimation (estimated in the training set only), i.e

$$\text{MSE}_{\boldsymbol{\beta}} = ||\hat{\boldsymbol{\beta}} - \boldsymbol{\beta}||_2^2\,. \tag{27}$$

The boxplots and the medians of both $\text{PE}_{\boldsymbol{y}}$ and $\text{MSE}_{\boldsymbol{\beta}}$ are then used to compare the methods. Across simulation settings, we vary the the level of sparsity, the level of correlation among the covariates as well as the SNR of the slopes.

**Simulation 5.1.** In this simulation we consider a sparse setting where we expect the $\widehat{\text{PDC}}_{j,j+1}^{\lambda_n}$ criteria to perform well. We consider a linear model with

$$\boldsymbol{\beta} = (1, \underbrace{0, ..., 0}_{58}, 1) \tag{28}$$

Table 3: *Model selection evaluation criteria*

| Criteria | Description |
|---|---|
| Cor. [%] | Proportion of times the correct model is selected. |
| Inc. [%] | Proportion of times the correct model is nested within the selected model. |
| true+ | Average number of selected significant variables (true positives). |
| false+ | Average number of selected non-significant variables (false positives). |
| NbReg | Average number of regressors in the selected model. |
| Med ($\mathrm{PE}_{\boldsymbol{y}}$) | Median of $\mathrm{PE}_{\boldsymbol{y}}$ (see (26)) computed on test samples. |
| Med ($\mathrm{MSE}_{\boldsymbol{\beta}}$) | Median of $\mathrm{MSE}_{\boldsymbol{\beta}}$ (see (27)) computed on test samples. |

and $\sigma_\epsilon^2 = 1$. The covariates are standard normal realizations with pairwise correlations between $\mathbf{x}_j$ and $\mathbf{x}_k$ (i.e. $j^{\text{th}}$ and $k^{\text{th}}$ columns of $\mathbf{X}$) arbitrarily set to $\text{corr}(\mathbf{x}_j, \mathbf{x}_k) = 0.5^{|j-k|}$. This situation corresponds to a theoretical $R^2$ of 66.7% and to a SNR for the significant slope parameter of 2.0. We chose $n = 80$ for the training and $n^\star = 800$ for the test samples.

**Simulation 5.2.** This simulation considers a relatively "dense" setting where we expect the $\widehat{\text{PDC}}_{j,j+1}^{\lambda_n}$ criteria to perform poorly. We consider a linear model with

$$\boldsymbol{\beta} = (0.3, 0, 0.3, 0, 0.3, 0, 0.3, 0, 0.3, 0) \tag{29}$$

and $\sigma_\epsilon^2 = 1$. The covariates are standard normal realizations with pairwise correlations between $\mathbf{x}_j$ and $\mathbf{x}_k$ arbitrarily set to $\text{corr}(\mathbf{x}_j, \mathbf{x}_k) = 0.75^{|j-k|}$. This situation corresponds to a theoretical $R^2$ of 52.6% and to a SNR for the significant slope parameter of 1.1. We used $n = 100$ for the training and $n^\star = 10^3$ for the test samples.

**Simulation 5.3.** In this simulation we consider a relatively sparse setting but where some of the elements of $\boldsymbol{\beta}$ are small and thus difficult to identify. We expect the $\widehat{\text{PDC}}_{j,j+1}^{\lambda_n}$ criteria to perform well in terms of prediction and estimation error but not necessarily well in terms of model identification. We consider a linear model with

$$\boldsymbol{\beta} = (2, 0, 1, 2, 0, 1, \underbrace{0, ..., 0}_{16}, \underbrace{0.1, ..., 0.1}_{6}, \underbrace{0, ..., 0}_{16}, 2, 0, 1, 2, 0, 1) \tag{30}$$

and $\sigma_\epsilon^2 = 4$. The covariates are standard normal realizations with pairwise correlations between $\mathbf{x}_j$ and $\mathbf{x}_k$ arbitrarily set to $\text{corr}(\mathbf{x}_j, \mathbf{x}_k) = 0.5^{|j-k|}$. This situation corresponds to a theoretical $R^2$ of 88.5% and to a SNR of approximately 7.7. We choose $n = 100$ for the training and $n^\star = 10^3$ for the test samples.

Table 4 presents the performance of some model selection criteria using the measures in Table 3 for respectively the three different settings. The complete results are given in Tables A-6 to A-8 in Appendix I. Figures 2 to 4 show the performance of all criteria in terms of prediction and estimation error.

When considering the mean squared error of prediction and of estimation in Figures 2 to 4, one notices that in the very sparse setting (Simulation 5.1, Figure 2) the PDC has a comparable performance to the adaptive methods, which are the best overall. The lasso follows closely, while the other criteria have overall a worse performance. In the dense setting (Simulation 5.2, Figure 3), the lasso, elastic-net and adaptive methods have a slight better performance compared to all other criteria. It should however be stressed, that in this case, the LSE is at least as good as the latter, and just slightly worse than the more sophisticated methods. In the sparse and low signal setting (Simulation 5.3, Figure 4), the performances of the more sophisticated methods are comparable to the one of the PDC and the BIC. Hence, overall, the PDC selects models that have a comparable performance in terms of prediction and estimation error, to the regularized and adaptive methods.

Table 4: *Performance of model selection criteria using the measures in Table 3 for Simulations 5.1 to 5.3. These include the* $\widehat{\text{PDC}}_{j,j+1}$ *(*PDC*) in (16), the AIC (*AIC*), the BIC (*BIC*) and the the full model with the LSE (*LSE*), together with the lasso (*lasso*), elastic-net (*enet*), adaptive lasso (a-lasso), MCP (MCP) and SCAD (SCAD) (see Appendix F for more details on the chosen criteria).* 500 *samples are simulated under the correct model as presented in simulations 5.1 to 5.3. The numbers in parentheses for the columns* $Med(PE_{\boldsymbol{y}})$ *and* $Med(MSE_{\boldsymbol{\beta}})$ *are the corresponding standard errors estimated by using the bootstrap with* $B = 500$ *resampling.*

| | $\text{Med}\,(\text{PE}_{\boldsymbol{y}})$ | $\text{Med}\,(\text{MSE}_{\boldsymbol{\beta}})$ | Cor. [%] | Inc. [%] | true+ | false+ | NbReg |
|---|---|---|---|---|---|---|---|
| | Simulation 5.1 | $(p = 2, K = 60)$ | | | | | |
| LS | $4.29\ (8.0 \cdot 10^{-2})$ | $5.59 \cdot 10^{0}\ (1.2 \cdot 10^{-1})$ | 0.0 | 100.0 | 2.0 | 58.0 | 60.0 |
| AIC | $2.30\ (3.6 \cdot 10^{-2})$ | $1.90 \cdot 10^{0}\ (8.0 \cdot 10^{-2})$ | 0.0 | 100.0 | 2.0 | 24.0 | 26.0 |
| BIC | $1.25\ (1.5 \cdot 10^{-2})$ | $2.71 \cdot 10^{-1}\ (1.2 \cdot 10^{-2})$ | 9.4 | 100.0 | 2.0 | 3.5 | 5.5 |
| lasso | $1.16\ (7.4 \cdot 10^{-3})$ | $1.46 \cdot 10^{-1}\ (5.0 \cdot 10^{-3})$ | 3.2 | 100.0 | 2.0 | 9.3 | 11.3 |
| PDC | $1.06\ (3.8 \cdot 10^{-3})$ | $5.45 \cdot 10^{-2}\ (6.0 \cdot 10^{-3})$ | 56.4 | 100.0 | 2.0 | 0.6 | 2.6 |
| enet | $1.19\ (2.1 \cdot 10^{-2})$ | $1.68 \cdot 10^{-1}\ (2.1 \cdot 10^{-2})$ | 18.2 | 64.0 | 1.3 | 3.6 | 4.9 |
| a-lasso | $1.04\ (4.2 \cdot 10^{-3})$ | $3.17 \cdot 10^{-2}\ (1.6 \cdot 10^{-3})$ | 65.6 | 100.0 | 2.0 | 0.9 | 2.9 |
| MCP | $1.04\ (3.4 \cdot 10^{-3})$ | $2.37 \cdot 10^{-2}\ (1.7 \cdot 10^{-3})$ | 67.2 | 100.0 | 2.0 | 1.1 | 3.1 |
| SCAD | $1.04\ (3.7 \cdot 10^{-3})$ | $2.60 \cdot 10^{-2}\ (1.5 \cdot 10^{-3})$ | 33.8 | 100.0 | 2.0 | 2.6 | 4.6 |
| | Simulation 5.2 | $(p = 5, K = 10)$ | | | | | |
| LS | $1.12\ (4.3 \cdot 10^{-3})$ | $3.44 \cdot 10^{-1}\ (1.7 \cdot 10^{-2})$ | 0.0 | 100.0 | 5.0 | 5.0 | 10.0 |
| AIC | $1.14\ (4.0 \cdot 10^{-3})$ | $4.70 \cdot 10^{-1}\ (1.5 \cdot 10^{-2})$ | 2.2 | 7.4 | 2.9 | 1.8 | 4.7 |
| BIC | $1.18\ (5.9 \cdot 10^{-3})$ | $5.85 \cdot 10^{-1}\ (1.6 \cdot 10^{-2})$ | 0.0 | 0.0 | 1.9 | 1.3 | 3.2 |
| lasso | $5.06\ (2.6 \cdot 10^{-2})$ | $1.20 \cdot 10^{0}\ (3.7 \cdot 10^{-2})$ | 0.0 | 1.0 | 10.2 | 9.9 | 20.0 |
| PDC | $1.18\ (6.0 \cdot 10^{-3})$ | $5.75 \cdot 10^{-1}\ (1.8 \cdot 10^{-2})$ | 0.0 | 0.0 | 1.9 | 1.3 | 3.2 |
| enet | $1.11\ (4.6 \cdot 10^{-3})$ | $2.79 \cdot 10^{-1}\ (9.5 \cdot 10^{-3})$ | 1.8 | 40.6 | 4.0 | 2.5 | 6.5 |
| a-lasso | $1.11\ (4.2 \cdot 10^{-3})$ | $2.90 \cdot 10^{-1}\ (1.1 \cdot 10^{-2})$ | 5.0 | 35.0 | 3.9 | 2.0 | 5.8 |
| MCP | $1.14\ (5.4 \cdot 10^{-3})$ | $4.23 \cdot 10^{-1}\ (2.0 \cdot 10^{-2})$ | 5.2 | 35.8 | 3.5 | 2.2 | 5.7 |
| SCAD | $1.14\ (5.6 \cdot 10^{-3})$ | $4.16 \cdot 10^{-1}\ (1.9 \cdot 10^{-2})$ | 3.8 | 38.0 | 3.7 | 2.3 | 6.1 |
| | Simulation 5.3 | $(p = 14, K = 50)$ | | | | | |
| LS | $7.76\ (8.3 \cdot 10^{-2})$ | $6.27 \cdot 10^{0}\ (1.5 \cdot 10^{-1})$ | 0.0 | 100.0 | 14.0 | 36.0 | 50.0 |
| AIC | $6.17\ (5.2 \cdot 10^{-2})$ | $3.26 \cdot 10^{0}\ (1.1 \cdot 10^{-1})$ | 0.0 | 0.4 | 9.9 | 9.5 | 19.3 |
| BIC | $5.00\ (3.3 \cdot 10^{-2})$ | $1.25 \cdot 10^{0}\ (3.1 \cdot 10^{-2})$ | 0.0 | 0.0 | 8.5 | 1.8 | 10.4 |
| lasso | $5.06\ (2.6 \cdot 10^{-2})$ | $1.20 \cdot 10^{0}\ (3.7 \cdot 10^{-2})$ | 0.0 | 1.0 | 10.2 | 9.9 | 20.0 |
| PDC | $4.83\ (3.0 \cdot 10^{-2})$ | $9.45 \cdot 10^{-1}\ (4.3 \cdot 10^{-2})$ | 0.0 | 0.0 | 7.9 | 0.5 | 8.5 |
| enet | $5.13\ (2.3 \cdot 10^{-2})$ | $1.32 \cdot 10^{0}\ (3.3 \cdot 10^{-2})$ | 0.0 | 0.2 | 8.8 | 3.9 | 12.7 |
| a-lasso | $4.77\ (2.4 \cdot 10^{-2})$ | $9.35 \cdot 10^{-1}\ (3.1 \cdot 10^{-2})$ | 0.0 | 0.0 | 8.6 | 2.2 | 10.9 |
| MCP | $4.87\ (3.0 \cdot 10^{-2})$ | $1.08 \cdot 10^{0}\ (4.1 \cdot 10^{-2})$ | 0.0 | 0.0 | 8.8 | 2.7 | 11.4 |
| SCAD | $4.89\ (2.9 \cdot 10^{-2})$ | $1.17 \cdot 10^{0}\ (5.5 \cdot 10^{-2})$ | 0.0 | 0.0 | 9.6 | 5.3 | 14.8 |

When considering other performance criteria in Table 4 (see also Tables A-6 to A-8 in Appendix I), in the very sparse setting (Simulation 5.1) the proportion of times the PDC selects the correct model is of 56%, and the adaptive lasso and MCP have a slightly better performance. The elastic net, SCAD and lasso are significantly worse, and the other selection criteria are just way out. By considering the proportion of times the correct model is nested within the selected model (Inc. [%]), one can see that at least all methods always include the correct model, with the notable exception of the elastic net. The number of extra noisy variables that are included in the selected model (false+) show that the PDC, adaptive lasso and also the MCP are the most accurate model selection procedures in catching the correct model when the setting is very sparse. In the dense setting (Simulation 5.2), the proportion of times the correct model is selected is very low overall. The enet and adaptive methods include the true model less than half the times, but perform better than all other methods. In this setting, we can say that none of the methods have a really satisfactory performance, and the LSE at the full model seems to be sufficient. Finally, in the low signal setting (Simulation 5.3), without surprise, no method selects the correct model most of the times. The average number of true positives (significant variables) included in the selected model is comparable across methods, while the average number of false negative (non significant variables) included in the model is the lowest for the PDC.

We hence can conclude that in sparse settings, the PDC, with a simple ordering rule, tends to have a similar performance not only in terms of out-of-sample prediction error but also in choosing the correct variables when compared to the other best methods, while it includes the smallest number of noisy variables. This might suggest that with a real dataset, the PDC will pick up at least as many of the significant variables as the other methods, but will pick up only a few of the noisy ones. This is a probable explanation of the conclusions drawn in the data analysis of the Childhood Malnutrition in Zambia in Section 2. Moreover, the PDC is simple to implement and hence can easily be used on very large datasets, without prior screening or subset selection.

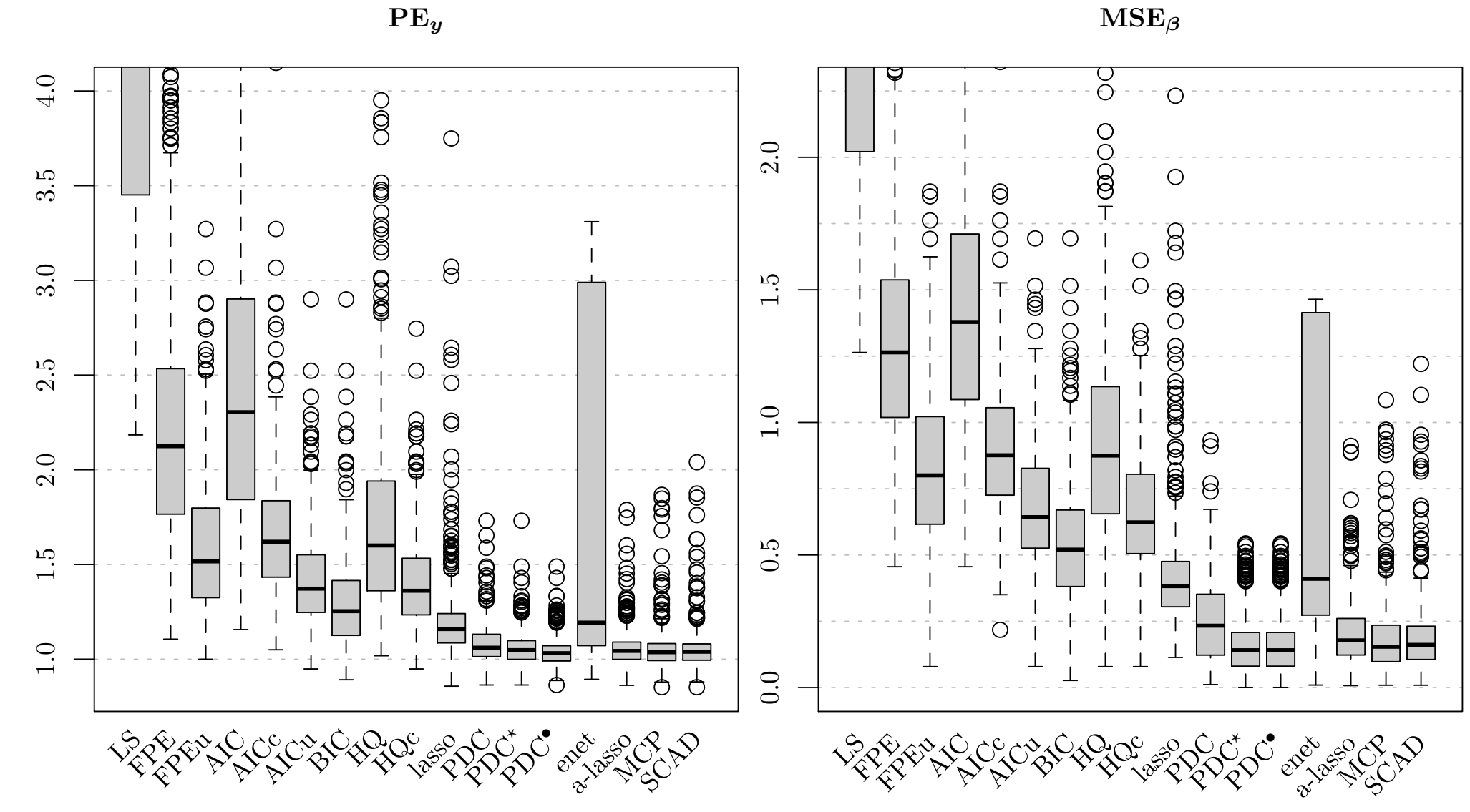

FIGURE 2: *Empirical distributions of* $PE_{\boldsymbol{y}}$ *and* $MSE_{\boldsymbol{\beta}}$ *as defined in (26) and (27) of* $\widehat{\mathrm{PDC}}_{j,j+1}^{\lambda_n}$ *with* $\lambda_n = 2$ *(PDC),* $\lambda_n = \log(n)$ *(*$\mathrm{PDC}^{\bullet}$*),* $\lambda_n = \log(\log(n))$ *(*$\mathrm{PDC}^{\star}$*), the stepwise forward FPE (*FPE*), FPEu (*FPEu*), AIC (*AIC*), AICc (*AICc*), AICu (*AICu*), BIC (*BIC*), HQ (*HQ*), HQc (*HQc*), and the full model with the LSE (*LSE*), together with the lasso (*lasso*), elastic-net (*enet*), adaptive lasso (a-lasso), MCP (MCP) and SCAD (SCAD) (see Appendix F for more details on the chosen criteria).* 500 *samples are simulated under the correct model as presented in Simulation 5.1.*

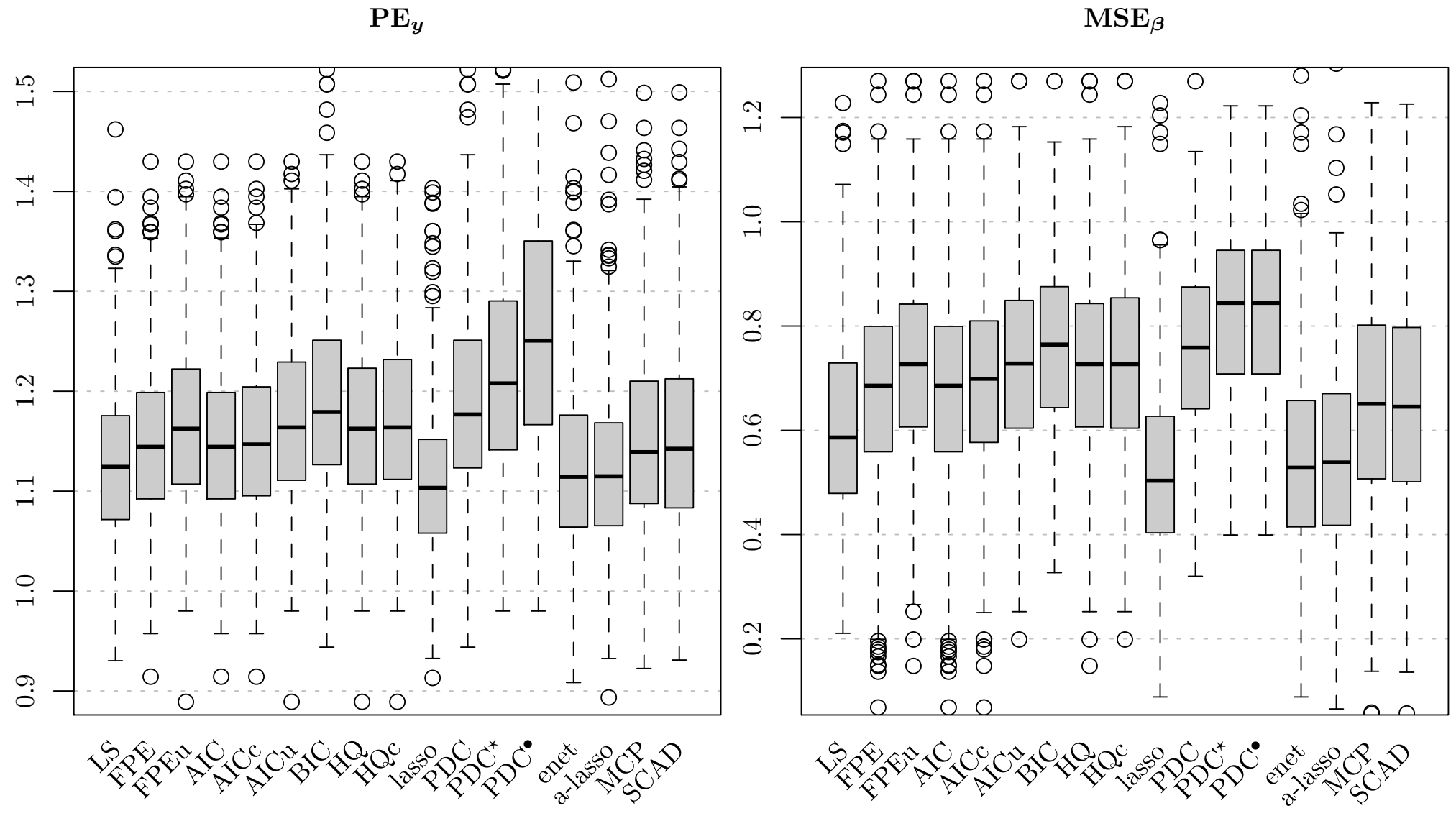

FIGURE 3: *Empirical distributions of $PE_{\boldsymbol{y}}$ and $MSE_{\boldsymbol{\beta}}$ as defined in (26) and (27) of $\widehat{\text{PDC}}_{j,j+1}^{\lambda_n}$ with $\lambda_n = 2$ (PDC), $\lambda_n = \log(n)$ ($\text{PDC}^{\bullet}$), $\lambda_n = \log(\log(n))$ ($\text{PDC}^{\star}$), the stepwise forward FPE (*FPE*), FPEu (*FPEu*), AIC (*AIC*), AICc (*AICc*), AICu (*AICu*), BIC (*BIC*), HQ (*HQ*), HQc (*HQc*), and the full model with the LSE (*LSE*), together with the lasso (*lasso*), elastic-net (*enet*), adaptive lasso (a-lasso), MCP (MCP) and SCAD (SCAD) (see Appendix F for more details on the chosen criteria).* 500 *samples are simulated under the correct model as presented in Simulation 5.2.*

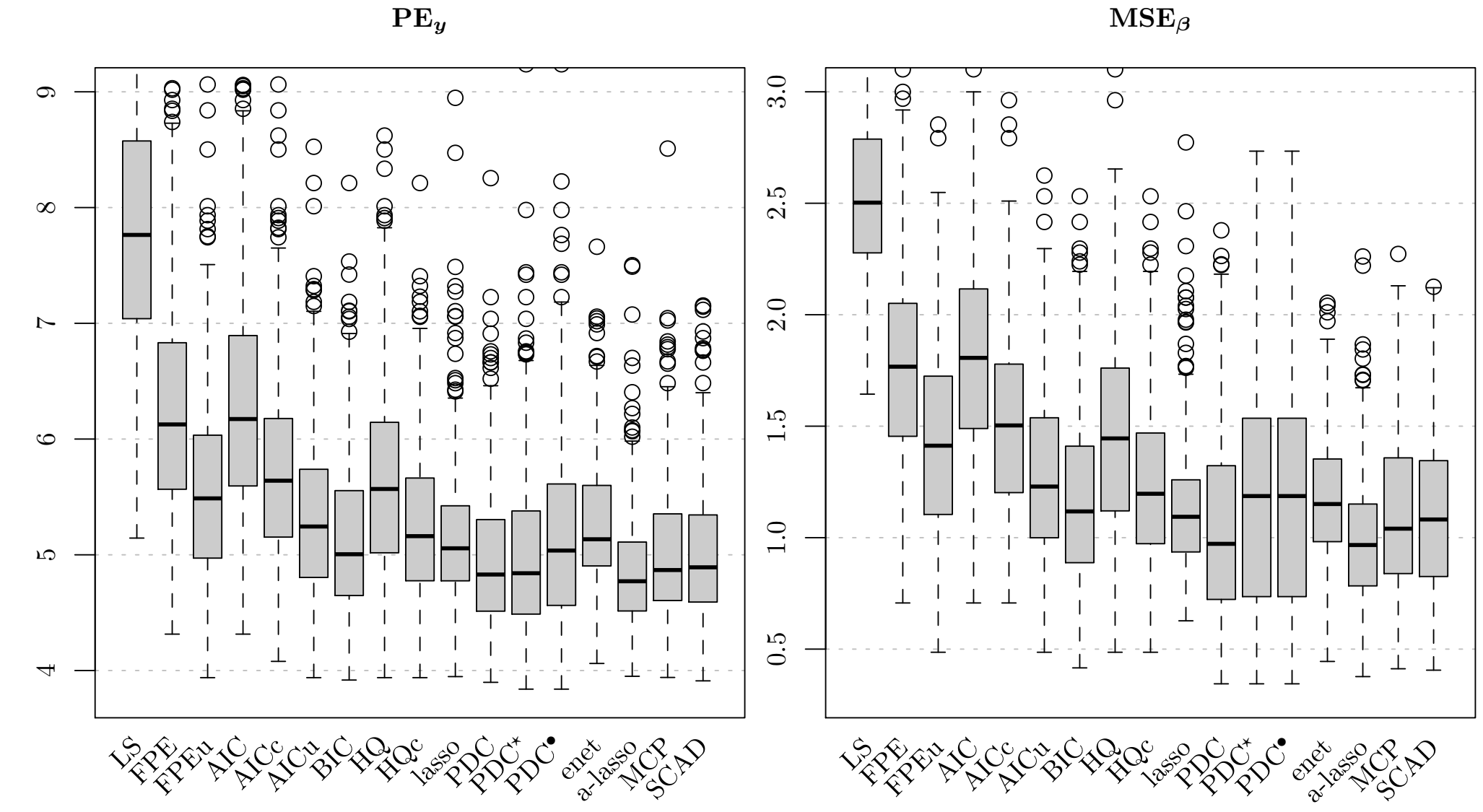

FIGURE 4: *Empirical distributions of* $PE_{\boldsymbol{y}}$ *and* $MSE_{\boldsymbol{\beta}}$ *as defined in (26) and (27) of* $\widehat{\mathrm{PDC}}_{j,j+1}^{\lambda_n}$ *with* $\lambda_n = 2$ *(PDC),* $\lambda_n = \log(n)$ *(*$\mathrm{PDC}^{\bullet}$*),* $\lambda_n = \log(\log(n))$ *(*$\mathrm{PDC}^{\star}$*), the stepwise forward FPE (*FPE*), FPEu (*FPEu*), AIC (*AIC*), AICc (*AICc*), AICu (*AICu*), BIC (*BIC*), HQ (*HQ*), HQc (*HQc*), and the full model with the LSE (*LSE*), together with the lasso (*lasso*), elastic-net (*enet*), adaptive lasso (a-lasso), MCP (MCP) and SCAD (SCAD) (see Appendix F for more details on the chosen criteria).* 500 *samples are simulated under the correct model as presented in Simulation 5.3.*

# Appendices

## A Model selection methods

A common model selection procedure consists in computing a criterion, say $C$, associated to a suitable sequence of potential models, and in choosing the one(s) that optimize this criterion. Many criteria have been proposed and the most popular ones include Mallow's $C_p$ (Mallows, 1973) based on prediction error, Akaike's Information Criterion (AIC) (Akaike, 1974), based on the Kullback-Leibler divergence, and the Bayesian Information Criterion (BIC) (Schwarz, 1978).

Beside the $C_p$, the AIC and the BIC a great number of criteria have been proposed in the literature (see e.g. Hannan and Quinn, 1979; Foster and George, 1994; Zheng and Loh, 1997; Tibshirani and Knight, 1999; George (2000) and the references therein). Kashid and Kulkarni (2002) consider a criterion based on the difference in prediction error. Alternative formulas for the AIC penalty have been proposed (see e.g. Hurvich and Tsai, 1989; McQuarrie et al., 1997; McQuarrie and Tsai, 1998 and Table A-5 in Appendix F). An alternative to explicit model selection criteria is to estimate the expected out-of-sample prediction error by simulation methods such as bootstrap (see e.g. Efron, 1983) or cross validation (see e.g. Shao, 1993; Reiss et al., 2012). These methods are strongly linked to explicit model selection criteria; for example, Shao (1993) showed that the leave-one-out cross validation is asymptotically equivalent to the AIC. Moreover, Efron (2004) demonstrated that model-based selection criteria such as the AIC have better model selection performance compared to nonparametric methods like cross validation, assuming that the model is believable.

Other methods that are nowadays well established are based on penalized estimation and include the bridge penalty type (Frank and Friedman, 1993), the nonnegative garrote (Breiman, 1995), the Smoothly Clipped Absolute Deviation or SCAD (Fan and Li, 2001), the elastic-net (Zou and Hastie, 2005), the Dantzig Selector (Candès and Tao, 2007), a correlation based penalty (Tutz and Ulbricht, 2009) or the Minimax Concave Penalty or MCP (Zhang, 2010).

Some recent work propose extensions of these methods to improve in some sense their performance. Zou (2006) propose an adaptive lasso, where adaptive weights are used for penalizing different coefficients, so that the resulting model selection method enjoys the oracle property. Meinshausen (2007) proposes the relaxed lasso, a two-stage procedure that provides a continuum of solutions that include both soft- and hard-thresholding of estimators. Zou and Zhang (2009) propose the adaptive elastic-net, that combines the strengths of the quadratic regularization and the adaptively weighted lasso shrinkage. Meinshausen and Bühlmann (2010) uses a very generic subsampling approach to determine the amount of regularization in the lasso such that a certain familywise type I error rate in multiple testing can be conservatively controlled for finite sample size. Sun and Zhang (2012), based on Staädler et al. (2010), propose to jointly estimate the slope and the noise level in the regression model using a penalized estimating equation (scaled lasso). Adaptive methods have also been proposed for computing the penalty of explicit selection criteria. Shen and Ye (2002) propose an adaptive model selection procedure that uses a data-adaptive complexity penalty based on a concept of generalized degrees of freedom. Müller and Welsh (2010) consider penalty curves, in which the penalty multiplier is varied over a given interval, for improving the performance of loss based criteria such as the AIC.

Finally, model selection tools for the regression model that are robust to small departures from the model (e.g. outlying observations) have also been proposed as alternatives to either models selection criteria or to robustify penalized regression method. Ronchetti (1982) propose a robust AIC, Machado (1993) a robust BIC, Ronchetti (1997) a robust $C_p$ and Ronchetti et al. (1997) a robust criterion based on cross-validation. Müller and Welsh combine a robust penalized estimation criterion and a robust prediction error loss function, and Khan et al. (2007) propose a robust version of the LARS algorithm (for the Lasso). Dupuis and Victoria-Feser (2011) propose the use of a forward search procedure together with adjusted robust estimators when there is a large number of potential covariates, and based on Lin et al. (2011), Dupuis and Victoria-Feser (2013) propose a fast model selection using a robust variance inflation factor criterion.

# B Childhood Malnutrition in Zambia

The selected explanatory variables to explain the response score (1) in the Child Malnutrition in Zambia dataset, taken from `www.measuredhs.com`, are the following:

- breastfeeding duration (month);
- age of the child (month);
- age of the mother (years);
- Body Mass Index (BMI) of the mother (kg/meter$^2$);
- height of the mother (meter);
- weight of the mother (kg);
- region of residence (9 levels: Central, Copperbelt, Eastern, Luapula, Lusaka, Northern, Northwestern, Southern and Western);
- mother's highest education level attended (4 levels: No education, Primary, Secondary and Higher);
- wealth index factor score;
- weight of child at birth (kg) ;
- sex of the child;
- interval between the current birth and the previous birth (month); and
- main source of drinking water (8 levels: Piped into dwelling, Piped to yard/plot, Public tap/standpipe, Protected well, Unprotected well, River/dam/lake/ponds/stream/canal/irrigation channel, Bottled water, Other).

# C Discussion: the PDC Class and Other Model Section Criteria

Although we defined the $d$-class in a general manner for two models $F_{\boldsymbol{\theta}_1}$ and $F_{\boldsymbol{\theta}_2}$, most model selection criteria are based on a chosen divergence between a given candidate model $F_{\boldsymbol{\theta}}$ and the true model $F_0$. In such a setting, we may simplify our definition given in (5) and write instead

$$\mathrm{C}^{\star} = \mathbb{E}\left[\mathbb{E}_0\left[D\left(\boldsymbol{g}_1\left(\boldsymbol{Y}^0\right), \boldsymbol{g}_2\left(\boldsymbol{Y}, \hat{\boldsymbol{\theta}}\right)\right)\right]\right] \tag{A-1}$$

which leads to the following (consistent) estimator

$$\widehat{\mathrm{C}}^{\star} = D\left(\boldsymbol{g}_1\left(\boldsymbol{y}\right), \boldsymbol{g}_2\left(\boldsymbol{y}, \hat{\boldsymbol{\theta}}\right)\right) + \mathrm{tr}\left\{\widehat{\mathrm{cov}}\left[\boldsymbol{g}_1\left(\boldsymbol{y}\right), \nabla\psi\left(\boldsymbol{g}_2\left(\boldsymbol{y}, \hat{\boldsymbol{\theta}}\right)\right)\right]\right\}. \tag{A-2}$$

Mallow's $C_p$ can be derived from (A-2). Indeed, consider the linear model given in Setting A. Suppose that we wish to find an estimator for the following criterion:

$$\mathrm{C} = \mathbb{E}\left[\mathbb{E}_0\left[||\boldsymbol{Y}^0 - \hat{\boldsymbol{Y}}||_2^2\right]\right].$$

Clearly, this model selection criterion belongs to the $d$-class error (and to the $q$-class as well) with $\boldsymbol{g}_1(\boldsymbol{Y}^0, \hat{\boldsymbol{\beta}}_1) = \boldsymbol{Y}^0$, $\boldsymbol{g}_2(\boldsymbol{Y}, \hat{\boldsymbol{\beta}}_2) = \mathbf{X}\hat{\boldsymbol{\beta}}$ and $\psi\left(\mathbf{z}\right) = \mathbf{z}^T\mathbf{z}$. By applying Theorem 1 we obtain

$$\Delta = \mathrm{tr}\left[\mathrm{cov}\left(\boldsymbol{Y}, 2\mathbf{X}\hat{\boldsymbol{\beta}}\right)\right] = 2\sigma_{\varepsilon}^2\,\mathrm{tr}\left(\mathbf{S}\right) = 2\sigma_{\varepsilon}^2 p$$

where $\mathbf{S}$ denotes the "hat" matrix of $\mathbf{X}$, i.e. $\mathbf{S} = \mathbf{X}\left(\mathbf{X}^T\mathbf{X}\right)^{-1}\mathbf{X}^T$. Using (A-2) we get

$$\widehat{\mathrm{C}} = D\left(\boldsymbol{g}_1\left(\boldsymbol{y}, \hat{\boldsymbol{\theta}}_1\right), \boldsymbol{g}_2\left(\boldsymbol{y}, \hat{\boldsymbol{\theta}}_2\right)\right) + \Delta = ||\boldsymbol{y} - \mathbf{X}\hat{\boldsymbol{\beta}}||_2^2 + 2\sigma_{\varepsilon}^2 p$$

which is unsurprisingly equal to Mallow's $C_p$.

At this point, it must be pointed out that many criteria can be defined using (A-1). However, only a few of them are meaningful for the task of model selection. Indeed, an estimator of $\mathrm{C}^{\star}$ in (A-1) is used in practice to compare several models, and in a stepwise procedure, two models at the time. Hence, $\mathrm{C}^{\star}$ should ideally satisfy (at least) two properties. Consider two candidate models, say $F_{\boldsymbol{\theta}_1}$ and $F_{\boldsymbol{\theta}_2}$ such that $F_{\boldsymbol{\theta}_1}$ is nested within $F_{\boldsymbol{\theta}_2}$, then, $\widehat{\mathrm{C}}^{\star}$ should satisfy the following two properties:

(Pr.1) For nested models, the apparent divergence is strictly non-increasing with respect to model complexity. More formally, we have that for any nested models $F_{\boldsymbol{\theta}_1}$ in $F_{\boldsymbol{\theta}_2}$ the apparent divergence is such that:

$$D\left(\boldsymbol{g}_1\left(\boldsymbol{y}\right), \boldsymbol{g}_2\left(\boldsymbol{y}, \hat{\boldsymbol{\theta}}_1\right)\right) \geq D\left(\boldsymbol{g}_1\left(\boldsymbol{y}\right), \boldsymbol{g}_2\left(\boldsymbol{y}, \hat{\boldsymbol{\theta}}_2\right)\right).$$

(Pr.2) For nested models, the divergence optimism is strictly non-decreasing with respect to model complexity. More formally, we have that for any nested models $F_{\boldsymbol{\theta}_1}$ in $F_{\boldsymbol{\theta}_2}$ the divergence optimism is such that:

$$\mathrm{tr}\left\{\widehat{\mathrm{cov}}\left[\boldsymbol{g}_1\left(\boldsymbol{y}\right), \nabla\psi\left(\boldsymbol{g}_2\left(\boldsymbol{y}, \hat{\boldsymbol{\theta}}_1\right)\right)\right]\right\} \leq \mathrm{tr}\left\{\widehat{\mathrm{cov}}\left[\boldsymbol{g}_1\left(\boldsymbol{y}\right), \nabla\psi\left(\boldsymbol{g}_2\left(\boldsymbol{y}, \hat{\boldsymbol{\theta}}_2\right)\right)\right]\right\}.$$

If Property (Pr.1) is not satisfied, it would mean that a larger model (i.e. $F_{\boldsymbol{\theta}_2}$) can provide a poorer (apparent) fit than $F_{\boldsymbol{\theta}_1}$. Such situations are extremely unlikely. In Corollary A-1 (below) we relate Property (Pr.1) to the estimator $\hat{\boldsymbol{\theta}}$. This enables to easily construct a criterion which verifies (Pr.1) given an estimator $\hat{\boldsymbol{\theta}}$. The proof of this result is straightforward and therefore shall be omitted.

Corollary A-1: *The estimator* $\widehat{\mathrm{C}}^{\star}$ *as defined in (A-2) satisfies Property (Pr.1) if the estimator* $\hat{\boldsymbol{\theta}}$ *is the result of the following minimization problem*

$$\hat{\boldsymbol{\theta}} = \underset{\boldsymbol{\theta}\in\boldsymbol{\Theta}}{\operatorname{argmin}}\; D\left(\boldsymbol{g}_1\left(\boldsymbol{y}\right), \boldsymbol{g}_2\left(\boldsymbol{y}, \boldsymbol{\theta}\right)\right).$$

The rational behind Property (Pr.2) is the following. Suppose that $F_{\boldsymbol{\theta}_1}$ is the true model (or is nested within the true model) and assume that models $F_{\boldsymbol{\theta}_j}$ with $j = 1, ..., K$ are models with increasing complexity such that model $F_{\boldsymbol{\theta}_j}$ is nested within model $F_{\boldsymbol{\theta}_k}$ if $1 \leq j < k \leq K$. Then, if a criterion satisfies Property (Pr.1), the apparent divergence will be smaller as the model complexity increases. Therefore, the optimism should have exactly the opposite property in order to allow a model "close" to $F_{\boldsymbol{\theta}_1}$ to be selected by the criterion at hand.

# D Proofs

## D.1 Proof of Theorem 1

Proof: Since the divergence $D(\cdot, \cdot)$ belongs to the $d$-class we have that $D(\cdot, \cdot)$ is a valid Bregman divergence, we may express (1) as:

$$\begin{aligned}\mathbb{E}\left[\mathbb{E}_0\left[D\left(\boldsymbol{g}_1\left(\boldsymbol{Y}^0, \hat{\boldsymbol{\theta}}_1\right), \boldsymbol{g}_2\left(\boldsymbol{Y}, \hat{\boldsymbol{\theta}}_2\right)\right)\right]\right] &= \mathbb{E}_0\left[\psi\left(\boldsymbol{g}_1\left(\boldsymbol{Y}^0, \hat{\boldsymbol{\theta}}_1\right)\right)\right] - \mathbb{E}\left[\psi\left(\boldsymbol{g}_2\left(\boldsymbol{Y}, \hat{\boldsymbol{\theta}}_2\right)\right)\right] \\ &- \mathbb{E}\left[\left(\mathbb{E}_0\left[\boldsymbol{g}_1\left(\boldsymbol{Y}^0, \hat{\boldsymbol{\theta}}_1\right)\right] - \boldsymbol{g}_2\left(\boldsymbol{Y}, \hat{\boldsymbol{\theta}}_2\right)\right)^T \nabla\psi\left(\boldsymbol{g}_2\left(\boldsymbol{Y}, \hat{\boldsymbol{\theta}}_2\right)\right)\right]\end{aligned}$$

Consider

$$\Delta = \mathbb{E}\left[\mathbb{E}_0\left[D\left(\boldsymbol{g}_1\left(\boldsymbol{Y}^0, \hat{\boldsymbol{\theta}}_1\right), \boldsymbol{g}_2\left(\boldsymbol{Y}, \hat{\boldsymbol{\theta}}_2\right)\right)\right]\right] - \mathbb{E}\left[D\left(\boldsymbol{g}_1\left(\boldsymbol{Y}, \hat{\boldsymbol{\theta}}_1\right), \boldsymbol{g}_2\left(\boldsymbol{Y}, \hat{\boldsymbol{\theta}}_2\right)\right)\right]$$

with

$$\begin{aligned}\mathbb{E}\left[D\left(\boldsymbol{g}_1\left(\boldsymbol{Y}, \hat{\boldsymbol{\theta}}_1\right), \boldsymbol{g}_2\left(\boldsymbol{Y}, \hat{\boldsymbol{\theta}}_2\right)\right)\right] &= \mathbb{E}\left[\psi\left(\boldsymbol{g}_1\left(\boldsymbol{Y}, \hat{\boldsymbol{\theta}}_1\right)\right)\right] - \mathbb{E}\left[\psi\left(\boldsymbol{g}_2\left(\boldsymbol{Y}, \hat{\boldsymbol{\theta}}_2\right)\right)\right] \\ &- \mathbb{E}\left[\left(\boldsymbol{g}_1\left(\boldsymbol{Y}, \hat{\boldsymbol{\theta}}_1\right) - \boldsymbol{g}_2\left(\boldsymbol{Y}, \hat{\boldsymbol{\theta}}_2\right)\right)^T \nabla\psi\left(\boldsymbol{g}_2\left(\boldsymbol{Y}, \hat{\boldsymbol{\theta}}_2\right)\right)\right].\end{aligned}$$

So by subtracting the two terms we obtain

$$\begin{aligned}\Delta &= \mathbb{E}_0\left[\psi\left(\boldsymbol{g}_1\left(\boldsymbol{Y}^0, \hat{\boldsymbol{\theta}}_1\right)\right)\right] - \mathbb{E}\left[\psi\left(\boldsymbol{g}_1\left(\boldsymbol{Y}, \hat{\boldsymbol{\theta}}_1\right)\right)\right] \\ &+ \mathbb{E}\left[\left(\boldsymbol{g}_1\left(\boldsymbol{Y}, \hat{\boldsymbol{\theta}}_1\right) - \mathbb{E}_0\left[\boldsymbol{g}_1\left(\boldsymbol{Y}^0, \hat{\boldsymbol{\theta}}_1\right)\right]\right)^T \nabla\psi\left(\boldsymbol{g}_2\left(\boldsymbol{Y}, \hat{\boldsymbol{\theta}}_2\right)\right)\right].\end{aligned}$$

Since $\mathbb{E}_0\left[\psi\left(\boldsymbol{g}_1\left(\boldsymbol{Y}^0,\hat{\boldsymbol{\theta}}_1\right)\right)\right]=\mathbb{E}\left[\psi\left(\boldsymbol{g}_1\left(\boldsymbol{Y},\hat{\boldsymbol{\theta}}_1\right)\right)\right]$ it follows that

$$\Delta=\mathbb{E}\left[\left(\boldsymbol{g}_1\left(\boldsymbol{Y},\hat{\boldsymbol{\theta}}_1\right)-\mathbb{E}_0\left[\boldsymbol{g}_1\left(\boldsymbol{Y}^0,\hat{\boldsymbol{\theta}}_1\right)\right]\right)^T\nabla\psi\left(\boldsymbol{g}_2\left(\boldsymbol{Y},\hat{\boldsymbol{\theta}}_2\right)\right)\right]$$

Let $\mathbf{x}$ and $\mathbf{z}$ be two real-valued vectors of the same dimension such that $\mathbb{E}\left[\mathbf{x}^T\mathbf{x}\right]<\infty$ and $\mathbb{E}\left[\mathbf{z}^T\mathbf{z}\right]<\infty$. Then we can always write

$$\mathbb{E}\left[\mathbf{x}^T\mathbf{z}\right]=\mathbb{E}\left[\operatorname{tr}\left(\mathbf{x}^T\mathbf{z}\right)\right]=\operatorname{tr}\left(\mathbb{E}\left[\mathbf{z}\mathbf{x}^T\right]\right)=\operatorname{tr}\left(\operatorname{cov}\left(\mathbf{x},\mathbf{z}\right)\right)+\operatorname{tr}\left(\mathbb{E}\left[\mathbf{x}\right]\mathbb{E}^T\left[\mathbf{z}\right]\right). \tag{A-3}$$

Using (A-3) we can express $\Delta$ as

$$\begin{aligned}\Delta=&\operatorname{tr}\left\{\operatorname{cov}\left[\boldsymbol{g}_1\left(\boldsymbol{Y},\hat{\boldsymbol{\theta}}_1\right)-\mathbb{E}_0\left[\boldsymbol{g}_1\left(\boldsymbol{Y}^0,\hat{\boldsymbol{\theta}}_1\right)\right],\nabla\psi\left(\boldsymbol{g}_2\left(\boldsymbol{Y},\hat{\boldsymbol{\theta}}_2\right)\right)\right]\right\}\\&+\operatorname{tr}\left\{\mathbb{E}\left[\boldsymbol{g}_1\left(\boldsymbol{Y},\hat{\boldsymbol{\theta}}_1\right)-\mathbb{E}_0\left[\boldsymbol{g}_1\left(\boldsymbol{Y}^0,\hat{\boldsymbol{\theta}}_1\right)\right]\right]\mathbb{E}^T\left[\nabla\psi\left(\boldsymbol{g}_2\left(\boldsymbol{Y},\hat{\boldsymbol{\theta}}_2\right)\right)\right]\right\}.\end{aligned}$$

Since $\mathbb{E}_0\left[\boldsymbol{g}_1\left(\boldsymbol{Y}^0,\hat{\boldsymbol{\theta}}_1\right)\right]$ is a non-stochastic quantity, it follows that

$$\begin{aligned}&\mathbb{E}\left[\mathbb{E}_0\left[D\left(\boldsymbol{g}_1\left(\boldsymbol{Y}^0,\hat{\boldsymbol{\theta}}_1\right),\boldsymbol{g}_2\left(\boldsymbol{Y},\hat{\boldsymbol{\theta}}_2\right)\right)\right]\right]=\\&\mathbb{E}\left[D\left(\boldsymbol{g}_1\left(\boldsymbol{Y},\hat{\boldsymbol{\theta}}_1\right),\boldsymbol{g}_2\left(\boldsymbol{Y},\hat{\boldsymbol{\theta}}_2\right)\right)\right]+\Delta=\\&\mathbb{E}\left[D\left(\boldsymbol{g}_1\left(\boldsymbol{Y},\hat{\boldsymbol{\theta}}_1\right),\boldsymbol{g}_2\left(\boldsymbol{Y},\hat{\boldsymbol{\theta}}_2\right)\right)\right]+\operatorname{tr}\left\{\operatorname{cov}\left[\boldsymbol{g}_1\left(\boldsymbol{Y},\hat{\boldsymbol{\theta}}_1\right),\nabla\psi\left(\boldsymbol{g}_2\left(\boldsymbol{Y},\hat{\boldsymbol{\theta}}_2\right)\right)\right]\right\}\end{aligned}$$

which verifies the result of Theorem 1. □

## D.2 Proof of Theorem 2

To prove Theorem 2, we first state and prove the following Lemma.

Lemma A-1: *Suppose that we have three competing models which can be written as:*

$$\begin{aligned}&\mathcal{M}_1:\ \boldsymbol{y}=\mathbf{X}^{(1)}\boldsymbol{\beta}_1+\boldsymbol{\varepsilon}_1\\&\mathcal{M}_2:\ \boldsymbol{y}=\mathbf{X}^{(1)}\boldsymbol{\beta}_1+\mathbf{X}^{(2)}\boldsymbol{\beta}_2+\boldsymbol{\varepsilon}_2\\&\mathcal{M}_3:\ \boldsymbol{y}=\mathbf{X}^{(1)}\boldsymbol{\beta}_1+\mathbf{X}^{(2)}\boldsymbol{\beta}_2+\mathbf{X}^{(3)}\boldsymbol{\beta}_3+\boldsymbol{\varepsilon}_3\end{aligned}$$

*where $\mathbf{X}^{(1)}$, $\mathbf{X}^{(2)}$ and $\mathbf{X}^{(3)}$ are respectively $n\times r_1$, $n\times r_2$ and $n\times r_3$ matrices and the dimensions of $\boldsymbol{\beta}_1$, of $\boldsymbol{\beta}_2$ and of $\boldsymbol{\beta}_3$ are defined accordingly. We assume that the matrix $\mathbf{X}^\star=\left[\mathbf{X}^{(1)}\ \ \mathbf{X}^{(2)}\ \ \mathbf{X}^{(3)}\right]$ is a non-random $n\times(r_1+r_2+r_3)$ full rank matrix. Moreover, under model $\mathcal{M}_l$ we suppose that $\boldsymbol{\varepsilon}_l\sim\mathcal{N}\left(\mathbf{0},\sigma_\varepsilon^2\mathbf{I}\right)$, $l=1,2,3$. Let also $\hat{\mathbf{Y}}_l$ denote the prediction function associated to model $\mathcal{M}_l$, $l=1$, 2, 3. Then the following results hold under the conditions of Theorem 1:*

1. *Assuming $\mathcal{M}_1$ or one of its sub-models to be the correct model, we have*

$$\mathbb{E}\left[||\hat{\mathbf{Y}}_1-\hat{\mathbf{Y}}_2||_2^2\right]=\sigma_\varepsilon^2 r_2. \tag{A-4}$$

*2. Assuming $\mathcal{M}_2$ to be the correct model, we have*

$$\mathbb{E}\left[||\hat{\boldsymbol{Y}}_1 - \hat{\boldsymbol{Y}}_2||_2^2\right] = \sigma_\varepsilon^2 r_2 + \mathbf{B}^T\mathbf{B} \tag{A-5}$$

*where*

$$\mathbf{B} = \left(\mathbf{I} - \mathbf{S}^{(1)}\right)\mathbf{X}^{(2)}\boldsymbol{\beta}_2$$
$$\mathbf{S}^{(1)} = \mathbf{X}^{(1)}\left(\mathbf{X}^{(1)T}\mathbf{X}^{(1)}\right)^{-1}\mathbf{X}^{(1)T}.$$

*3. Assuming $\mathcal{M}_3$ to be the correct model, we have*

$$\mathbb{E}\left[||\hat{\boldsymbol{Y}}_1 - \hat{\boldsymbol{Y}}_2||_2^2\right] = \sigma_\varepsilon^2 r_2 + \mathbf{B}^T\mathbf{B} + \mathbf{C}^T\mathbf{C} \tag{A-6}$$

*where*

$$\mathbf{C} = \left(\mathbf{S}^{(12)} - \mathbf{S}^{(1)}\right)\mathbf{X}^{(3)}\boldsymbol{\beta}_3$$
$$\mathbf{S}^{(12)} = \mathbf{X}^{(12)}\left(\mathbf{X}^{(12)T}\mathbf{X}^{(12)}\right)^{-1}\mathbf{X}^{(12)T}$$
$$\mathbf{X}^{(12)} = \left[\mathbf{X}^{(1)} \quad \mathbf{X}^{(2)}\right].$$

Proof: By definition $||\hat{\boldsymbol{Y}}_1 - \hat{\boldsymbol{Y}}_2||_2^2$ is a divergence measure. So by using (6) for the squared loss function, we have that

$$\mathbb{E}\left[||\hat{\boldsymbol{Y}}_1 - \hat{\boldsymbol{Y}}_2||_2^2\right] = \mathbb{E}\left[\hat{\boldsymbol{Y}}_1^T\hat{\boldsymbol{Y}}_1\right] - \mathbb{E}\left[\hat{\boldsymbol{Y}}_2^T\hat{\boldsymbol{Y}}_2\right] - 2\mathbb{E}\left[\left(\hat{\boldsymbol{Y}}_1 - \hat{\boldsymbol{Y}}_2\right)^T\hat{\boldsymbol{Y}}_2\right]$$

and under the conditions of Theorem 1 we have that $\mathbb{E}\left[\hat{\boldsymbol{Y}}_j^T\hat{\boldsymbol{Y}}_j\right] < \infty, j = 1, 2$. So we can write

$$\mathbb{E}\left[\hat{\boldsymbol{Y}}_j^T\hat{\boldsymbol{Y}}_k\right] = \operatorname{tr}\left[\operatorname{cov}\left(\hat{\boldsymbol{Y}}_j, \hat{\boldsymbol{Y}}_k\right)\right] + \operatorname{tr}\left(\mathbb{E}\left[\hat{\boldsymbol{Y}}_j\right]\mathbb{E}^T\left[\hat{\boldsymbol{Y}}_k\right]\right) \quad j,\, k = 1,\, 2$$

and we obtain

$$\begin{aligned}\mathbb{E}\left[||\hat{\boldsymbol{Y}}_1 - \hat{\boldsymbol{Y}}_2||_2^2\right] =& \operatorname{tr}\left(\operatorname{cov}\left(\hat{\boldsymbol{Y}}_1, \hat{\boldsymbol{Y}}_1\right)\right) + \operatorname{tr}\left(\operatorname{cov}\left(\hat{\boldsymbol{Y}}_2, \hat{\boldsymbol{Y}}_2\right)\right) \\ &- 2\operatorname{tr}\left(\operatorname{cov}\left(\hat{\boldsymbol{Y}}_1, \hat{\boldsymbol{Y}}_2\right)\right) + \operatorname{tr}\left(\mathbb{E}\left[\hat{\boldsymbol{Y}}_1\right]\mathbb{E}^T\left[\hat{\boldsymbol{Y}}_1\right]\right) \\ &+ \operatorname{tr}\left(\mathbb{E}\left[\hat{\boldsymbol{Y}}_2\right]\mathbb{E}^T\left[\hat{\boldsymbol{Y}}_2\right]\right) - 2\operatorname{tr}\left(\mathbb{E}\left[\hat{\boldsymbol{Y}}_1\right]\mathbb{E}^T\left[\hat{\boldsymbol{Y}}_2\right]\right).\end{aligned} \tag{A-7}$$

Since $\mathbf{X}^\star$ is full rank the matrices $\mathbf{S}^{(1)}$ and $\mathbf{S}^{(2)}$ exist and the first three elements of (A-7) can be simplified under model $\mathcal{M}_l$, $l = 1,\, 2,\, 3$, as

$$\begin{aligned}&\operatorname{tr}\left(\operatorname{cov}\left(\hat{\boldsymbol{Y}}_1, \hat{\boldsymbol{Y}}_1\right)\right) + \operatorname{tr}\left(\operatorname{cov}\left(\hat{\boldsymbol{Y}}_2, \hat{\boldsymbol{Y}}_2\right)\right) - 2\operatorname{tr}\left(\operatorname{cov}\left(\hat{\boldsymbol{Y}}_1, \hat{\boldsymbol{Y}}_2\right)\right) \\ &= \sigma_\varepsilon^2\operatorname{tr}\left(\mathbf{S}^{(1)} + \mathbf{S}^{(2)} - 2\mathbf{S}^{(1)}\mathbf{S}^{(2)}\right) = \sigma_\varepsilon^2\operatorname{tr}\left(\mathbf{S}^{(2)} - \mathbf{S}^{(1)}\right) = \sigma_\varepsilon^2 r_2.\end{aligned} \tag{A-8}$$

However, the last three elements of (A-7) depend on the assumed true model. Indeed, if $\mathcal{M}_1$ is the correct model, we have

$$\mathbb{E}\left[\hat{\boldsymbol{Y}}_1\right] = \mathbb{E}\left[\hat{\boldsymbol{Y}}_2\right]$$

and, therefore

$$\mathbb{E}\left[\hat{\mathbf{Y}}_1\right]\mathbb{E}^T\left[\hat{\mathbf{Y}}_1\right]+\mathbb{E}\left[\hat{\mathbf{Y}}_2\right]\mathbb{E}^T\left[\hat{\mathbf{Y}}_2\right]-2\mathbb{E}\left[\hat{\mathbf{Y}}_1\right]\mathbb{E}^T\left[\hat{\mathbf{Y}}_2\right]=0. \tag{A-9}$$

By using (A-8) and (A-9), we have that (A-7) simplifies to

$$\mathbb{E}\left[||\hat{\mathbf{Y}}_1-\hat{\mathbf{Y}}_2||_2^2\right]=\sigma_\varepsilon^2 r_2$$

which verifies (A-4). If $\mathcal{M}_2$ is the correct model, we have

$$\mathbb{E}\left[\hat{\mathbf{Y}}_1\right]=\mathbf{S}^{(1)}\left(\mathbf{X}^{(1)}\boldsymbol{\beta}_1+\mathbf{X}^{(2)}\boldsymbol{\beta}_2\right)=\mathbf{X}^{(1)}\boldsymbol{\beta}_1+\mathbf{S}^{(1)}\mathbf{X}^{(2)}\boldsymbol{\beta}_2$$

and we obtain

$$\begin{aligned}
&\operatorname{tr}\left(\mathbb{E}\left[\hat{\mathbf{Y}}_1\right]\mathbb{E}^T\left[\hat{\mathbf{Y}}_1\right]+\mathbb{E}\left[\hat{\mathbf{Y}}_2\right]\mathbb{E}^T\left[\hat{\mathbf{Y}}_2\right]-2\mathbb{E}\left[\hat{\mathbf{Y}}_1\right]\mathbb{E}^T\left[\hat{\mathbf{Y}}_2\right]\right)\\
&=\operatorname{tr}\left(\mathbf{S}^{(1)}\mathbf{X}^{(2)}\boldsymbol{\beta}_2\boldsymbol{\beta}_2^T\mathbf{X}^{(2)T}\mathbf{S}^{(1)}\right)+\operatorname{tr}\left(\mathbf{X}^{(2)}\boldsymbol{\beta}_2\boldsymbol{\beta}_2^T\mathbf{X}^{(2)T}\right)\\
&-2\operatorname{tr}\left(\mathbf{X}^{(2)}\boldsymbol{\beta}_2\boldsymbol{\beta}_2^T\mathbf{X}^{(2)T}\mathbf{S}^{(1)}\right)=\operatorname{tr}\left(\left(\mathbf{I}-\mathbf{S}^{(1)}\right)\mathbf{X}^{(2)}\boldsymbol{\beta}_2\boldsymbol{\beta}_2^T\mathbf{X}^{(2)T}\right)\\
&=\boldsymbol{\beta}_2^T\mathbf{X}^{(2)T}\left(\mathbf{I}-\mathbf{S}^{(1)}\right)\mathbf{X}^{(2)}\boldsymbol{\beta}_2=\mathbf{B}^T\mathbf{B}.
\end{aligned} \tag{A-10}$$

Using (A-8) and (A-10) simplifies (A-7) to

$$\mathbb{E}\left[||\hat{\mathbf{Y}}_1-\hat{\mathbf{Y}}_2||_2^2\right]=\sigma_\varepsilon^2 r_2+\mathbf{B}^T\mathbf{B},$$

thus verifying (A-5). Finally, if $\mathcal{M}_3$ is the correct model and since $\mathbf{X}^\star$ is full rank, we have

$$\begin{aligned}
\mathbb{E}\left[\hat{\mathbf{Y}}_1\right]&=\mathbf{X}^{(1)}\boldsymbol{\beta}_1+\mathbf{S}^{(1)}\mathbf{X}^{(23)}\boldsymbol{\beta}_{23}\\
\mathbb{E}\left[\hat{\mathbf{Y}}_2\right]&=\mathbf{X}^{(1)}\boldsymbol{\beta}_1+\mathbf{X}^{(2)}\boldsymbol{\beta}_2+\mathbf{S}^{(12)}\mathbf{X}^{(3)}\boldsymbol{\beta}_3
\end{aligned}$$

where

$$\begin{aligned}
\mathbf{X}^{(23)}&=\begin{bmatrix}\mathbf{X}^{(2)} & \mathbf{X}^{(3)}\end{bmatrix}\\
\boldsymbol{\beta}_{23}&=\begin{bmatrix}\boldsymbol{\beta}_2^T & \boldsymbol{\beta}_3^T\end{bmatrix}^T
\end{aligned}$$

and we obtain

$$\begin{aligned}
&\mathbb{E}\left[\hat{\mathbf{Y}}_1\right]\mathbb{E}^T\left[\hat{\mathbf{Y}}_1\right]+\mathbb{E}\left[\hat{\mathbf{Y}}_2\right]\mathbb{E}^T\left[\hat{\mathbf{Y}}_2\right]-2\mathbb{E}\left[\hat{\mathbf{Y}}_1\right]\mathbb{E}^T\left[\hat{\mathbf{Y}}_2\right]\\
&=\boldsymbol{\beta}_2^T\mathbf{X}^{(2)T}\mathbf{X}^{(2)}\boldsymbol{\beta}_2-\boldsymbol{\beta}_2^T\mathbf{X}^{(2)T}\mathbf{S}^{(1)}\mathbf{X}^{(2)}\boldsymbol{\beta}_2+\boldsymbol{\beta}_3^T\mathbf{X}^{(3)T}\mathbf{S}^{(12)}\mathbf{X}^{(3)}\boldsymbol{\beta}_3\\
&+\boldsymbol{\beta}_3^T\mathbf{X}^{(3)T}\mathbf{S}^{(1)}\mathbf{X}^{(3)}\boldsymbol{\beta}_3-2\boldsymbol{\beta}_3^T\mathbf{X}^{(3)T}\mathbf{S}^{(12)}\mathbf{S}^{(1)}\mathbf{X}^{(3)}\boldsymbol{\beta}_3\\
&=\mathbf{B}^T\mathbf{B}+\boldsymbol{\beta}_3^T\mathbf{X}^{(3)T}\mathbf{S}^{(12)}\mathbf{X}^{(3)}\boldsymbol{\beta}_3-\boldsymbol{\beta}_3^T\mathbf{X}^{(3)T}\mathbf{S}^{(1)}\mathbf{X}^{(3)}\boldsymbol{\beta}_3\\
&=\mathbf{B}^T\mathbf{B}+\boldsymbol{\beta}_3^T\mathbf{X}^{(3)T}\left(\mathbf{S}^{(12)}-\mathbf{S}^{(1)}\right)\mathbf{X}^{(3)}\boldsymbol{\beta}_3=\mathbf{B}^T\mathbf{B}+\mathbf{C}^T\mathbf{C}.
\end{aligned} \tag{A-11}$$

By using (A-8) and (A-11) we have that (A-7) simplifies to

$$\mathbb{E}\left[||\hat{\mathbf{Y}}_1-\hat{\mathbf{Y}}_2||_2^2\right]=\sigma_\varepsilon^2 r_2+\mathbf{B}^T\mathbf{B}+\mathbf{C}^T\mathbf{C}$$

thus verifying (A-6). □

Using Lemma A-1, the proof of Theorem 2 is the following.

PROOF: Using the definition of the $\widehat{\text{PDC}}$ and since $\sigma_\varepsilon^2$ is either known or estimated by an unbiased estimator, we have that

$$\mathbb{E}\left[\widehat{\text{PDC}}_{j,j+m}\right] = \mathbb{E}\left[||\hat{\boldsymbol{Y}}_j - \hat{\boldsymbol{Y}}_{j+m}||_2^2\right] + 2\sigma_\varepsilon^2 j. \tag{A-12}$$

Since

$$\mathbb{E}\left[\hat{\boldsymbol{Y}}_{\jmath_0}\right] = \mathbb{E}\left[\hat{\boldsymbol{Y}}_{\jmath_0+1}\right],$$

it follows, using (A-4), that

$$\mathbb{E}\left[\widehat{\text{PDC}}_{\jmath_0,\jmath_0+1}\right] = (2\jmath_0 + 1)\,\sigma_\varepsilon^2. \tag{A-13}$$

In order to demonstrate the results of Theorem 2, we have to show that $\mathbb{E}\left[\widehat{\text{PDC}}_{j,j+m}\right]$ is minimized for $j = \jmath_0$ and $m = 1$. We consider three different cases:

1. $0 \leq j + m \leq \jmath_0$: In this case the models considered for estimating $\hat{\boldsymbol{\beta}}_j$ and $\hat{\boldsymbol{\beta}}_{j+m}$ can be rewritten as:

$$\begin{aligned}
\boldsymbol{Y} &= \mathbf{X}_j \boldsymbol{\beta}_j^\star + \boldsymbol{\varepsilon}_j \\
\boldsymbol{Y} &= \mathbf{X}_{j+m}\boldsymbol{\beta}_{j+m}^{'} + \boldsymbol{\varepsilon}_k = \mathbf{X}_j\boldsymbol{\beta}_j^{'} + \mathbf{X}_{j+1:j+m}\boldsymbol{\beta}_{j+1:j+m}^{'} + \boldsymbol{\varepsilon}_k
\end{aligned}$$

where $\mathbf{X}_{j+1:j+m}$ is a $n \times m$ matrix that contains the columns $j+1$ to $j+m$ of the matrix $\mathbf{X}$. Under this setting we can use (A-6) of Lemma A-1 and obtain

$$\mathbb{E}\left[\widehat{\text{PDC}}_{j,j+m}\right] = (m + 2j)\,\sigma_\varepsilon^2 + \left(\mathbf{B}^T\mathbf{B} + \mathbf{C}^T\mathbf{C}\right)$$

where

$$\begin{aligned}
\mathbf{B} &= \left(\mathbf{I} - \mathbf{S}^{(1)}\right)\mathbf{X}_{j+1:j+m}\boldsymbol{\beta}_{j+1:j+m}^{'} \\
\mathbf{S}^{(1)} &= \mathbf{X}_j\left(\mathbf{X}_j^T\mathbf{X}_j\right)^{-1}\mathbf{X}_j^T \\
\mathbf{C} &= \left(\mathbf{S}^{(12)} - \mathbf{S}^{(1)}\right)\mathbf{X}_{j+m+1:\jmath_0}\boldsymbol{\beta}_{j+m+1:\jmath_0}^{'} \\
\mathbf{S}^{(12)} &= \mathbf{X}_{j+m}\left(\mathbf{X}_{j+m}^T\mathbf{X}_{j+m}\right)^{-1}\mathbf{X}_{j+m}^T
\end{aligned}$$

and where $\mathbf{X}_{j+m+1:\jmath_0}$ is a matrix that contains the columns $j+m+1$ to $\jmath_0$ of the matrix $\mathbf{X}$. Compared to (A-13), we have that for sufficiently large $n$

$$(2j + m)\,\sigma_\varepsilon^2 + \left(\mathbf{B}^T\mathbf{B} + \mathbf{C}^T\mathbf{C}\right) > (2\jmath_0 + 1)\,\sigma_\varepsilon^2 \tag{A-14}$$

since both $\mathbf{B}^T\mathbf{B}$ and $\mathbf{C}^T\mathbf{C}$ are positive strictly increasing functions of $n$. Hence for sufficiently large $n$ and $j + m \leq \jmath_0$, (17) is verified.

2. $0 < j < \jmath_0$ and $\jmath_0 < j + m \leq K + 1$: In this case we consider the following models for estimating $\boldsymbol{\beta}_j$ and $\boldsymbol{\beta}_{j+m}$:

$$\begin{aligned}
\boldsymbol{Y} &= \mathbf{X}_j\boldsymbol{\beta}_j^\star + \boldsymbol{\varepsilon}_j \\
\boldsymbol{Y} &= \mathbf{X}_{j+m}\boldsymbol{\beta}_{j+m}^{'} + \boldsymbol{\varepsilon}_{j+m} = \mathbf{X}_j\boldsymbol{\beta}_j^{'} + \mathbf{X}_{j+1:\jmath_0}\boldsymbol{\beta}_{j+1:\jmath_0}^{'} + \mathbf{X}_{\jmath_0+1:j+m}\boldsymbol{\beta}_{\jmath_0+1:j+m}^{'} + \boldsymbol{\varepsilon}_{j+m}.
\end{aligned}$$

It can be noted that in this case we have that

$$\mathbb{E}\left[\hat{\mathbf{Y}}_{j+m}\right] = \mathbb{E}\left[\hat{\mathbf{Y}}_{\jmath_0}\right] \tag{A-15}$$

so using (A-7) and (A-15) leads to

$$\begin{aligned}\mathbb{E}\left[\widehat{\mathrm{PDC}}_{j,j+m}\right] = \mathrm{tr}\Big(&\mathbb{E}\left[\hat{\mathbf{Y}}_j\right]\mathbb{E}^T\left[\hat{\mathbf{Y}}_j\right] + \mathbb{E}\left[\hat{\mathbf{Y}}_{\jmath_0}\right]\mathbb{E}^T\left[\hat{\mathbf{Y}}_{\jmath_0}\right] - 2\mathbb{E}\left[\hat{\mathbf{Y}}_j\right]\mathbb{E}^T\left[\hat{\mathbf{Y}}_{\jmath_0}\right]\Big)\\ &+ (2j+m)\,\sigma_\varepsilon^2.\end{aligned}$$

Moreover, using (A-10) in the proof of Lemma A-1 yields

$$\mathbb{E}\left[\widehat{\mathrm{PDC}}_{j,j+m}\right] = (2j+m)\,\sigma_\varepsilon^2 + \mathbf{B}^T\mathbf{B}$$

where

$$\begin{aligned}\mathbf{B} &= \left(\mathbf{I} - \mathbf{S}^{(1)}\right)\mathbf{X}_{j+1,\jmath_0}\boldsymbol{\beta}^\star_{j+1,\jmath_0}\\ \mathbf{S}^{(1)} &= \mathbf{X}_j\left(\mathbf{X}_j^T\mathbf{X}_j\right)^{-1}\mathbf{X}_j^T.\end{aligned}$$

Compared to (A-13), we have that, for sufficiently large $n$,

$$(2j+m)\,\sigma_\varepsilon^2 + \mathbf{B}^T\mathbf{B} > (2\jmath_0+1)\,\sigma_\varepsilon^2 \tag{A-16}$$

since $\mathbf{B}^T\mathbf{B}$ is a positive strictly increasing function of $n$. Hence for sufficiently large $n$, (17) is also verified when $0 < j < \jmath_0$ and $\jmath_0 < j + m \leq K + 1$.

3. $\jmath_0 \leq j < j+m$, $j+m \leq K+1$ and $m \neq 1$ if $j = \jmath_0$: In this case, we have that

$$\mathbb{E}\left[\hat{\mathbf{Y}}_j\right] = \mathbb{E}\left[\hat{\mathbf{Y}}_{j+m}\right].$$

Therefore,

$$\mathbb{E}\left[\widehat{\mathrm{PDC}}_{j,j+m}\right] = (2j+m)\,\sigma_\varepsilon^2$$

and we obtain

$$(2j+m)\,\sigma_\varepsilon^2 > (2\jmath_0+1)\,\sigma_\varepsilon^2 \tag{A-17}$$

since $\jmath_0 < j + \frac{m-1}{2}$. This verifies (17) when $\jmath_0 \leq j < j+m$, $j+m \leq K+1$ and $m \neq 1$ if $j = \jmath_0$.

Therefore, by combining (A-14), (A-16) and (A-17) we have that for sufficiently large $n$,

$$\mathbb{E}\left[\widehat{\mathrm{PDC}}_{\jmath_0,\jmath_0+1}\right] < \mathbb{E}\left[\widehat{\mathrm{PDC}}_{j,j+m}\right]$$

for $\forall j:\ 0 < j < K;\ \ \forall m:\ m > 0;\ j+m \leq K+1\ ;\ j = \jmath_0\ \Rightarrow m \neq 1$ which verifies (17) and therefore Theorem 2. □

## D.3 Proof of Theorem 3

Before considering the proof of Theorem 3, we state and prove the following lemmas.

Lemma A-2: *Assuming* $\mathbf{X}_{j+1}$ *is* $n \times (j+1)$ *of rank* $j+1$*, under Setting A we have that the vector* $\boldsymbol{\mu}$ *as defined in (12) is such that* $\forall j \notin \mathscr{J}_0$

$$\begin{aligned} \boldsymbol{\mu}^T \left(\mathbf{S}_{j+1} - \mathbf{S}_j\right) \boldsymbol{\mu} &= \mathcal{O}(n) \\ \boldsymbol{\mu}^T \left(\mathbf{S}_{j+1} - \mathbf{S}_j\right) \boldsymbol{\mu} &> 0 \end{aligned} \tag{A-18}$$

*and* $\forall j \in \mathscr{J}_0$

$$\begin{aligned} \boldsymbol{\mu}^T \left(\mathbf{S}_{j} - \mathbf{S}_{\jmath_0}\right) \boldsymbol{\mu} &= \mathcal{O}(n) \\ \boldsymbol{\mu}^T \left(\mathbf{S}_{j} - \mathbf{S}_{\jmath_0}\right) \boldsymbol{\mu} &\geq 0 \end{aligned} \tag{A-19}$$

Proof: We have that

$$\mathbf{H}_{k_2} = \mathbf{H}_{k_1} \left( \mathbf{I} - \frac{\mathbf{x}_{k_2}\mathbf{x}_{k_2}^T \mathbf{H}_{k_1}^T}{\mathbf{x}_{k_2}^T \mathbf{H}_{k_1} \mathbf{x}_{k_2}} \right) \tag{A-20}$$

with $k_1 < k_2 \leq j+1$, $\mathbf{H}_k = \mathbf{I} - \mathbf{S}_k$ and $\mathbf{x}_k$ denotes the $k$th column of $\mathbf{X}_{j+1}$. We also have, using (A-20) that

$$\begin{aligned} \boldsymbol{\mu}^T(\mathbf{S}_{j+1} - \mathbf{S}_j)\boldsymbol{\mu} &= \boldsymbol{\mu}^T(\mathbf{H}_j - \mathbf{H}_{j+1})\boldsymbol{\mu} && \text{(A-21)} \\ &= \frac{(\boldsymbol{\mu}^T \mathbf{H}_j \mathbf{x}_{j+1})(\mathbf{x}_{j+1}^T \mathbf{H}_j^T \boldsymbol{\mu})}{\mathbf{x}_{j+1}^T \mathbf{H}_j \mathbf{x}_{j+1}} = \mathcal{O}(n^{a-b}) && \text{(A-22)} \end{aligned}$$

with $a = 2, b = 1$. This proves the first equation in (A-18). The proof of the first equation in (A-19) is obtained by the same argument. When $j \notin \mathscr{J}_0$, $\boldsymbol{\mu}^T \mathbf{H}_j \mathbf{x}_{j+1} = 0$ if and only if $\mathbf{x}_{j+1} = \mathbf{0}$ which by assumption cannot be. Hence, since from (A-22) $\boldsymbol{\mu}^T(\mathbf{S}_{j+1} - \mathbf{S}_j)\boldsymbol{\mu}$ is the ratio of quadratic forms, this proves the second equation in (A-18). When $j \in \mathscr{J}_0$, $\boldsymbol{\mu}^T(\mathbf{S}_j - \mathbf{S}_{\jmath_0})\boldsymbol{\mu} = 0$ when $j = \jmath_0$ and positive otherwise, which proves the second equation of (A-19). □

Lemma A-3: *Assuming* $j \notin \mathscr{J}_0$*, then under Setting A and Assumptions (A.1) and (A.2) we have that*

$$\lim_{n \to \infty} \Pr\left(\hat{\jmath}_{\lambda_n} = j\right) = 0.$$

Proof: The model $j \notin \mathscr{J}_0$ will be selected if

$$\widehat{\text{PDC}}_{j,j+1}^{\lambda_n} < \min_{\substack{i \in \mathscr{J}_0 \\ i \neq j}} \widehat{\text{PDC}}_{i,i+1}^{\lambda_n}.$$

Therefore, we have

$$\begin{aligned} \Pr\left(\hat{\jmath}_{\lambda_n} = j\right) &= \Pr\left( \widehat{\text{PDC}}_{j,j+1}^{\lambda_n} < \min_{\substack{i \in \mathscr{J}_0 \\ i \neq j}} \widehat{\text{PDC}}_{i,i+1}^{\lambda_n} \right) \\ &\leq \Pr\left( \widehat{\text{PDC}}_{j,j+1}^{\lambda_n} < \widehat{\text{PDC}}_{\jmath_0,\jmath_0+1}^{\lambda_n} \right) = \Pr\left( X_j^{\lambda_n} \leq 0 \right) \end{aligned} \tag{A-23}$$

where

$$X_j^{\lambda_n} = \frac{||\hat{\boldsymbol{Y}}_j - \hat{\boldsymbol{Y}}_{j+1}||_2^2}{\hat{\sigma}_\varepsilon^2} - \frac{||\hat{\boldsymbol{Y}}_{\jmath_0} - \hat{\boldsymbol{Y}}_{\jmath_0+1}||_2^2}{\hat{\sigma}_\varepsilon^2} + \lambda_n (j - \jmath_0)$$
$$= \frac{\sigma_\varepsilon^2}{\hat{\sigma}_\varepsilon^2} \left( \frac{||\hat{\boldsymbol{Y}}_j - \hat{\boldsymbol{Y}}_{j+1}||_2^2}{\sigma_\varepsilon^2} - \frac{||\hat{\boldsymbol{Y}}_{\jmath_0} - \hat{\boldsymbol{Y}}_{\jmath_0+1}||_2^2}{\sigma_\varepsilon^2} \right) + \lambda_n (j - \jmath_0).$$

The term $\sigma_\varepsilon^{-2} ||\hat{\boldsymbol{Y}}_j - \hat{\boldsymbol{Y}}_{j+1}||_2^2$ follows a non-central $\chi^2$ under the assumption of normally distributed errors since the matrix $(\mathbf{S}_{j+1} - \mathbf{S}_j)$ is both symmetric and idempotent. So we have (see (A-41) in Appendix H for details)

$$\frac{||\hat{\boldsymbol{Y}}_j - \hat{\boldsymbol{Y}}_{j+1}||_2^2}{\sigma_\varepsilon^2} \sim \chi_1^2 \left(\gamma_n^2\right)$$

where $\gamma_n^2 = \sigma_\varepsilon^{-2} \boldsymbol{\mu}^T (\mathbf{S}_{j+1} - \mathbf{S}_j) \boldsymbol{\mu}$ and using (A-18) in Lemma A-2 we have that $\gamma_n^2 = \mathcal{O}(n)$ since $j \notin \mathscr{J}_0$. Let $Y_{1,\gamma_n} \sim \chi_1^2 \left(\gamma_n^2\right)$ and $Z_1 \sim \mathcal{N}(0,1)$, we can always write that $Y_{1,\gamma_n} = (Z_1 + \gamma_n)^2 = Z_1^2 + \gamma_n^2 + 2Z_1\gamma_n$. The second term $\sigma_\varepsilon^{-2} ||\hat{\boldsymbol{Y}}_{\jmath_0} - \hat{\boldsymbol{Y}}_{\jmath_0+1}||_2^2$ also follows a $\chi_1^2$ distribution and is independent from $\sigma^{-2} ||\hat{\boldsymbol{Y}}_j - \hat{\boldsymbol{Y}}_{j+1}||_2^2$ by (Craig, 1943, Theorem 2) (see also Appendix H) since $(\mathbf{S}_{j+1} - \mathbf{S}_j)(\mathbf{S}_{j_0+1} - \mathbf{S}_{j_0}) = \mathbf{0}$ and therefore is also independent from $Z_1$. Let $Z_2 \sim \mathcal{N}(0,1)$ be independent from $Z_1$ and let

$$U = \frac{Z_1^2}{\sqrt{n}} - \frac{Z_2^2}{\sqrt{n}}.$$

Using Chebyshev's inequality we can show that $U = o_p(1)$. Indeed, for any given $\epsilon > 0$, we have

$$\lim_{n\to\infty} \Pr\left(|U - \mathbb{E}[U]| \geq \epsilon\right) = \lim_{n\to\infty} \Pr\left(|U| \geq \epsilon\right) \leq \lim_{n\to\infty} \frac{\text{var}(U)}{\epsilon^2} = \lim_{n\to\infty} \frac{4}{\epsilon^2 n} = 0.$$

Moreover, we also have that

$$\frac{2\gamma_n Z_1}{\sqrt{n}} \xrightarrow{\mathcal{L}} c\, Z_1$$

where $c$ is a positive constant since $\gamma_n > 0$ and $\gamma_n = \mathcal{O}(\sqrt{n})$ for $j \notin \mathscr{J}_0$ using (A-18) in Lemma A-2. By Slutsky's theorem we have that

$$\frac{\sigma^2}{\hat{\sigma}^2} \left( \frac{2\gamma_n Z_1}{\sqrt{n}} + \frac{Z_1^2}{\sqrt{n}} - \frac{Z_2^2}{\sqrt{n}} \right) \xrightarrow{\mathcal{L}} c\, Z_1$$

since by Assumption (A.1) we have that

$$\frac{\sigma_\varepsilon^2}{\hat{\sigma}_\varepsilon^2} \xrightarrow{\mathcal{P}} 1.$$

We can rewrite $X_j^{\lambda_n}$ as follows

$$X_j^{\lambda_n} = \sqrt{n} \left[ \frac{\sigma_\varepsilon^2}{\hat{\sigma}_\varepsilon^2} \left( \frac{2\gamma_n Z_1}{\sqrt{n}} + \frac{Z_1^2}{\sqrt{n}} - \frac{Z_2^2}{\sqrt{n}} \right) \right] + \frac{\sigma_\varepsilon^2}{\hat{\sigma}_\varepsilon^2} \gamma_n^2 + \lambda_n (j - \jmath_0).$$

Since $n^{-1/2} X_j^{\lambda_n} \neq o_p(1)$, we have that

$$\begin{aligned}
\lim_{n\to\infty} \Pr\left(X_j^{\lambda_n} \leq 0\right) &= \lim_{n\to\infty} \Pr\left(\frac{X_j^{\lambda_n}}{\sqrt{n}} \leq 0\right) \\
&= \lim_{n\to\infty} \Pr\left(cZ_1 + \frac{\gamma_n^2}{\sqrt{n}} + \frac{\lambda_n}{\sqrt{n}}(j - \jmath_0) \leq 0\right) \\
&= \lim_{n\to\infty} \Pr\left(Z_1 \leq -\frac{\gamma_n^2}{c\sqrt{n}} + \frac{\lambda_n(\jmath_0 - j)}{c\sqrt{n}}\right) \\
&\overset{(A.2)}{=} \lim_{n\to\infty} \lim_{n\to\infty} \Pr\left(Z_1 \leq -\frac{\gamma_n^2}{c\sqrt{n}} + d\right) \\
&= \lim_{n\to\infty} \Pr\left(Z_1 \leq -c^\star \sqrt{n} + d\right) = 0
\end{aligned} \tag{A-24}$$

where $c^\star$ is a positive constant, $d$ a non-negative constant and the last equality in (A-24) is obtained using (A-18) in Lemma A-2. This verifies the result of Lemma A-3. □

Using Lemma A-3, the proof of Theorem 3 is the following.

PROOF: As the events $\{\hat{\jmath}_{\lambda_n} = j\}$, $\{\hat{\jmath}_{\lambda_n} = k\}$ with $j \neq k$, $j \in \mathscr{J}$ and $k \in \mathscr{J}$ are mutually exclusive, we have that

$$\Pr\left(\hat{\jmath}_{\lambda_n} \notin \mathscr{J}_0\right) = \Pr\left(\hat{\jmath}_{\lambda_n} = 0 \cup \hat{\jmath}_{\lambda_n} = 1 \cup ... \cup \hat{\jmath}_{\lambda_n} = (\jmath_0 - 1)\right) = \sum_{j=1}^{\jmath_0} \Pr\left(\hat{\jmath}_{\lambda_n} = j - 1\right).$$

From the proof of Lemma A-3 and by combining (A-23) and (A-24) we get

$$\Pr\left(\hat{\jmath}_{\lambda_n} = j\right) \leq \lim_{n\to\infty} \Pr\left(Z \leq -\frac{\gamma_n^2}{c\sqrt{n}} + \frac{\lambda_n(\jmath_0 - j)}{c\sqrt{n}}\right)$$

where $Z \sim \mathcal{N}(0,1)$ and $c$ is a positive constant. Moreover, by the tail probability of the standard normal distribution we get the following inequality:

$$\Pr\left(|Z| \geq t\right) \leq \exp\left(-\tfrac{t^2}{2}\right).$$

Thus, we have

$$\begin{aligned}
\lim_{n\to\infty} \Pr\left(Z \leq -\frac{\gamma_n^2}{c\sqrt{n}} + \frac{\lambda_n(\jmath_0 - j)}{c\sqrt{n}}\right) &\leq \lim_{n\to\infty} \exp\left(-\frac{1}{2}\left(-\frac{\gamma_n^2}{c\sqrt{n}} + \frac{\lambda_n(\jmath_0 - j)}{c\sqrt{n}}\right)^2\right) \\
= \lim_{n\to\infty} \exp\left(-\frac{\gamma_n^4}{2c^2 n}\right) \exp\left(-\frac{\lambda_n^2(\jmath_0 - j)^2}{2c^2 n}\right) &\exp\left(-\frac{\gamma_n^2\lambda_n j}{c^2 n}\right) \exp\left(\frac{\gamma_n^2\lambda_n \jmath_0}{c^2 n}\right).
\end{aligned}$$

Let

$$a_j = \exp\left(-\frac{\gamma_n^4}{2c^2 n}\right) \exp\left(-\frac{\lambda_n^2(\jmath_0 - j)^2}{2c^2 n}\right) \exp\left(-\frac{\gamma_n^2\lambda_n j}{c^2 n}\right) \exp\left(\frac{\gamma_n^2\lambda_n \jmath_0}{c^2 n}\right).$$

Then we can write

$$\lim_{n\to\infty} \Pr\left(\hat{\jmath}_{\lambda_n} \notin \mathscr{J}_0\right) = \lim_{n\to\infty} \sum_{j=1}^{\jmath_0} \Pr\left(\hat{\jmath}_{\lambda_n} = j - 1\right) \leq \lim_{n\to\infty} \sum_{j=1}^{\jmath_0} a_j.$$

Let

$$b_j = 0$$
$$c_j = \exp\left(-\frac{\gamma_n^4}{2c^2 n}\right)\exp\left(\frac{\gamma_n^2 \lambda_n \jmath_0}{c^2 n}\right).$$

Then we have that $b_j \le a_j \le c_j$, $\forall j$, $\forall n$ since $\lambda_n > 0$ by (A.2), $\gamma_n > 0$ by (A-18) in Lemma A-2 and $j \ge 0$. Clearly,

$$\lim_{n\to\infty}\sum_{j=1}^{\jmath_0} b_j = 0.$$

In addition, we have that

$$\lim_{n\to\infty}\sum_{j=1}^{\jmath_0} c_j = \lim_{n\to\infty} \jmath_0 \exp\left(-\frac{\gamma_n^4}{2c^2 n}\right)\exp\left(\frac{\gamma_n^2 \lambda_n \jmath_0}{c^2 n}\right).$$

Using (A-18) in Lemma A-2 we have that $\gamma_n^2 = \mathcal{O}(n)$ and by Assumption (A.2) we have that $\lambda_n \jmath_0 = \mathcal{O}(\sqrt{n})$ and consequently that $\jmath_0 = \mathcal{O}(\sqrt{n})$. Thus, we may write

$$\lim_{n\to\infty} \jmath_0 \exp\left(-\frac{\gamma_n^4}{2c^2 n}\right)\exp\left(\frac{\gamma_n^2 \lambda_n \jmath_0}{c^2 n}\right) = \lim_{n\to\infty} c_1\sqrt{n}\exp\left(-c_2\, n + c_3\sqrt{n}\right)$$
$$= c_1 \lim_{n\to\infty} \frac{\sqrt{n}}{\exp\left(c_2\, n - c_3\sqrt{n}\right)} = \tfrac{1}{2}\, c_1 \lim_{n\to\infty} \frac{n^{-1/2}}{\exp\left(c_2\, n - c_3\sqrt{n}\right)\left(c_2 - \frac{c_3}{2\sqrt{n}}\right)}$$
$$= c_4 \lim_{n\to\infty} \exp\left(-c_2\, n\right) = 0$$

where $c_1$, $c_3$ and $c_4$ are non-negative constants and $c_2$ is a positive constant. Since we have the following inequality

$$\lim_{n\to\infty}\sum_{j=1}^{\jmath_0} b_j \le \lim_{n\to\infty}\sum_{j=1}^{\jmath_0} a_j \le \lim_{n\to\infty}\sum_{j=1}^{\jmath_0} c_j$$

we also have that

$$\lim_{n\to\infty} \Pr\left(\hat{\jmath}_{\lambda_n} \notin \mathscr{J}_0\right) \le \lim_{n\to\infty}\sum_{j=1}^{\jmath_0} a_j = 0$$

which concludes the proof. □

## D.4 Proof of Theorem 4

Proof: We express the inequality $\widehat{\mathrm{PDC}}^{\lambda_n}_{j,j+m} \le \widehat{\mathrm{PDC}}^{\lambda_n}_{\jmath_0,\jmath_0+m}$ as

$$\frac{\sigma_\varepsilon^2}{\hat{\sigma}_\varepsilon^2}\left(\frac{||\hat{\boldsymbol{Y}}_j - \hat{\boldsymbol{Y}}_{j+m}||_2^2}{\sigma_\varepsilon^2} - \frac{||\hat{\boldsymbol{Y}}_{\jmath_0} - \hat{\boldsymbol{Y}}_{\jmath_0+m}||_2^2}{\sigma_\varepsilon^2}\right) \le \lambda_n\, m.$$

We also have that

$$||\hat{\boldsymbol{Y}}_j - \hat{\boldsymbol{Y}}_{j+m}||_2^2 = \boldsymbol{Y}^T\left(\mathbf{S}_j - \mathbf{S}_{j+m}\right)^T\left(\mathbf{S}_j - \mathbf{S}_{j+m}\right)\boldsymbol{Y} = \boldsymbol{Y}^T\left(\mathbf{S}_{j+m} - \mathbf{S}_j\right)\boldsymbol{Y}$$
$$||\hat{\boldsymbol{Y}}_{\jmath_0} - \hat{\boldsymbol{Y}}_{\jmath_0+m}||_2^2 = \boldsymbol{Y}^T\left(\mathbf{S}_{\jmath_0} - \mathbf{S}_{\jmath_0+m}\right)^T\left(\mathbf{S}_{\jmath_0} - \mathbf{S}_{\jmath_0+m}\right)\boldsymbol{Y} = \boldsymbol{Y}^T\left(\mathbf{S}_{\jmath_0+m} - \mathbf{S}_{\jmath_0}\right)\boldsymbol{Y}$$

and since the matrices $\mathbf{S}_{j+m} - \mathbf{S}_j$ and $\mathbf{S}_{\jmath_0+m} - \mathbf{S}_{\jmath_0}$ are symmetric and idempotent, we have

$$\frac{||\hat{\boldsymbol{Y}}_j - \hat{\boldsymbol{Y}}_{j+m}||_2^2}{\sigma_\varepsilon^2} \sim \chi_m^2,$$
$$\frac{||\hat{\boldsymbol{Y}}_{\jmath_0} - \hat{\boldsymbol{Y}}_{\jmath_0+m}||_2^2}{\sigma_\varepsilon^2} \sim \chi_m^2.$$

Moreover, we remark that $\boldsymbol{Y}^T\left(\mathbf{S}_{j+m} - \mathbf{S}_j\right)\boldsymbol{Y}$ and $\boldsymbol{Y}^T\left(\mathbf{S}_{\jmath_0+m} - \mathbf{S}_{\jmath_0}\right)\boldsymbol{Y}$ are independent since $\left(\mathbf{S}_{j+m} - \mathbf{S}_j\right)\left(\mathbf{S}_{\jmath_0+m} - \mathbf{S}_{\jmath_0}\right) = \mathbf{0}$ (see Craig, 1943, Theorem 2 or Appendix H). Now, let

$$X_1 = \frac{||\hat{\boldsymbol{Y}}_j - \hat{\boldsymbol{Y}}_{j+m}||_2^2}{\sigma_\varepsilon^2},$$
$$X_2 = \frac{||\hat{\boldsymbol{Y}}_{\jmath_0} - \hat{\boldsymbol{Y}}_{\jmath_0+m}||_2^2}{\sigma_\varepsilon^2},$$
$$\mathcal{G}_m = X_1 - X_2.$$

Then, the Moment Generating Function (MGF) associated to $X_1$ and $X_2$ is

$$M_{X_1}(t) = M_{X_2}(t) = \left(\frac{1}{1-2t}\right)^{\frac{m}{2}}, \quad t < \frac{1}{2}$$

and the MGF of $\mathcal{G}_m$ is

$$M_{\mathcal{G}_m}(t) = M_{X_1}(t)\, M_{X_2}(-t) = \left(\frac{1}{1-4t^2}\right)^{\frac{m}{2}} = \left(\frac{\left(\frac{1}{2}\right)^2}{\left(\frac{1}{2}\right)^2 - t^2}\right)^{\frac{m}{2}}, \quad t < \frac{1}{2}.$$

Let $Y$ follow a variance-gamma distribution. From (A-37) we know that the MGF associated to $Y$ is given by

$$M_Y(t) = \exp(\mu t)\left(\frac{\alpha^2 - \beta^2}{\alpha^2 - (\beta + t)^2}\right)^{\xi}.$$

Clearly, with the set of parameters $\mu = 0, \beta = 0, \alpha = \frac{1}{2}$ and $\xi = \frac{m}{2}$ the MGFs of $Y$ and $\mathcal{G}_m$ are the same for all $t$ in some neighborhood of 0. This implies that $F_{\mathcal{G}_m}(u) = F_Y(u)$, $\forall u$. Thus, $\mathcal{G}_m$ follows a modified variance-gamma distribution whose PDF is given in (A-38). Therefore, when $\sigma_\varepsilon^2$ is known, we have

$$\Pr\left(\widehat{\text{PDC}}^{\lambda_n}_{j,j+m} \leq \widehat{\text{PDC}}^{\lambda_n}_{\jmath_0,\jmath_0+m}\right) = \Pr\left(\mathcal{G}_m \leq \lambda_n\, m\right). \tag{A-25}$$

In the case where $\sigma_\varepsilon^2$ is unknown, Assumption (A.1) enables to write

$$\frac{\sigma_\varepsilon^2}{\hat{\sigma}_\varepsilon^2} \xrightarrow{\mathcal{P}} 1.$$

So by Slutsky's theorem we obtain that

$$\frac{\sigma_\varepsilon^2}{\hat{\sigma}_\varepsilon^2}\left(\frac{||\hat{\boldsymbol{Y}}_j - \hat{\boldsymbol{Y}}_{j+m}||_2^2}{\sigma_\varepsilon^2} - \frac{||\hat{\boldsymbol{Y}}_{\jmath_0} - \hat{\boldsymbol{Y}}_{\jmath_0+m}||_2^2}{\sigma_\varepsilon^2}\right) \xrightarrow{\mathcal{L}} \mathcal{G}_m.$$

Therefore,

$$\lim_{n\to\infty} \Pr\left(\widehat{\mathrm{PDC}}_{j,j+m}^{\lambda_n} \leq \widehat{\mathrm{PDC}}_{\jmath_0,\jmath_0+m}^{\lambda_n}\right) = \lim_{n\to\infty} \Pr\left(\mathcal{G}_m \leq \lambda_n\, m\right)$$

which concludes the proof. □

## D.5 Asymptotic Probability of Overfitting

Let $Z_l^2 \overset{iid}{\sim} \chi_1^2,\;\; l = 1, ..., k^\star+1$. Then we define the random variable $\mathcal{W}_l$ as $\mathcal{W}_l = Z_l^2 - Z_{l+1}^2,\; l = 1, 2, ..., k^\star$. In addition let $a \in \mathbb{N}$ and $b \in \mathbb{N}$ be such that $k^\star = a + b$. We define the following probability:

$$\begin{aligned}\alpha_{a,b}^{\lambda_n} = \Pr\big(&\mathcal{W}_1 \leq -\lambda_n a,\, \mathcal{W}_2 \leq -\lambda_n\,(a-1)\,,\, ...,\, \mathcal{W}_a \leq -\lambda_n,\\ &\mathcal{W}_{a+1} \leq \lambda_n,\, \mathcal{W}_{a+2} \leq 2\lambda_n,\, ...,\, \mathcal{W}_{a+b} \leq \lambda_n b\big).\end{aligned} \tag{A-26}$$

Then, the asymptotic probability of overfitting for $\widehat{\mathrm{PDC}}_{j,j+1}^{\lambda_n}$ is given in the following theorem.

Theorem A-7: *Let $j \in \mathscr{J}$, then under Setting A and Assumptions (A.1) and (A.2) we have that*

$$\lim_{n\to\infty} \Pr\left(\hat{\jmath}_{\lambda_n} = j\right) = \lim_{n\to\infty} \mathbb{1}_{j\in\mathscr{J}_0}\; \alpha_{j-\jmath_0,K-\jmath_0}^{\lambda_n}$$

*where $\mathbb{1}$ denotes the indicator function.*

To prove Theorem A-7 we first state and prove the following lemmas.

Lemma A-4: *Let*

$$\begin{aligned}Z_1^2 &= \frac{||\hat{\mathbf{Y}}_{\jmath_0} - \hat{\mathbf{Y}}_{\jmath_0+1}||_2^2}{\sigma_\varepsilon^2} = \frac{\boldsymbol{Y}^T\left(\mathbf{S}_{\jmath_0+1} - \mathbf{S}_{\jmath_0}\right)\boldsymbol{Y}}{\sigma_\varepsilon^2} \sim \chi_1^2\\ &\vdots\\ Z_{K-\jmath_0+1}^2 &= \frac{||\hat{\mathbf{Y}}_K - \hat{\mathbf{Y}}_{K+1}||_2^2}{\sigma_\varepsilon^2} = \frac{\boldsymbol{Y}^T\left(\mathbf{S}_{K+1} - \mathbf{S}_K\right)\boldsymbol{Y}}{\sigma_\varepsilon^2} \sim \chi_1^2.\end{aligned}$$

*Then all $Z_l^2,\; l = 1, ..., K - \jmath_0 + 1$ are all mutually independent.*

Proof: Let $\mathbf{A}$ and $\mathbf{B}$ be two idempotent $n\times n$ matrices and let $\boldsymbol{Y} \sim \mathcal{N}\left(\boldsymbol{\mu}, \sigma_\varepsilon^2\mathbf{I}\right)$. By Craig (1943, Theorem 2) (see also Appendix H) we have that two quadratic forms, say $q_1 = \boldsymbol{Y}^T\mathbf{A}\boldsymbol{Y}$ and $q_2 = \boldsymbol{Y}^T\mathbf{B}\boldsymbol{Y}$, are independently distributed as $\chi^2$ if $\mathbf{AB} = \mathbf{0}$. Therefore, all $Z_l^2,\; l = 1, ..., K - \jmath_0 + 1$ are all mutually independent in the case where $\mathbf{S}_{i,i'} = \left(\mathbf{S}_{i+1} - \mathbf{S}_i\right)\left(\mathbf{S}_{i'+1} - \mathbf{S}_{i'}\right) = \mathbf{0}$ for any $i \in \mathscr{J}_0$, $i' \in \mathscr{J}_0$ and $i > i'$. Consider the term $\mathbf{S}_{i,i'}$. We have that

$$\begin{aligned}\mathbf{S}_{i,i'} &= \mathbf{S}_{\min\{i,i'\}+1} - \mathbf{S}_{\min\{i+1,i'\}} - \mathbf{S}_{\min\{i,i'+1\}} + \mathbf{S}_{\min\{i,i'\}}\\ &\overset{i>i'}{=} \mathbf{S}_{i'+1} - \mathbf{S}_{i'} - \mathbf{S}_{i'+1} + \mathbf{S}_{i'} = \mathbf{0}\end{aligned}$$

and thus we verify that all $Z_l^2,\; l = 1, ..., K - \jmath_0 + 1$ are all mutually independent and conclude the proof. □

Lemma A-5: *Let the model $j \in \mathscr{J}_0$. Then if $\sigma_\varepsilon^2$ is known we have that*

$$\Pr\left(\hat{\jmath}_{\lambda_n} = j\right) = \alpha^{\lambda_n}_{j-\jmath_0, K-\jmath_0}.$$

Proof: The probability of selecting model $j \in \mathscr{J}_0$ can be expressed as:

$$\begin{aligned}
\Pr\left(\hat{\jmath}_{\lambda_n} = j | j \in \mathscr{J}_0\right) = &\Pr\left(\widehat{\text{PDC}}^{\lambda_n}_{j,j+1} < \min_{\substack{i \in \mathscr{J}_0 \\ i \neq j}} \widehat{\text{PDC}}^{\lambda_n}_{i,i+1}\right) \\
= &\Pr\left(\frac{||\hat{\mathbf{Y}}_j - \hat{\mathbf{Y}}_{j+1}||_2^2}{\sigma_\varepsilon^2} - \frac{||\hat{\mathbf{Y}}_{\jmath_0} - \hat{\mathbf{Y}}_{\jmath_0+1}||_2^2}{\sigma_\varepsilon^2} < \lambda\left(\jmath_0 - j\right), \ldots, \right. \\
&\frac{||\hat{\mathbf{Y}}_j - \hat{\mathbf{Y}}_{j+1}||_2^2}{\sigma_\varepsilon^2} - \frac{||\hat{\mathbf{Y}}_{j-1} - \hat{\mathbf{Y}}_{j}||_2^2}{\sigma_\varepsilon^2} < -\lambda_n, \\
&\frac{||\hat{\mathbf{Y}}_j - \hat{\mathbf{Y}}_{j+1}||_2^2}{\sigma_\varepsilon^2} - \frac{||\hat{\mathbf{Y}}_{j+1} - \hat{\mathbf{Y}}_{j+2}||_2^2}{\sigma_\varepsilon^2} < \lambda_n, \ldots, \\
&\left.\frac{||\hat{\mathbf{Y}}_j - \hat{\mathbf{Y}}_{j+1}||_2^2}{\sigma_\varepsilon^2} - \frac{||\hat{\mathbf{Y}}_{K} - \hat{\mathbf{Y}}_{K+1}||_2^2}{\sigma_\varepsilon^2} < \lambda_n\left(K - j\right)\right).
\end{aligned}$$

Let $Z_l^2$, $l = 1, \ldots, K - \jmath_0 + 1$ be defined as in Lemma A-4 so that we have that $Z_l^2 \overset{iid}{\sim} \chi_1^2$, $l = 1, \ldots, K - \jmath_0 + 1$ and let $\mathcal{W}_l = Z_l^2 - Z_{l+1}^2$, $l = 1, 2, \ldots, j-1, j+1, \ldots, K - \jmath_0$. Then, we may rewrite $\Pr\left(\hat{\jmath}_{\lambda_n} = j | j \in \mathscr{J}_0\right)$ as

$$\begin{aligned}
\Pr\left(\hat{\jmath}_{\lambda_n} = j | j \in \mathscr{J}_0\right) &= \Pr\left(\mathcal{W}_1 < -\lambda_n\left(j - \jmath_0\right), \ldots, \mathcal{W}_{K-\jmath_0} < \lambda_n\left(K - \jmath_0\right)\right) \\
&= \alpha^{\lambda_n}_{j-\jmath_0, K-\jmath_0}
\end{aligned}$$

which concludes the proof. □

The proof of Theorem A-7 can now be given.

Proof: This proof is straightforward from Lemma A-5 and Theorem 3. Nevertheless, we provide here a proof for completeness. By applying the law of total probability we have the following equality:

$$\begin{aligned}
\lim_{n\to\infty} \Pr\left(\hat{\jmath}_{\lambda_n} = j\right) = &\lim_{n\to\infty} \Pr\left(j \in \mathscr{J}_0\right) \Pr\left(\hat{\jmath}_{\lambda_n} = j | \, j \in \mathscr{J}_0\right) \\
&+ \lim_{n\to\infty} \Pr\left(j \notin \mathscr{J}_0\right) \Pr\left(\hat{\jmath}_{\lambda_n} = j | \, j \notin \mathscr{J}_0\right) \\
= &\mathbb{1}_{j \in \mathscr{J}_0} \lim_{n\to\infty} \Pr\left(\hat{\jmath}_{\lambda_n} = j | \, j \in \mathscr{J}_0\right).
\end{aligned}$$

Since by Assumption (A.1) we have that

$$\frac{\sigma^2}{\hat{\sigma}^2} \overset{\mathcal{P}}{\longrightarrow} 1$$

then by Slutsky's theorem,

$$\frac{\sigma^2}{\hat{\sigma}^2}\mathcal{W}_i \overset{\mathcal{L}}{\longrightarrow} \mathcal{W}_i, \; i = 1, 2, \ldots, j-1, j+1, \ldots, K - \jmath$$

which implies that

$$\lim_{n\to\infty} \Pr\left(\hat{\jmath}_{\lambda} = j \middle| j \in \mathscr{J}_0\right) = \lim_{n\to\infty} \alpha^{\lambda_n}_{j-\jmath_0, K-\jmath_0}.$$

Finally, we obtain

$$\lim_{n\to\infty} \Pr\left(\hat{\jmath}_{\lambda_n} = j\right) = \lim_{n\to\infty} \mathbb{1}_{j\in\mathscr{J}_0}\, \alpha^{\lambda_n}_{j-\jmath_0, K-\jmath_0}$$

which concludes the proof. □

## D.6 Proof of Theorem 5

To prove Theorem 5, we first need to state and prove the corrolaries.

Corollary A-2: *Let $\boldsymbol{\mu} = \mathbf{X}\boldsymbol{\beta}_{\jmath_0} + \boldsymbol{\delta}$. Then if $\mathscr{J} = \mathscr{J}_0$ and if $\sigma^2_\varepsilon$ is known, we have that*

$$\mathbb{E}\left[L_n\left(\hat{\jmath}_{\lambda_n}\right)\right] = \frac{\boldsymbol{\delta}^T\boldsymbol{\delta}}{n} + \frac{\sigma^2_\varepsilon}{n}\left(\jmath_0 + \sum_{j=\jmath_0+1}^{K} j\, \alpha^{\lambda_n}_{j-\jmath_0, K-\jmath_0}\right).$$

*and under the assumptions of Theorem A-7, we have that*

$$\lim_{n\to\infty} \mathbb{E}\left[n\, L_n\left(\hat{\jmath}_{\lambda_n}\right)\right] = \lim_{n\to\infty}\left[\boldsymbol{\delta}^T\boldsymbol{\delta} + \sigma^2_\varepsilon\left(\jmath_0 + \sum_{j=1}^{K-\jmath_0} j\, \alpha^{\lambda_n}_{j-\jmath_0, K-\jmath_0}\right)\right].$$

Proof: Let $R_n(j) = \mathbb{E}\left[L_n(j)\right]$. Then if $j \in \mathscr{J}_0$ we have that

$$R_n(j) = \frac{\sigma^2_\varepsilon j}{n} + \frac{\boldsymbol{\delta}^T\boldsymbol{\delta}}{n}.$$

Since $\mathscr{J} = \mathscr{J}_0$ we have that

$$\begin{aligned} R_n\left(\hat{\jmath}_{\lambda_n}\right) &= \sum_{j\in\mathscr{J}_0} \Pr\left(\hat{\jmath}_{\lambda_n} = j\right) R_n(j) = \frac{\boldsymbol{\delta}^T\boldsymbol{\delta}}{n} + \frac{\sigma^2_\varepsilon}{n}\left(\jmath_0 + \sum_{j=\jmath_0+1}^{K} j\, \Pr\left(\hat{\jmath}_{\lambda_n} = j\right)\right) \\ &= \frac{\boldsymbol{\delta}^T\boldsymbol{\delta}}{n} + \frac{\sigma^2_\varepsilon}{n}\left(\jmath_0 + \sum_{j=\jmath_0+1}^{K} j\, \alpha^{\lambda_n}_{j-\jmath_0, K-\jmath_0}\right) \end{aligned}$$

Then let $R(j) = \mathbb{E}\left[n\, L_n(j)\right]$, we have

$$\begin{aligned} \lim_{n\to\infty} \mathbb{E}\left[n\, L_n\left(\hat{\jmath}_{\lambda_n}\right)\right] &= \lim_{n\to\infty} \sum_{j\in\mathscr{J}} \Pr\left(\hat{\jmath}_{\lambda_n} = j\right) R(j) = \lim_{n\to\infty} \sum_{j\in\mathscr{J}} \mathbb{1}_{j\in\mathscr{J}_0}\, \alpha^{\lambda_n}_{j-\jmath_0, K-\jmath_0}\, R(j) \\ &= \lim_{n\to\infty} \sum_{j\in\mathscr{J}_0} \alpha^{\lambda_n}_{j-\jmath_0, K-\jmath_0}\left(\sigma^2_\varepsilon j + \boldsymbol{\delta}^T\boldsymbol{\delta}\right) \\ &= \lim_{n\to\infty}\left[\boldsymbol{\delta}^T\boldsymbol{\delta} + \sigma^2_\varepsilon\left(\jmath_0 + \sum_{j=1}^{K-\jmath_0} j\, \alpha^{\lambda_n}_{j-\jmath_0, K-\jmath_0}\right)\right] \end{aligned}$$

□

From Theorem A-7 we have the following Corollary

Corollary A-3: *Under the assumptions of Theorem A-7 we have that*

$$\lim_{n\to\infty} \mathbb{E}\left[\jmath_{\lambda_n}\right] = \jmath_0 + \lim_{n\to\infty} \sum_{j=1}^{K-\jmath_0} j\, \alpha_{j-\jmath_0,K-\jmath_0}^{\lambda_n}.$$

Proof: The proof is straightforward from Theorem A-7:

$$\lim_{n\to\infty} \mathbb{E}\left[\hat{\jmath}_{\lambda_n}\right] = \lim_{n\to\infty} \sum_{j\in\mathscr{J}} \Pr\left(\hat{\jmath}_{\lambda_n} = j\right) j = \lim_{n\to\infty} \sum_{j\in\mathscr{J}} \mathbb{1}_{j\in\mathscr{J}_0}\, \alpha_{j-\jmath_0,K-\jmath_0}^{\lambda_n}\, j$$

$$= \lim_{n\to\infty} \sum_{j\in\mathscr{J}_0} \alpha_{j-\jmath_0,K-\jmath_0}^{\lambda_n}\, j = \jmath_0 + \lim_{n\to\infty} \sum_{j=1}^{K-\jmath_0} j\, \alpha_{j-\jmath_0,K-\jmath_0}^{\lambda_n}.$$

□

Remark G: *In the situation where $\lambda_n = 2$, we obtain from Corollary A-3 that*

$$\lim_{n\to\infty} \mathbb{E}\left[\hat{\jmath}_2\right] \leq \lim_{K-\jmath_0\to\infty} \lim_{n\to\infty} \mathbb{E}\left[\hat{\jmath}_2\right] \approx \jmath_0 + 0.111,$$

*while for the $C_p$ (and the* AIC*) we have that*

$$\lim_{n\to\infty} \mathbb{E}\left[\hat{\jmath}_{C_p}\right] \leq \lim_{K-\jmath_0\to\infty} \lim_{n\to\infty} \mathbb{E}\left[\hat{\jmath}_{C_p}\right] \approx \jmath_0 + 0.946$$

*(see Woodroofe, 1982 and Zhang, 1992 for details).*

From Corollary A-2 and Corollary A-3, we may define the asymptotic efficiency as

$$\lim_{n\to\infty} \frac{\mathbb{E}\left[n\, L_n\left(\hat{\jmath}_{\lambda_n}\right)\right]}{\mathbb{E}\left[n\, L_n\left(\jmath_0\right)\right]} = \lim_{n\to\infty} \frac{\boldsymbol{\delta}^T\boldsymbol{\delta} + \sigma_\varepsilon^2 \mathbb{E}\left[\hat{\jmath}_{\lambda_n}\right]}{\boldsymbol{\delta}^T\boldsymbol{\delta} + \sigma_\varepsilon^2 \jmath_0}.$$

In the case where $\boldsymbol{\delta} = \mathbf{0}$, we simply obtain

$$\lim_{n\to\infty} \frac{\mathbb{E}\left[n\, L_n\left(\hat{\jmath}_{\lambda_n}\right)\right]}{\mathbb{E}\left[n\, L_n\left(\jmath_0\right)\right]} = 1 + \lim_{n\to\infty} \frac{1}{\jmath_0} \sum_{j=1}^{K-\jmath_0} j\, \alpha_{j-\jmath_0,K-\jmath_0}^{\lambda_n}.$$

Clearly if $\lambda_n$ is such that it tends to infinity with $n$, the asymptotic efficiency tends to 1.

Remark H: *In the case where $\lambda_n = 2$, we have approximately that*

$$\lim_{n\to\infty} \frac{\mathbb{E}\left[n\, L_n\left(\hat{\jmath}_{\lambda_n}\right)\right]}{\mathbb{E}\left[n\, L_n\left(\jmath_0\right)\right]} \approx 1 + \lim_{n\to\infty} \frac{0.111}{\jmath_0}$$

*which is approximately equal to 1 for large values of $\jmath_0$.*

Proof of Theorem 5: Note that some elements of this theorem are closely related to Theorem 5.3 of Müller (2008).

First, we define for notational simplicity $B_n(j)$ as

$$B_n(j) = \frac{L_n(j)}{L_n(\jmath_0)}.$$

Second, by applying the law of total probability, we have for any given $\epsilon > 0$ that

$$\begin{aligned}\Pr\left(\left|B_n\left(\hat{\jmath}_{\lambda_n}\right)-1\right|\geq\epsilon\right) &= \Pr\left(\left|B_n\left(\hat{\jmath}_{\lambda_n}\right)-1\right|\geq\epsilon|\hat{\jmath}_{\lambda_n}\in\mathscr{J}_0\right)\Pr\left(\hat{\jmath}_{\lambda_n}\in\mathscr{J}_0\right)\\ &\quad+\Pr\left(\left|B_n\left(\hat{\jmath}_{\lambda_n}\right)-1\right|\geq\epsilon|\hat{\jmath}_{\lambda_n}\notin\mathscr{J}_0\right)\Pr\left(\hat{\jmath}\notin\mathscr{J}_0\right)\\ &\leq \Pr\left(\hat{\jmath}_{\lambda_n}\notin\mathscr{J}_0\right)+\Pr\left(\left|B_n\left(\hat{\jmath}_{\lambda_n}\right)-1\right|\geq\epsilon|\hat{\jmath}_{\lambda_n}\in\mathscr{J}_0\right).\end{aligned}$$

The result of Theorem 3 implies that $\lim_{n\to\infty}\Pr\left(\hat{\jmath}_{\lambda_n}\notin\mathscr{J}_0\right)=0$. So we only need to show that $\lim_{n\to\infty}\Pr\left(\left|B_n\left(\hat{\jmath}_{\lambda_n}\right)-1\right|\geq\epsilon|\hat{\jmath}_{\lambda_n}\in\mathscr{J}_0\right)=0$ to prove that $B_n\left(\hat{\jmath}_{\lambda_n}\right)\xrightarrow{\mathcal{P}}1$:

$$\begin{aligned}&\Pr\left(\left|B_n\left(\hat{\jmath}_{\lambda_n}\right)-1\right|\geq\epsilon\middle|\,\hat{\jmath}_{\lambda_n}\in\mathscr{J}_0\right)\\ &=\Pr\left(\frac{1}{nL_n\left(\jmath_0\right)}\left|\boldsymbol{\varepsilon}^T\left(\mathbf{S}_{\hat{\jmath}_{\lambda_n}}-\mathbf{S}_{\jmath_0}\right)\boldsymbol{\varepsilon}-\boldsymbol{\mu}^T\left(\mathbf{S}_{\hat{\jmath}_{\lambda_n}}-\mathbf{S}_{\jmath_0}\right)\boldsymbol{\mu}\right|\geq\epsilon\middle|\hat{\jmath}_{\lambda_n}\in\mathscr{J}_0\right)\\ &\leq\Pr\left(\boldsymbol{\varepsilon}^T\left(\mathbf{S}_{\hat{\jmath}_{\lambda_n}}-\mathbf{S}_{\jmath_0}\right)\boldsymbol{\varepsilon}+\boldsymbol{\mu}^T\left(\mathbf{S}_{\hat{\jmath}_{\lambda_n}}-\mathbf{S}_{\jmath_0}\right)\boldsymbol{\mu}\geq n\epsilon L_n\left(\jmath_0\right)\middle|\,\hat{\jmath}_{\lambda_n}\in\mathscr{J}_0\right)\\ &=\Pr\left(\boldsymbol{\varepsilon}^T\left(\mathbf{S}_{\hat{\jmath}_{\lambda_n}}-\mathbf{S}_{\jmath_0}\right)\boldsymbol{\varepsilon}\geq n\,a_j\middle|\,\hat{\jmath}_{\lambda_n}\in\mathscr{J}_0\right)\end{aligned}$$

where $a_j=\epsilon L_n\left(\jmath_0\right)-\frac{\upsilon}{n}$ and where $\upsilon=\boldsymbol{\mu}^T\left(\mathbf{S}_{\hat{\jmath}_{\lambda_n}}-\mathbf{S}_{\jmath_0}\right)\boldsymbol{\mu}$. Note that using (A-19) in Lemma A-2 we have that $\upsilon=\mathcal{O}\left(n\right)$. Using Markov's inequality yields to

$$\begin{aligned}\Pr\left(\left|B_n\left(\hat{\jmath}_{\lambda_n}\right)-1\right|\geq\epsilon\middle|\,\hat{\jmath}_{\lambda_n}\in\mathscr{J}_0\right)&\leq\Pr\left(\boldsymbol{\varepsilon}^T\left(\mathbf{S}_{\hat{\jmath}_{\lambda_n}}-\mathbf{S}_{\jmath_0}\right)\boldsymbol{\varepsilon}\geq na_j\middle|\,\hat{\jmath}_{\lambda_n}\in\mathscr{J}_0\right)\\ &\leq\frac{\mathbb{E}\left[\boldsymbol{\varepsilon}^T\left(\mathbf{S}_{\hat{\jmath}_{\lambda_n}}-\mathbf{S}_{\jmath_0}\right)\boldsymbol{\varepsilon}\middle|\,\hat{\jmath}_{\lambda_n}\in\mathscr{J}_0\right]}{na_j}\\ &=\frac{\sigma^2\left(\hat{\jmath}_{\lambda_n}-\jmath_0\right)}{n\left(\epsilon L_n\left(\jmath_0\right)-\frac{\upsilon}{n}\right)}.\end{aligned}$$

Finally, by taking the limit we can write

$$\begin{aligned}\lim_{n\to\infty}\Pr\left(\left|B_n\left(\hat{\jmath}_{\lambda_n}\right)-1\right|\geq\epsilon\right)&\leq\lim_{n\to\infty}\Pr\left(\hat{\jmath}_{\lambda_n}\notin\mathscr{J}_0\right)+\lim_{n\to\infty}\Pr\left(\left|B_n\left(\hat{\jmath}_{\lambda_n}\right)\right|\geq\epsilon|\hat{\jmath}_{\lambda_n}\in\mathscr{J}_0\right)\\ &\leq\lim_{n\to\infty}\frac{\sigma^2\left(\hat{\jmath}_{\lambda_n}-\jmath_0\right)}{n\left(\epsilon L_n\left(\jmath_0\right)-\frac{\upsilon}{n}\right)}\\ &=\lim_{n\to\infty}\frac{\sigma^2\left(\hat{\jmath}_{\lambda_n}-\jmath_0\right)}{n\left(\epsilon L_n\left(\jmath_0\right)-c\right)}=0\end{aligned}$$

where $c$ is a positive constant. Therefore, we have that $B_n\left(\hat{\jmath}_{\lambda_n}\right)\xrightarrow{\mathcal{P}}1$.

Next, we consider the case where $\lim_{n\to\infty}\lambda_n=\infty$. In such situations, we have that

$$\lim_{n\to\infty}\Pr\left(\hat{\jmath}_{\lambda_n}=j\right)=\lim_{n\to\infty}\mathbb{1}_{j\in\mathscr{J}_0}\,\alpha^{\lambda_n}_{j-\jmath_0,K-\jmath_0}=\mathbb{1}_{j=\jmath_0}$$

and thus $\Pr\left(\hat{\jmath}_{\lambda_n}=\jmath_0\right)\xrightarrow{\mathcal{P}}1$. □

## D.7 Proof of Theorem 6

Proof: At step $j$, the variable indexed by $k$ will be selected if

$$k=\underset{i=1,\ldots,K-j}{\operatorname{argmax}}\;||\hat{\mathbf{Y}}_j-\hat{\mathbf{Y}}^{(i)}_{j+1}||_2^2\,.$$

Let $\mathcal{K}_j = \{1, ..., K-j\}$ and $\mathcal{K}_j^\star$ denotes the set of indices in $\mathcal{K}_j$ corresponding to the $K-j$ unordered variables such that these variables are all associated to one of the first $\jmath_0$ columns of the matrix $\mathbf{X}^\star$. In order to verify Theorem 6 we have to prove that at step $j$, and assuming that $\mathcal{K}_j^\star$ is not empty, $k$ asymptotically belongs to $\mathcal{K}_j^\star$.

Let $k^\star \in \mathcal{K}_j^\star$, $k^\bullet \in \mathcal{K}_j \setminus \mathcal{K}_j^\star$ and

$$W = \frac{||\hat{\mathbf{Y}}_j - \hat{\mathbf{Y}}_{j+1}^{(k^\star)}||_2^2}{\sigma_\varepsilon^2} - \frac{||\hat{\mathbf{Y}}_j - \hat{\mathbf{Y}}_{j+1}^{(k^\bullet)}||_2^2}{\sigma_\varepsilon^2}.$$

Clearly, we have that

$$\frac{||\hat{\mathbf{Y}}_j - \hat{\mathbf{Y}}_{j+1}^{(k^\star)}||_2^2}{\sigma_\varepsilon^2} \sim \chi_1^2\left(\gamma_n^2\right) \text{ where } \gamma_n^2 = \frac{\boldsymbol{\mu}^T\left(\mathbf{S}_{j+1}^{(k^\star)} - \mathbf{S}_j\right)\boldsymbol{\mu}}{\sigma_\varepsilon^2}$$
$$\frac{||\hat{\mathbf{Y}}_j - \hat{\mathbf{Y}}_{j+1}^{(k^\bullet)}||_2^2}{\sigma_\varepsilon^2} \sim \chi_1^2$$

where $\mathbf{S}_{j+1}^{(k^\star)}$ denotes the "hat" matrix associated to $\hat{\mathbf{Y}}_{j+1}^{(k^\star)}$. Note that by (A-18) in Lemma A-2 we have that $\gamma_n^2 = \mathcal{O}(n)$. Let $Y_{1,\gamma_n} \sim \chi_1^2\left(\gamma_n^2\right)$ and $Z_1 \sim \mathcal{N}(0,1)$. We can always write that $Y_{1,\gamma_n} = \left(Z_1 + \gamma_n\right)^2 = Z_1^2 + \gamma_n^2 + 2Z_1\gamma_n$. The term $\sigma_\varepsilon^{-2}\,||\hat{\mathbf{Y}}_j - \hat{\mathbf{Y}}_{j+1}^{(k^\bullet)}||_2^2$ also follows a $\chi_1^2$ distribution and is independent from $\sigma^{-2}\,||\hat{\mathbf{Y}}_j - \hat{\mathbf{Y}}_{j+1}^{(k^\star)}||_2^2$ by Craig (1943, Theorem 2) (see also Appendix H) and therefore is also independent from $Z_1$. Let $Z_2 \sim \mathcal{N}(0,1)$ be independent from $Z_1$ and let

$$U = \frac{Z_1^2}{\sqrt{n}} - \frac{Z_2^2}{\sqrt{n}}.$$

By means of Chebyshev's inequality we can show that $U = o_p(1)$ (see Lemma A-3). Moreover, we also have that

$$\frac{2\gamma_n Z_1}{\sqrt{n}} \xrightarrow{\mathcal{L}} c\,Z_1.$$

where $c$ is a positive constant since $\gamma_n > 0$ and $\gamma_n = \mathcal{O}(\sqrt{n})$. By Slutsky's theorem we have that

$$\frac{2\gamma_n Z_1}{\sqrt{n}} + \frac{Z_1^2}{\sqrt{n}} - \frac{Z_2^2}{\sqrt{n}} \xrightarrow{\mathcal{L}} c\,Z_1.$$

We can rewrite $W$ as follows:

$$W = \sqrt{n}\left(\frac{2\gamma_n Z_1}{\sqrt{n}} + \frac{Z_1^2}{\sqrt{n}} - \frac{Z_2^2}{\sqrt{n}}\right) + \gamma_n^2.$$

Since $0 < \sigma_\varepsilon^2 < \infty$ we get

$$\lim_{n\to\infty} \Pr\left(||\hat{\mathbf{Y}}_j - \hat{\mathbf{Y}}_{j+1}^{(k^\star)}||_2^2 > ||\hat{\mathbf{Y}}_j - \hat{\mathbf{Y}}_{j+1}^{(k^\bullet)}||_2^2\right) = \lim_{n\to\infty} \Pr\left(W > 0\right).$$

Finally, since $n^{-1/2}W \neq o_p(1)$ and we also have that

$$\lim_{n\to\infty} \Pr\left(W > 0\right) = \lim_{n\to\infty} \Pr\left(\frac{W}{\sqrt{n}} > 0\right) = \lim_{n\to\infty} \Pr\left(cZ_1 + \frac{\gamma_n^2}{\sqrt{n}} > 0\right)$$
$$= \lim_{n\to\infty} \Pr\left(Z_1 > -c^\star\sqrt{n}\right) = 1$$

where $c^\star$ is a positive constant. Therefore, we have that, asymptotically, the probability that $k$ belongs to $\mathcal{K}_j^\star$ is one, which proves Theorem 6. □

# E Signal to noise ratio Theorem

Suppose that one wishes to compare two nested models $\mathcal{M}_1$ (with $k$ variables) and $\mathcal{M}_2$ (with $m \geq 1$ additional variables) using an estimator of the criterion C. We define the $\Delta\,\mathrm{C}$ as

$$\Delta\,\mathrm{C} = \widehat{\mathrm{C}}_{\mathcal{M}_2} - \widehat{\mathrm{C}}_{\mathcal{M}_1} \tag{A-27}$$

where $\widehat{\mathrm{C}}_{\mathcal{M}_j}$, $j = 1, 2$ denotes the values of $\widehat{\mathrm{C}}$ for model $\mathcal{M}_j$. Clearly, model $\mathcal{M}_1$ will be considered better in the case where $\Delta\,\mathrm{C} \leq 0$ while if $\Delta\,\mathrm{C} > 0$ model $\mathcal{M}_2$ will be considered preferable. The random variable $\Delta\,\mathrm{C}$ depends largely on the strength of the penalty function (i.e. the divergence optimism) and the SNR is a possible tool that can be used to study the decision rule (A-27). Following McQuarrie and Tsai (1998) we define the SNR as

$$\mathrm{SNR}\,(\mathrm{C}) = \frac{\mathbb{E}\,[\Delta\,\mathrm{C}]}{\sqrt{\mathrm{var}\,(\Delta\,\mathrm{C})}} \tag{A-28}$$

where the numerator of the SNR corresponds to the *signal* and the denominator to the *noise*. As explained in McQuarrie and Tsai (1998), while the signal depends primarily on the difference of divergence optimism of the criterion used, the noise depends mainly on the distribution of the difference of apparent errors. A model selection procedure with a small penalty function will tend to have a weak signal, a weak SNR and will be prone to overfit. Although a procedure with a larger SNR might overcome this difficulty, such an approach might tend to underfit. The AIC is an example of the former while the BIC is one of the latter. In the following theorem, the SNR of $\widehat{\mathrm{PDC}}$ is compared to the SNR of some of the model selection criteria presented in Table A-5 in Appendix F, assuming $\sigma_\varepsilon^2$ known. More general results (in the overfitting case) can be found in McQuarrie and Tsai (1998, Section 2.3).

Theorem A-8: *Consider the following three competing models from Setting A*

$$\begin{aligned}
\mathcal{M}_1 &: \; \boldsymbol{y} = \mathbf{X}^{(1)}\boldsymbol{\beta}_1 + \boldsymbol{\varepsilon}^\star \\
\mathcal{M}_2 &: \; \boldsymbol{y} = \mathbf{X}^{(1)}\boldsymbol{\beta}_1 + \mathbf{X}^{(2)}\boldsymbol{\beta}_2 + \boldsymbol{\varepsilon}' \\
\mathcal{M}_3 &: \; \boldsymbol{y} = \mathbf{X}^{(1)}\boldsymbol{\beta}_1 + \mathbf{X}^{(2)}\boldsymbol{\beta}_2 + \mathbf{X}^{(3)}\boldsymbol{\beta}_3 + \boldsymbol{\varepsilon}
\end{aligned}$$

*where $\mathbf{X}^{(1)}$, $\mathbf{X}^{(2)}$ and $\mathbf{X}^{(3)}$ are respectively $n \times r_1$, $n \times r_2$ and $n \times r_3$ matrices and the dimensions of $\boldsymbol{\beta}_1$, of $\boldsymbol{\beta}_2$ and of $\boldsymbol{\beta}_3$ are defined accordingly. We assume that the matrix $\begin{bmatrix}\mathbf{X}^{(1)} & \mathbf{X}^{(2)} & \mathbf{X}^{(3)}\end{bmatrix}$ is a non-random $n \times (r_1 + r_2 + r_3)$ full rank matrix. Let $\boldsymbol{Y} \sim \mathcal{N}\left(\mathbf{X}\boldsymbol{\beta}, \sigma_\varepsilon^2 \mathbf{I}\right)$ where $\sigma_\varepsilon^2$ is known. Let respectively $C_{p_j}$, $\mathrm{AIC}_j$, $\mathrm{BIC}_j$ and $\mathrm{HQ}_j$, $j = 1, 2, 3$ be the $C_p$, AIC, BIC and HQ for models $\mathcal{M}_j$, and $\widehat{\mathrm{PDC}}^{\lambda_n}_{j,j+1}$, $j = 1, 2$ be the $\widehat{\mathrm{PDC}}^{\lambda_n}$ based on models $\mathcal{M}_j$ and $\mathcal{M}_{j+1}$. Define the following criterion's differences for respectively the $C_p$, AIC, BIC, HQ and $\widehat{\mathrm{PDC}}$*

$$\begin{aligned}
\Delta C_{p_{2,1}} &= C_{p_2} - C_{p_1} \\
\Delta\,\mathrm{AIC}_{2,1} &= \mathrm{AIC}_2 - \mathrm{AIC}_1 \\
\Delta\,\mathrm{BIC}_{2,1} &= \mathrm{BIC}_2 - \mathrm{BIC}_1 \\
\Delta\,\mathrm{HQ}_{2,1} &= \mathrm{HQ}_2 - \mathrm{HQ}_1 \\
\Delta\widehat{\mathrm{PDC}}^{\lambda_n}_{2,1} &= \widehat{\mathrm{PDC}}^{\lambda_n}_{2,3} - \widehat{\mathrm{PDC}}^{\lambda_n}_{1,2}.
\end{aligned}$$

*Let also* $\kappa = \frac{\mathbf{d}^T\mathbf{d}}{\sigma_\varepsilon^2}$, $\rho = \frac{\mathbf{h}^T\mathbf{h}}{\sigma_\varepsilon^2}$, *with*

$$\begin{aligned}
\mathbf{d} &= \left(\mathbf{S}^{(12)} - \mathbf{S}^{(1)}\right)\mathbf{X}\boldsymbol{\beta},\\
\mathbf{h} &= \left(\mathbf{S}^{(123)} - \mathbf{S}^{(12)}\right)\mathbf{X}\boldsymbol{\beta},\\
\mathbf{S}^{(12)} &= \mathbf{X}^{(12)}\left(\mathbf{X}^{(12)T}\mathbf{X}^{(12)}\right)^{-1}\mathbf{X}^{(12)T},\\
\mathbf{S}^{(123)} &= \mathbf{X}^{(123)}\left(\mathbf{X}^{(123)T}\mathbf{X}^{(123)}\right)^{-1}\mathbf{X}^{(123)T},\\
\mathbf{X}^{(12)} &= \begin{bmatrix}\mathbf{X}^{(1)} & \mathbf{X}^{(2)}\end{bmatrix} \quad \textit{and}\\
\mathbf{X}^{(123)} &= \begin{bmatrix}\mathbf{X}^{(1)} & \mathbf{X}^{(2)} & \mathbf{X}^{(3)}\end{bmatrix}.
\end{aligned}$$

*Then*

$$\text{SNR}\left(\Delta C_{p_{2,1}}\right) = \frac{\sqrt{2}}{2}\frac{r_2 - \kappa}{\sqrt{r_2 + 2\kappa}} \tag{A-29}$$

$$\text{SNR}\left(\Delta\,\text{AIC}_{2,1}\right) = \frac{\sqrt{2}}{2}\frac{r_2 - \kappa}{\sqrt{r_2 + 2\kappa}} \tag{A-30}$$

$$\text{SNR}\left(\Delta\,\text{BIC}_{2,1}\right) = \frac{\sqrt{2}}{2}\frac{r_2\left(\log(n) - 1\right) - \kappa}{\sqrt{r_2 + 2\kappa}} \tag{A-31}$$

$$\text{SNR}\left(\Delta\,\text{HQ}_{2,1}\right) = \frac{\sqrt{2}}{2}\frac{r_2\left(2\log\left(\log(n)\right) - 1\right) - \kappa}{\sqrt{r_2 + 2\kappa}} \tag{A-32}$$

$$\text{SNR}\left(\Delta\widehat{\text{PDC}}_{2,1}^{\lambda_n}\right) = \frac{\sqrt{2}}{2}\frac{(r_2(\lambda_n - 1)r_3 + \rho - \kappa}{\sqrt{r_2 + r_3 + 2\left(\rho + \kappa\right)}}. \tag{A-33}$$

Proof: For Mallow's $C_p$, we can write

$$\begin{aligned}
\Delta C_{p_{2,1}} &= C_{p_2} - C_{p_1} = ||\boldsymbol{Y} - \hat{\boldsymbol{Y}}_2||_2^2 - ||\boldsymbol{Y} - \hat{\boldsymbol{Y}}||_2^2 + 2\sigma_\varepsilon^2 r_2\\
&= \boldsymbol{Y}^T\left(\mathbf{I} - \mathbf{S}^{(12)}\right)\boldsymbol{Y} - \boldsymbol{Y}^T\left(\mathbf{I} - \mathbf{S}^{(1)}\right)\boldsymbol{Y} + 2\sigma_\varepsilon^2 r_2\\
&= -\boldsymbol{Y}^T\left(\mathbf{S}^{(12)} - \mathbf{S}^{(1)}\right)\boldsymbol{Y} + 2\sigma_\varepsilon^2 r_2.
\end{aligned}$$

Clearly, the matrix $\left(\mathbf{S}^{(12)} - \mathbf{S}^{(1)}\right)$ is symmetric and idempotent and thus using (A-42) from Appendix H we have

$$\begin{aligned}
\mathbb{E}\left[\boldsymbol{Y}^T\left(\mathbf{S}^{(12)} - \mathbf{S}^{(1)}\right)\boldsymbol{Y}\right] &= \sigma_\varepsilon^2\left(r_2 + \kappa\right)\\
\text{var}\left(\boldsymbol{Y}^T\left(\mathbf{S}^{(12)} - \mathbf{S}^{(1)}\right)\boldsymbol{Y}\right) &= 2\sigma_\varepsilon^4\left(r_2 + 2\kappa\right).
\end{aligned} \tag{A-34}$$

The relation (A-34) enables to write,

$$\begin{aligned}
\mathbb{E}\left[\Delta C_{p_{2,1}}\right] &= \sigma_\varepsilon^2\left(r_2 - \kappa\right)\\
\text{var}\left(\Delta C_{p_{2,1}}\right) &= 2\sigma_\varepsilon^4\left(r_2 + 2\kappa\right).
\end{aligned}$$

Therefore, the SNR of Mallow's $C_p$ is given by

$$\mathrm{SNR}\left(\Delta C_{p_{2,1}}\right) = \frac{\sqrt{2}}{2}\frac{r_2 - \kappa}{\sqrt{r_2 + 2\kappa}}$$

which verifies (A-29).

The AIC, BIC and HQ are based on log-likelhood, i.e $n\log(2\pi) + n\log\left(\sigma_\varepsilon^2\right) + \frac{1}{\sigma_\varepsilon^2}||\boldsymbol{Y} - \hat{\boldsymbol{Y}}||_2^2$ and can be expressed as

$$\mathrm{Crit} = ||\boldsymbol{Y} - \hat{\boldsymbol{Y}}||_2^2 + \sigma_\varepsilon^2 \,\mathrm{penalty} + C$$

where $C$ does not depend on the candidate model and can therefore be neglected.

Since the penalty term of the AIC is $2p$, it is equivalent to the $C_p$ and therefore it has the same SNR, thus verifying (A-30).

For the BIC, where the penalty term is $\log(n)p$. We can write

$$\Delta\,\mathrm{BIC}_{2,1} = -\boldsymbol{Y}^T\left(\mathbf{S}^{(12)} - \mathbf{S}^{(1)}\right)\boldsymbol{Y} + \log(n)\sigma_\varepsilon^2 r_2.$$

By using (A-34) we obtain,

$$\begin{aligned}\mathbb{E}\left[\Delta\,\mathrm{BIC}_{2,1}\right] &= \sigma_\varepsilon^2\left(r_2\left(\log(n) - 1\right) - \kappa\right)\\ \mathrm{var}\left(\Delta\,\mathrm{BIC}_{2,1}\right) &= 2\sigma_\varepsilon^4\left(r_2 + 2\kappa\right).\end{aligned}$$

Therefore, the SNR of the BIC is given by

$$\mathrm{SNR}\left(\Delta\,\mathrm{BIC}_{2,1}\right) = \frac{\sqrt{2}}{2}\frac{r_2\left(\log(n) - 1\right) - \kappa}{\sqrt{r_2 + 2\kappa}}$$

which verifies (A-31).

For the HQ, where the penalty term is $2\log(\log(n))p$. We can write

$$\Delta\,\mathrm{HQ}_{2,1} = -\boldsymbol{Y}^T\left(\mathbf{S}^{(12)} - \mathbf{S}^{(1)}\right)\boldsymbol{Y} + 2\log\left(\log(n)\right)\sigma_\varepsilon^2 r_2.$$

Using (A-34) yields

$$\begin{aligned}\mathbb{E}\left[\Delta\,\mathrm{HQ}_{2,1}\right] &= \sigma_\varepsilon^2\left(r_2\left(2\log\left(\log(n)\right) - 1\right) - \kappa\right)\\ \mathrm{var}\left(\Delta\,\mathrm{HQ}_{2,1}\right) &= 2\sigma_\varepsilon^4\left(r_2 + 2\kappa\right).\end{aligned}$$

Therefore, the SNR of the HQ is given by

$$\mathrm{SNR}\left(\Delta\,\mathrm{HQ}_{2,1}\right) = \frac{\sqrt{2}}{2}\frac{r_2\left(2\log\left(\log(n)\right) - 1\right) - \kappa}{\sqrt{r_2 + 2\kappa}}$$

which verifies (A-32).

For the $\widehat{\mathrm{PDC}}$, we can write that

$$\Delta\widehat{\mathrm{PDC}}_{2,1}^{\lambda_n} = \boldsymbol{y}^T\left(\mathbf{S}^{(123)} - \mathbf{S}^{(12)}\right)\boldsymbol{y} - \boldsymbol{y}^T\left(\mathbf{S}^{(12)} - \mathbf{S}^{(1)}\right)\boldsymbol{y} + \lambda_n\sigma_\varepsilon^2 r_2.$$

Using (A-42) and (A-43) from Appendix H we have

$$\begin{aligned}
\mathbb{E}\left[\boldsymbol{y}^T\left(\mathbf{S}^{(123)}-\mathbf{S}^{(12)}\right)\boldsymbol{y}\right] &= \sigma_\varepsilon^2\left(r_3+\rho\right)\\
\mathbb{E}\left[\boldsymbol{y}^T\left(\mathbf{S}^{(12)}-\mathbf{S}^{(1)}\right)\boldsymbol{y}\right] &= \sigma_\varepsilon^2\left(r_2+\kappa\right)\\
\operatorname{var}\left[\boldsymbol{y}^T\left(\mathbf{S}^{(123)}-\mathbf{S}^{(12)}\right)\boldsymbol{y}\right] &= 2\sigma_\varepsilon^4\left(r_3+2\rho\right)\\
\operatorname{var}\left[\boldsymbol{y}^T\left(\mathbf{S}^{(12)}-\mathbf{S}^{(1)}\right)\boldsymbol{y}\right] &= 2\sigma_\varepsilon^4\left(r_2+2\kappa\right)\\
\operatorname{cov}\left(\boldsymbol{y}^T\left(\mathbf{S}^{(123)}-\mathbf{S}^{(12)}\right)\boldsymbol{y},\boldsymbol{y}^T\left(\mathbf{S}^{(12)}-\mathbf{S}^{(1)}\right)\boldsymbol{y}\right) &= 0.
\end{aligned} \tag{A-35}$$

So, by combining the results of (A-35), we have

$$\begin{aligned}
\mathbb{E}\left[\Delta\widehat{\mathrm{PDC}}_{2,1}^{\lambda_n}\right] &= \sigma_\varepsilon^2\left(r_2(\lambda_n-1)r_3+\rho-\kappa\right)\\
\operatorname{var}\left[\Delta\widehat{\mathrm{PDC}}_{2,1}^{\lambda_n}\right] &= 2\sigma_\varepsilon^4\left(r_2+r_3+2\left(\rho+\kappa\right)\right).
\end{aligned}$$

Finally, we obtain

$$\mathrm{SNR}\left(\Delta\widehat{\mathrm{PDC}}_{2,1}^{\lambda_n}\right)=\frac{\sqrt{2}}{2}\frac{\left(r_2(\lambda_n-1)r_3+\rho-\kappa\right.}{\sqrt{r_2+r_3+2\left(\rho+\kappa\right)}}$$

which verifies (A-33) and concludes the proof. □

To illustrate the results of Theorem A-8, we compute the SNR for the $C_p$ (or AIC), BIC, HQ and the following three versions of the $\widehat{\mathrm{PDC}}$ :

$$\begin{aligned}
\widehat{\mathrm{PDC}}_{j,j+1} &= \|\hat{\boldsymbol{y}}_j-\hat{\boldsymbol{y}}_{j+1}\|_2^2+2j\sigma^2\\
\widehat{\mathrm{PDC}}^{\bullet}_{j,j+1} &= \|\hat{\boldsymbol{y}}_j-\hat{\boldsymbol{y}}_{j+1}\|_2^2+\log\left(n\right)j\sigma^2\\
\widehat{\mathrm{PDC}}^{\star}_{j,j+1} &= \|\hat{\boldsymbol{y}}_j-\hat{\boldsymbol{y}}_{j+1}\|_2^2+2\log\left(\log\left(n\right)\right)j\sigma^2
\end{aligned} \tag{A-36}$$

to compare the following models:

$$\begin{aligned}
&\mathcal{M}_1 \quad Y_i=\alpha+\varepsilon_i\\
&\mathcal{M}_2 \quad Y_i=\alpha+\beta x_i+\varepsilon_i\\
&\mathcal{M}_3 \quad Y_i=\alpha+\beta x_i+\gamma z_i+\varepsilon_i
\end{aligned}$$

where $\varepsilon_i \overset{iid}{\sim} \mathcal{N}\left(0,\sigma_\varepsilon^2\right)$ with $\sigma_\varepsilon^2$ having a known value of 1. The $n=100$ values of $z_i$ are generated from a $\mathcal{N}\left(0,1\right)$ and the values of $x_i$ are equidistant in $[0\,,\,1.3]$. For $\beta$, we consider 100 different values equally spaced between 0 and 1.3 and set $\alpha=1$ and $\gamma=0$. Therefore, when $\beta=0$ the true model is $\mathcal{M}_1$ while for $\beta>0$ the true model is $\mathcal{M}_2$.

It can be noted that $\widehat{\mathrm{PDC}}^{\bullet}_{j,j+1}$ and $\widehat{\mathrm{PDC}}^{\star}_{j,j+1}$ in (A-36) are developed based on, respectively, the BIC and the HQ information criterion and are, unlike $\widehat{\mathrm{PDC}}_{j,j+1}$, consistent criteria (see Theorem 5).

The left panel of Figure A-5 shows the SNR of the $C_p$ (which in this case is equivalent to the SNR of the AIC), HQ, BIC, $\widehat{\mathrm{PDC}}$, $\widehat{\mathrm{PDC}}^{\star}$ and $\widehat{\mathrm{PDC}}^{\bullet}$ for selection between $\mathcal{M}_1$ and $\mathcal{M}_2$, as a function of $\beta$. Globally, it can be observed that the $\widehat{\text{PDC}}$ has a larger SNR than the $C_p$ and the same can be said for $\widehat{\text{PDC}}^\star$ and $\widehat{\text{PDC}}^\bullet$ with respect to the HQ and the BIC. As it can be seen in the right panel of Figure A-5 that a criterion with a high SNR tends to choose the smallest model $\mathcal{M}_1$ more frequently. Therefore, criteria of the PDC class select on average smaller models than the "classical" counterpart. In this example, the $\widehat{\text{PDC}}$ has a behavior that is very similar to the HQ information criterion and is a compromise between the behavior of the BIC and $C_p$.

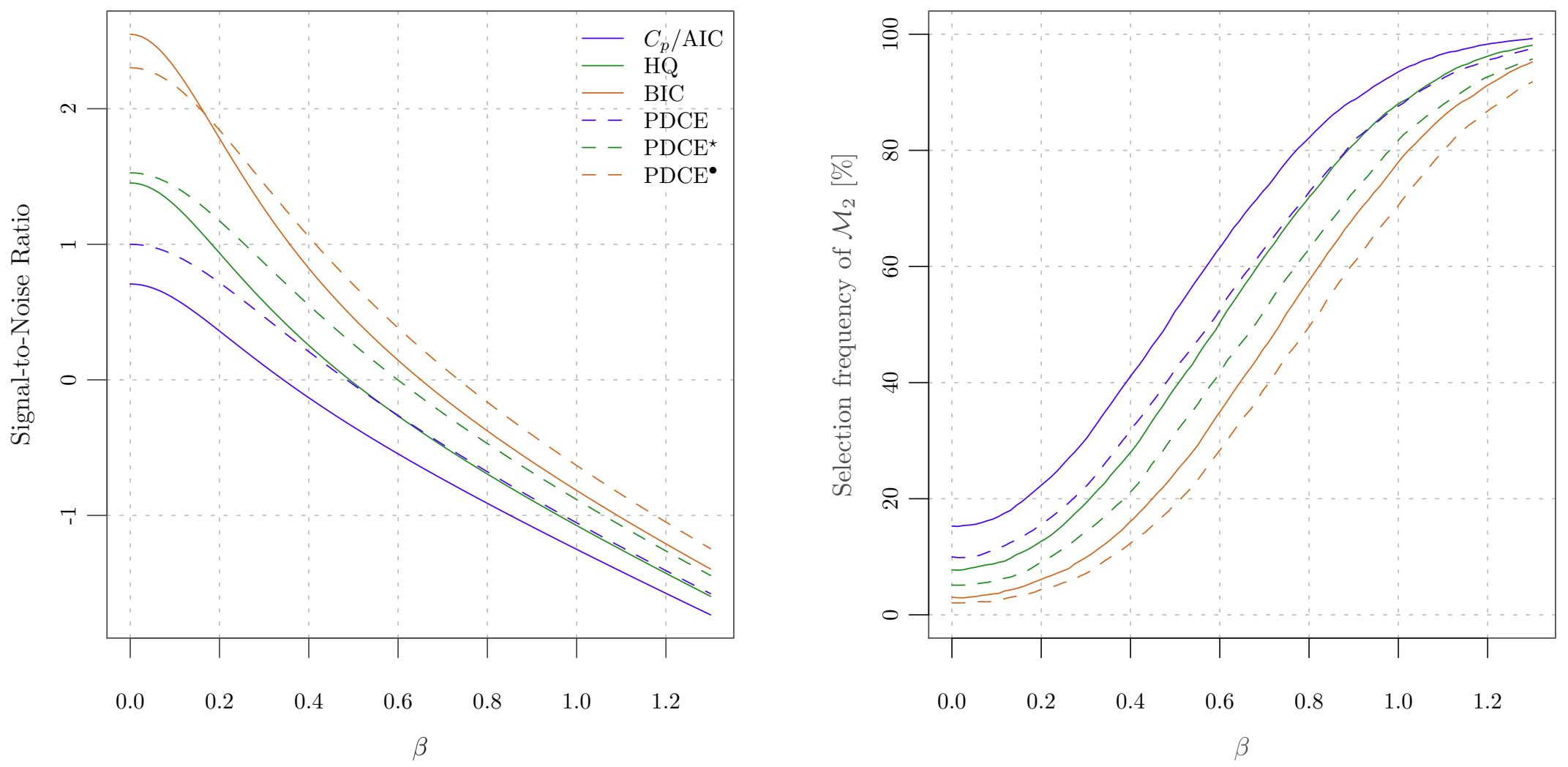

Figure A-5: (Left Panel:) SNR *of* $C_p$ *(or* AIC*),* HQ, BIC, $\widehat{\text{PDC}}$, $\widehat{\text{PDC}}^\star$ *and* $\widehat{\text{PDC}}^\bullet$ *for selection between* $\mathcal{M}_1$ *and* $\mathcal{M}_2$*, as a function of* $\beta$*.* (Right Panel:) *Percentage selection of model* $\mathcal{M}_2$ *using the* $C_p$ *(or* AIC*),* HQ*,* BIC*,* $\widehat{\text{PDC}}$, $\widehat{\text{PDC}}^\star$ *and* $\widehat{\text{PDC}}^\bullet$ *as a function of* $\beta$*, based on* $10^4$ *simulated samples.*

# F Criteria for model selection in linear regression

In Table A-5 are provided some of the most commonly used criteria for model selection in the context of linear regression models. For the lasso and adaptive methods, we used the following specifications:

- For the lasso we used the R function `lars` of the `lars` package. The shrinkage coefficient $\lambda$ was chosen by minimizing the $C_p$ statistic.
- For the elastic-net we used the R function `enet` of the `elasticnet` package. The shrinkage coefficient $\lambda_2$ ($L_2$ penalty) was chosen by tenfold cross validation and $\lambda_1$ ($L_1$ penalty) by minimizing the $C_p$ statistic.
- For the adaptive lasso we used the R function `adalasso` of the `parcor` which computes the solution based on tenfold cross validation. The initial weights for adaptive lasso are computed from a lasso fit.

Table A-5: *Some of the most commonly used criteria for model selection in linear regression models. $p$ and $n$ denote, respectively, the number of parameters in the candidate model and the sample size. $\hat{\sigma}_\varepsilon^2$ is the LSE and $\tilde{\sigma}_\varepsilon^2$ the MLE of $\sigma_\varepsilon^2$ for the candidate model. For Mallow's $C_p$, $\tilde{\sigma}_\star^2$ denotes one of the estimators of $\sigma_\varepsilon^2$ computed at the largest candidate model (or another "low-bias" model).*

| Criteria | Reference |
|---|---|
| $\text{FPE} = \hat{\sigma}_\varepsilon^2 \left(\frac{n+p}{n-p}\right)$ | Akaike (1969) |
| $\text{FPEu} = \tilde{\sigma}_\varepsilon^2 \left(\frac{n+p}{n-p}\right)$ | McQuarrie et al. (1997) |
| $C_p = \frac{n\,\hat{\sigma}_\varepsilon^2}{\tilde{\sigma}_\star^2} - n + 2p$ | Mallows (1973) |
| $\text{AIC} = \log\left(\hat{\sigma}_\varepsilon^2\right) + \frac{2(p+1)}{n}$ | Akaike (1974) |
| $\text{AICc} = \log\left(\hat{\sigma}_\varepsilon^2\right) + \frac{n+p}{n-p-2}$ | Hurvich and Tsai (1989) |
| $\text{AICu} = \log\left(\tilde{\sigma}_\varepsilon^2\right) + \frac{n+p}{n-p-2}$ | McQuarrie et al. (1997) |
| $\text{BIC} = \log\left(\hat{\sigma}_\varepsilon^2\right) + \frac{\log(n)\,p}{n}$ | Schwarz (1978) |
| $\text{HQ} = \log\left(\hat{\sigma}_\varepsilon^2\right) + \frac{2\,\log(\log(n))\,p}{n}$ | Hannan and Quinn (1979) |
| $\text{HQc} = \log\left(\hat{\sigma}_\varepsilon^2\right) + \frac{2\,\log(\log(n))\,p}{n-p-2}$ | McQuarrie and Tsai (1998) |

- For the MCP and SCAD penalized regression we used the R function `ncvreg` of the `ncvreg` package with its default settings. The coefficient $\lambda$ was chosen by tenfold cross validation using the function cv.ncvreg (of the package ncvreg)

# G The 4-parameters variance-gamma distribution

The PDF of a variance-gamma distribution is given by (see also e.g. Kotz et al. (2001) for a more detailed presentation of this distribution)

$$f_Z(z) = \exp\left(\beta\left(z-\mu\right)\right) \frac{\left(\alpha^2-\beta^2\right)^\xi \, |z-\mu|^{\xi-\frac{1}{2}} \, K_{\xi-\frac{1}{2}}\left(\alpha|x-\mu|\right)}{\sqrt{\pi}\,\Gamma\left(\lambda\right)\,\left(2\alpha\right)^{\xi-\frac{1}{2}}}$$

where $K_\xi\left(\cdot\right)$ denotes a modified Bessel function of the second kind and $\Gamma\left(\cdot\right)$ denotes the Gamma function. The parameters are such that $\mu \in \mathbb{R}$, $\alpha \in \mathbb{R}$, $\beta \in \mathbb{R}$ and $\xi \in \mathbb{R}^+$. The MGF of such distribution is given by:

$$M_Z(t) = \exp\left(\mu t\right)\left(\frac{\alpha^2-\beta^2}{\alpha^2-\left(\beta+t\right)^2}\right)^\xi. \tag{A-37}$$

We define the *modified* variance-gamma distribution as a standard variance-gamma but with the set of parameters $\mu = 0$, $\alpha = \frac{1}{2}$, $\beta = 0$ and $\xi = \frac{m}{2}$. Therefore the modified variance-gamma distribution only depends on the parameter $m$ and has the following PDF:

$$f_Z(z) = \frac{|z|^{\frac{1}{2}(m-1)} K_{\frac{1}{2}(m-1)}\left(\frac{1}{2}|z|\right)}{2^m\sqrt{\pi}\Gamma\left(\frac{m}{2}\right)}. \tag{A-38}$$

In the special case where $m = 1$, (A-38) can be simplified as

$$f_Z(z) = (2\pi)^{-1} K_0\left(\tfrac{1}{2}|z|\right). \tag{A-39}$$

Note that the function $K_0(z)$ has the following integral representation (see e.g. Abramowitz and Stegun, 1965, eq. 9.6.22, p. 376):

$$K_0(x) = \lim_{u\to\infty} \int_0^u \frac{\cos(xt)}{\sqrt{t^2+1}} dt, \quad x > 0. \tag{A-40}$$

# H Basic results on Quadratic Forms

Let $\boldsymbol{Y} \sim \mathcal{N}\left(\boldsymbol{\mu}, \sigma^2\mathbf{I}\right)$ be an $n$-vector with a spherical normal distribution and $\mathbf{A}$ be an $n \times n$ symmetric idempotent matrix with $\mathrm{rank}(\mathbf{A}) = r$. Then

$$\frac{\boldsymbol{Y}^T\mathbf{A}\boldsymbol{Y}}{\sigma^2} \sim \chi^2_r\left(\gamma^2\right) \quad \text{where } \gamma^2 = \frac{\boldsymbol{\mu}^T\mathbf{A}\boldsymbol{\mu}}{\sigma^2}. \tag{A-41}$$

A proof of (A-41) can, for example, be found in Bates (2011, proof of Theorem 7).

Moreover, if $\boldsymbol{Y} \sim \mathcal{N}(\mathbf{0}, \mathbf{V})$ and if there are two quadratic forms of $\boldsymbol{Y}$, say $\boldsymbol{Y}^T\mathbf{Q}\boldsymbol{Y}$ and $\boldsymbol{Y}^T\mathbf{P}\boldsymbol{Y}$ and both quadratic forms have a chi-square distribution as in (A-41), then the two quadratic forms are independent if $\mathbf{QVP} = \mathbf{0}$. The proof of this statement can be found in Craig (1943, Theorem 2).

If $\mathbf{Q}$ and $\mathbf{P}$ are symmetric $k \times k$ matrices and if $\boldsymbol{Y}$ is a random vector of length $k$ such that $\boldsymbol{\mu} = \mathbb{E}[\boldsymbol{Y}]$ and $\mathbf{V} = \mathrm{var}[\boldsymbol{Y}]$, then we have that

$$\begin{aligned} \mathbb{E}\left[\boldsymbol{Y}^T\mathbf{Q}\boldsymbol{Y}\right] &= \mathrm{tr}(\mathbf{QV}) + \boldsymbol{\mu}^T\mathbf{Q}\boldsymbol{\mu}, \\ \mathrm{var}\left[\boldsymbol{Y}^T\mathbf{Q}\boldsymbol{Y}\right] &= 2\,\mathrm{tr}(\mathbf{QVQV}) + 4\boldsymbol{\mu}^T\mathbf{QVQ}\boldsymbol{\mu}. \end{aligned} \tag{A-42}$$

In addition we have that

$$\mathrm{cov}\left(\boldsymbol{Y}^T\mathbf{Q}\boldsymbol{Y}, \boldsymbol{Y}^T\mathbf{P}\boldsymbol{Y}\right) = 2\,\mathrm{tr}(\mathbf{QVPV}) + 4\boldsymbol{\mu}^T\mathbf{QVP}\boldsymbol{\mu}. \tag{A-43}$$

# I Simulation results

We present here the results of the simulation studies 5.1, 5.2 and 5.3. The evaluation criteria are presented in Table 3 and the simulation results are reported in Tables A-6 to A-8.

Table A-6: *Performance of model selection criteria using the measures in Table 3. These include the* $\widehat{\text{PDC}}_{j,j+1}^{\lambda_n}$ *criteria in (16), with* $\lambda_n = 2$ *(PDC),* $\lambda_n = \log(n)$ *(*$\text{PDC}^{\bullet}$*),* $\lambda_n = \log(\log(n))$ *(*$\text{PDC}^{\star}$*), the stepwise forward FPE (*FPE*), FPEu (*FPEu*), AIC (*AIC*), AICc (*AICc*), AICu (*AICu*), BIC (*BIC*), HQ (*HQ*), HQc (*HQc*), and the full model with the LSE (*LSE*), together with the lasso (*lasso*), elastic-net (*enet*), adaptive lasso (a-lasso), MCP (MCP) and SCAD (SCAD) (see Appendix F for more details on the chosen criteria).* 500 *samples are simulated under the correct model as presented in Simulation 5.1 (*$p = 2, K = 60$*). The numbers in parentheses for the columns Med*$(PE_{\boldsymbol{y}})$ *and Med*$(MSE_{\boldsymbol{\beta}})$ *are the corresponding standard errors estimated by using the bootstrap with* $B = 500$ *resamplings. The numbers in superscript indicate the (average) ranked performance for each evaluation criterion (before rounding).*

| | Med $(\text{PE}_{\boldsymbol{y}})$ | Med $(\text{MSE}_{\boldsymbol{\beta}})$ | Cor. [%] | Inc. [%] | true+ | false+ | NbReg |
|---|---|---|---|---|---|---|---|
| LS | $4.29\ (8.0 \cdot 10^{-2})\ ^{13}$ | $5.59 \cdot 10^{0}\ (1.2 \cdot 10^{-1})\ ^{13}$ | $0.0\ ^{11.5}$ | $100\ ^{7}$ | $2.0\ ^{7}$ | $58.0\ ^{13}$ | $60.0\ ^{13}$ |
| FPE | $2.12\ (2.8 \cdot 10^{-2})\ ^{11}$ | $1.60 \cdot 10^{0}\ (5.4 \cdot 10^{-2})\ ^{11}$ | $0.0\ ^{11.5}$ | $100\ ^{7}$ | $2.0\ ^{7}$ | $20.4\ ^{11}$ | $22.4\ ^{11}$ |
| FPEu | $1.52\ (2.3 \cdot 10^{-2})\ ^{8}$ | $6.41 \cdot 10^{-1}\ (2.6 \cdot 10^{-2})\ ^{8}$ | $0.6\ ^{8.5}$ | $100\ ^{7}$ | $2.0\ ^{7}$ | $8.8\ ^{7}$ | $10.8\ ^{7}$ |
| AIC | $2.30\ (3.6 \cdot 10^{-2})\ ^{12}$ | $1.90 \cdot 10^{0}\ (8.0 \cdot 10^{-2})\ ^{12}$ | $0.0\ ^{11.5}$ | $100\ ^{7}$ | $2.0\ ^{7}$ | $24.0\ ^{12}$ | $26.0\ ^{12}$ |
| AICc | $1.62\ (2.0 \cdot 10^{-2})\ ^{10}$ | $7.68 \cdot 10^{-1}\ (1.9 \cdot 10^{-2})\ ^{10}$ | $0.0\ ^{11.5}$ | $100\ ^{7}$ | $2.0\ ^{7}$ | $10.2\ ^{9}$ | $12.2\ ^{9}$ |
| AICu | $1.37\ (1.1 \cdot 10^{-2})\ ^{7}$ | $4.13 \cdot 10^{-1}\ (1.6 \cdot 10^{-2})\ ^{7}$ | $0.8\ ^{6.5}$ | $100\ ^{7}$ | $2.0\ ^{7}$ | $5.6\ ^{6}$ | $7.6\ ^{6}$ |
| BIC | $1.25\ (1.5 \cdot 10^{-2})\ ^{5}$ | $2.71 \cdot 10^{-1}\ (1.2 \cdot 10^{-2})\ ^{5}$ | $9.4\ ^{4}$ | $100\ ^{7}$ | $2.0\ ^{7}$ | $3.5\ ^{4}$ | $5.5\ ^{4}$ |
| HQ | $1.60\ (2.9 \cdot 10^{-2})\ ^{9}$ | $7.66 \cdot 10^{-1}\ (3.7 \cdot 10^{-2})\ ^{9}$ | $0.6\ ^{8.5}$ | $100\ ^{7}$ | $2.0\ ^{7}$ | $11.2\ ^{10}$ | $13.2\ ^{10}$ |
| HQc | $1.36\ (9.7 \cdot 10^{-3})\ ^{6}$ | $3.89 \cdot 10^{-1}\ (1.5 \cdot 10^{-2})\ ^{6}$ | $0.8\ ^{6.5}$ | $100\ ^{7}$ | $2.0\ ^{7}$ | $5.3\ ^{5}$ | $7.3\ ^{5}$ |
| lasso | $1.16\ (7.4 \cdot 10^{-3})\ ^{4}$ | $1.46 \cdot 10^{-1}\ (5.0 \cdot 10^{-3})\ ^{4}$ | $3.2\ ^{5}$ | $100\ ^{7}$ | $2.0\ ^{7}$ | $9.3\ ^{8}$ | $11.3\ ^{8}$ |
| PDC | $1.06\ (3.8 \cdot 10^{-3})\ ^{3}$ | $5.45 \cdot 10^{-2}\ (6.0 \cdot 10^{-3})\ ^{3}$ | $56.4\ ^{3}$ | $100\ ^{7}$ | $2.0\ ^{7}$ | $0.6\ ^{3}$ | $2.6\ ^{3}$ |
| $\text{PDC}^{\star}$ | $1.05\ (3.4 \cdot 10^{-3})\ ^{2}$ | $1.99 \cdot 10^{-2}\ (1.4 \cdot 10^{-3})\ ^{1.5}$ | $73.8\ ^{2}$ | $100\ ^{7}$ | $2.0\ ^{7}$ | $0.3\ ^{2}$ | $2.3\ ^{2}$ |
| $\text{PDC}^{\bullet}$ | $1.03\ (3.6 \cdot 10^{-3})\ ^{1}$ | $1.99 \cdot 10^{-2}\ (1.5 \cdot 10^{-3})\ ^{1.5}$ | $89.0\ ^{1}$ | $100\ ^{7}$ | $2.0\ ^{7}$ | $0.1\ ^{1}$ | $2.1\ ^{1}$ |
| enet | $1.19\ (2.1 \cdot 10^{-2})$ | $1.68 \cdot 10^{-1}\ (2.1 \cdot 10^{-2})$ | 18.2 | 64 | 1.3 | 3.6 | 4.9 |
| a-lasso | $1.04\ (4.2 \cdot 10^{-3})$ | $3.17 \cdot 10^{-2}\ (1.6 \cdot 10^{-3})$ | 65.6 | 100 | 2.0 | 0.9 | 2.9 |
| MCP | $1.04\ (3.4 \cdot 10^{-3})$ | $2.37 \cdot 10^{-2}\ (1.7 \cdot 10^{-3})$ | 67.2 | 100 | 2.0 | 1.1 | 3.1 |
| SCAD | $1.04\ (3.7 \cdot 10^{-3})$ | $2.60 \cdot 10^{-2}\ (1.5 \cdot 10^{-3})$ | 33.8 | 100 | 2.0 | 2.6 | 4.6 |

Table A-7: *Performance of model selection criteria using the measures in Table 3. These include the* $\widehat{\mathrm{PDC}}_{j,j+1}^{\lambda_n}$ *criteria in (16), with* $\lambda_n = 2$ *(PDC),* $\lambda_n = \log(n)$ *(*$\mathrm{PDC}^{\bullet}$*),* $\lambda_n = \log(\log(n))$ *(*$\mathrm{PDC}^{\star}$*), the stepwise forward FPE (*FPE*), FPEu (*FPEu*), AIC (*AIC*), AICc (*AICc*), AICu (*AICu*), BIC (*BIC*), HQ (*HQ*), HQc (*HQc*), and the full model with the LSE (*LSE*), together with the lasso (*lasso*), elastic-net (*enet*), adaptive lasso (a-lasso), MCP (MCP) and SCAD (SCAD) (see Appendix F for more details on the chosen criteria).* 500 *samples are simulated under the correct model as presented in Simulation 5.2 (*$p = 5, K = 10$*). The numbers in parentheses for the columns Med*$(PE_{\boldsymbol{y}})$ *and Med*$(MSE_{\boldsymbol{\beta}})$ *are the corresponding standard errors estimated by using the bootstrap with* $B = 500$ *resamplings. The numbers in superscript indicate the (average) ranked performance for each evaluation criterion (before rounding).*

| | Med $(\mathrm{PE}_{\boldsymbol{y}})$ | Med $(\mathrm{MSE}_{\boldsymbol{\beta}})$ | Cor. [%] | Inc. [%] | true+ | false+ | NbReg |
|---|---|---|---|---|---|---|---|
| LS | $1.12\ (4.3 \cdot 10^{-3})\ ^{2}$ | $3.44 \cdot 10^{-1}\ (1.7 \cdot 10^{-2})\ ^{2}$ | $0.0\ ^{11}$ | $100.0\ ^{1}$ | $5.0\ ^{1}$ | $5.0\ ^{13}$ | $10.0\ ^{13}$ |
| FPE | $1.14\ (3.9 \cdot 10^{-3})\ ^{3.5}$ | $4.70 \cdot 10^{-1}\ (1.5 \cdot 10^{-2})\ ^{3.5}$ | $2.2\ ^{1.5}$ | $7.2\ ^{4}$ | $2.9\ ^{4}$ | $1.8\ ^{10}$ | $4.7\ ^{10}$ |
| FPEu | $1.16\ (5.7 \cdot 10^{-3})\ ^{6.5}$ | $5.29 \cdot 10^{-1}\ (1.3 \cdot 10^{-2})\ ^{7}$ | $0.6\ ^{5.5}$ | $1.8\ ^{6.5}$ | $2.3\ ^{6}$ | $1.5\ ^{8}$ | $3.9\ ^{8}$ |
| AIC | $1.14\ (4.0 \cdot 10^{-3})\ ^{3.5}$ | $4.70 \cdot 10^{-1}\ (1.5 \cdot 10^{-2})\ ^{3.5}$ | $2.2\ ^{1.5}$ | $7.4\ ^{3}$ | $2.9\ ^{3}$ | $1.8\ ^{11}$ | $4.7\ ^{11}$ |
| AICc | $1.15\ (4.5 \cdot 10^{-3})\ ^{5}$ | $4.88 \cdot 10^{-1}\ (1.2 \cdot 10^{-2})\ ^{5}$ | $1.0\ ^{4}$ | $3.8\ ^{5}$ | $2.7\ ^{5}$ | $1.7\ ^{9}$ | $4.4\ ^{9}$ |
| AICu | $1.16\ (5.6 \cdot 10^{-3})\ ^{8.5}$ | $5.30 \cdot 10^{-1}\ (1.4 \cdot 10^{-2})\ ^{9}$ | $0.4\ ^{7.5}$ | $1.4\ ^{8}$ | $2.3\ ^{8}$ | $1.5\ ^{6}$ | $3.8\ ^{6}$ |
| BIC | $1.18\ (5.9 \cdot 10^{-3})\ ^{11}$ | $5.85 \cdot 10^{-1}\ (1.6 \cdot 10^{-2})\ ^{11}$ | $0.0\ ^{11}$ | $0.0\ ^{11.5}$ | $1.9\ ^{11}$ | $1.3\ ^{3}$ | $3.2\ ^{3}$ |
| HQ | $1.16\ (5.2 \cdot 10^{-3})\ ^{6.5}$ | $5.29 \cdot 10^{-1}\ (1.4 \cdot 10^{-2})\ ^{7}$ | $0.6\ ^{5.5}$ | $1.8\ ^{6.5}$ | $2.3\ ^{7}$ | $1.5\ ^{7}$ | $3.9\ ^{7}$ |
| HQc | $1.16\ (5.5 \cdot 10^{-3})\ ^{8.5}$ | $5.29 \cdot 10^{-1}\ (1.5 \cdot 10^{-2})\ ^{7}$ | $0.4\ ^{7.5}$ | $1.0\ ^{9}$ | $2.2\ ^{9}$ | $1.5\ ^{5}$ | $3.7\ ^{5}$ |
| lasso | $1.10\ (4.3 \cdot 10^{-3})\ ^{1}$ | $2.53 \cdot 10^{-1}\ (1.1 \cdot 10^{-2})\ ^{1}$ | $2.0\ ^{3}$ | $46.0\ ^{2}$ | $4.2\ ^{2}$ | $2.6\ ^{12}$ | $6.9\ ^{12}$ |
| PDC | $1.18\ (6.0 \cdot 10^{-3})\ ^{10}$ | $5.75 \cdot 10^{-1}\ (1.8 \cdot 10^{-2})\ ^{10}$ | $0.0\ ^{11}$ | $0.0\ ^{11.5}$ | $1.9\ ^{10}$ | $1.3\ ^{4}$ | $3.2\ ^{4}$ |
| $\mathrm{PDC}^{\star}$ | $1.21\ (7.3 \cdot 10^{-3})\ ^{12}$ | $7.13 \cdot 10^{-1}\ (1.4 \cdot 10^{-2})\ ^{12.5}$ | $0.0\ ^{11}$ | $0.0\ ^{11.5}$ | $1.6\ ^{12}$ | $1.2\ ^{2}$ | $2.8\ ^{2}$ |
| $\mathrm{PDC}^{\bullet}$ | $1.25\ (8.9 \cdot 10^{-3})\ ^{13}$ | $7.13 \cdot 10^{-1}\ (1.4 \cdot 10^{-2})\ ^{12.5}$ | $0.0\ ^{11}$ | $0.0\ ^{11.5}$ | $1.3\ ^{13}$ | $1.1\ ^{1}$ | $2.4\ ^{1}$ |
| enet | $1.11\ (4.6 \cdot 10^{-3})$ | $2.79 \cdot 10^{-1}\ (9.5 \cdot 10^{-3})$ | 1.8 | 40.6 | 4.0 | 2.5 | 6.5 |
| a-lasso | $1.11\ (4.2 \cdot 10^{-3})$ | $2.90 \cdot 10^{-1}\ (1.1 \cdot 10^{-2})$ | 5.0 | 35.0 | 3.9 | 2.0 | 5.8 |
| MCP | $1.14\ (5.4 \cdot 10^{-3})$ | $4.23 \cdot 10^{-1}\ (2.0 \cdot 10^{-2})$ | 5.2 | 35.8 | 3.5 | 2.2 | 5.7 |
| SCAD | $1.14\ (5.6 \cdot 10^{-3})$ | $4.16 \cdot 10^{-1}\ (1.9 \cdot 10^{-2})$ | 3.8 | 38.0 | 3.7 | 2.3 | 6.1 |

TABLE A-8: *Performance of model selection criteria using the measures in Table 3. These include the* $\widehat{\mathrm{PDC}}^{\lambda_n}_{j,j+1}$ *criteria in (16), with* $\lambda_n = 2$ *(PDC),* $\lambda_n = \log(n)$ *(*$\mathrm{PDC}^{\bullet}$*),* $\lambda_n = \log(\log(n))$ *(*$\mathrm{PDC}^{\star}$*), the stepwise forward FPE (*FPE*), FPEu (*FPEu*), AIC (*AIC*), AICc (*AICc*), AICu (*AICu*), BIC (*BIC*), HQ (*HQ*), HQc (*HQc*), and the full model with the LSE (*LSE*), together with the lasso (*lasso*), elastic-net (*enet*), adaptive lasso (a-lasso), MCP (MCP) and SCAD (SCAD) (see Appendix F for more details on the chosen criteria).* 500 *samples are simulated under the correct model as presented in Simulation 5.3 (*$p = 14, K = 50$*). The numbers in parentheses for the columns Med(*$PE_{\boldsymbol{y}}$*) and Med(*$MSE_{\boldsymbol{\beta}}$*) are the corresponding standard errors estimated by using the bootstrap with* $B = 500$ *resamplings. The numbers in superscript indicate the (average) ranked performance for each evaluation criterion (before rounding).*

| | Med ($\mathrm{PE}_{\boldsymbol{y}}$) | Med ($\mathrm{MSE}_{\boldsymbol{\beta}}$) | Cor. [%] | Inc. [%] | true+ | false+ | NbReg |
|---|---|---|---|---|---|---|---|
| LS | $7.76\ (8.3 \cdot 10^{-2})^{13}$ | $6.27 \cdot 10^{0}\ (1.5 \cdot 10^{-1})^{13}$ | $0^{7}$ | $100.0^{1}$ | $14.0^{1}$ | $36.0^{13}$ | $50.0^{13}$ |
| FPE | $6.13\ (5.1 \cdot 10^{-2})^{11}$ | $3.12 \cdot 10^{0}\ (1.1 \cdot 10^{-1})^{11}$ | $0^{7}$ | $0.2^{4}$ | $9.8^{4}$ | $9.0^{10}$ | $18.8^{10}$ |
| FPEu | $5.49\ (4.5 \cdot 10^{-2})^{8}$ | $2.00 \cdot 10^{0}\ (9.0 \cdot 10^{-2})^{8}$ | $0^{7}$ | $0.0^{9}$ | $9.1^{7}$ | $4.3^{7}$ | $13.4^{7}$ |
| AIC | $6.17\ (5.2 \cdot 10^{-2})^{12}$ | $3.26 \cdot 10^{0}\ (1.1 \cdot 10^{-1})^{12}$ | $0^{7}$ | $0.4^{3}$ | $9.9^{3}$ | $9.5^{11}$ | $19.3^{11}$ |
| AICc | $5.64\ (4.2 \cdot 10^{-2})^{10}$ | $2.26 \cdot 10^{0}\ (7.1 \cdot 10^{-2})^{10}$ | $0^{7}$ | $0.0^{9}$ | $9.3^{5}$ | $5.3^{9}$ | $14.6^{9}$ |
| AICu | $5.24\ (4.7 \cdot 10^{-2})^{7}$ | $1.51 \cdot 10^{0}\ (5.0 \cdot 10^{-2})^{7}$ | $0^{7}$ | $0.0^{9}$ | $8.8^{8}$ | $2.8^{6}$ | $11.6^{6}$ |
| BIC | $5.00\ (3.3 \cdot 10^{-2})^{3}$ | $1.25 \cdot 10^{0}\ (3.1 \cdot 10^{-2})^{3}$ | $0^{7}$ | $0.0^{9}$ | $8.5^{10}$ | $1.8^{4}$ | $10.4^{4}$ |
| HQ | $5.57\ (4.6 \cdot 10^{-2})^{9}$ | $2.09 \cdot 10^{0}\ (7.8 \cdot 10^{-2})^{9}$ | $0^{7}$ | $0.0^{9}$ | $9.2^{6}$ | $4.8^{8}$ | $13.9^{8}$ |
| HQc | $5.16\ (4.8 \cdot 10^{-2})^{6}$ | $1.43 \cdot 10^{0}\ (4.9 \cdot 10^{-2})^{6}$ | $0^{7}$ | $0.0^{9}$ | $8.7^{9}$ | $2.5^{5}$ | $11.2^{5}$ |
| lasso | $5.06\ (2.6 \cdot 10^{-2})^{5}$ | $1.20 \cdot 10^{0}\ (3.7 \cdot 10^{-2})^{2}$ | $0^{7}$ | $1.0^{2}$ | $10.2^{2}$ | $9.9^{12}$ | $20.0^{12}$ |
| PDC | $4.83\ (3.0 \cdot 10^{-2})^{1}$ | $9.45 \cdot 10^{-1}\ (4.3 \cdot 10^{-2})^{1}$ | $0^{7}$ | $0.0^{9}$ | $7.9^{11}$ | $0.5^{3}$ | $8.5^{3}$ |
| $\mathrm{PDC}^{\star}$ | $4.84\ (4.0 \cdot 10^{-2})^{2}$ | $1.41 \cdot 10^{0}\ (4.8 \cdot 10^{-2})^{4.5}$ | $0^{7}$ | $0.0^{9}$ | $7.7^{12}$ | $0.2^{2}$ | $7.9^{2}$ |
| $\mathrm{PDC}^{\bullet}$ | $5.04\ (6.0 \cdot 10^{-2})^{4}$ | $1.41 \cdot 10^{0}\ (4.6 \cdot 10^{-2})^{4.5}$ | $0^{7}$ | $0.0^{9}$ | $7.2^{13}$ | $0.1^{1}$ | $7.4^{1}$ |
| enet | $5.13\ (2.3 \cdot 10^{-2})$ | $1.32 \cdot 10^{0}\ (3.3 \cdot 10^{-2})$ | 0 | 0.2 | 8.8 | 3.9 | 12.7 |
| a-lasso | $4.77\ (2.4 \cdot 10^{-2})$ | $9.35 \cdot 10^{-1}\ (3.1 \cdot 10^{-2})$ | 0 | 0.0 | 8.6 | 2.2 | 10.9 |
| MCP | $4.87\ (3.0 \cdot 10^{-2})$ | $1.08 \cdot 10^{0}\ (4.1 \cdot 10^{-2})$ | 0 | 0.0 | 8.8 | 2.7 | 11.4 |
| SCAD | $4.89\ (2.9 \cdot 10^{-2})$ | $1.17 \cdot 10^{0}\ (5.5 \cdot 10^{-2})$ | 0 | 0.0 | 9.6 | 5.3 | 14.8 |